\documentclass[11pt]{article}

\usepackage[preprint]{acl}

\usepackage{times}
\usepackage{latexsym}

\usepackage[T1]{fontenc}
\usepackage[utf8]{inputenc}

\usepackage{microtype}

\usepackage{inconsolata}

\usepackage{graphicx}

\usepackage{tikz}
\usepackage{amsmath}
\usepackage{algorithm}
\usepackage{algorithmic}
\usepackage{amssymb}
\usepackage{booktabs}
\usepackage{multirow} 
\usepackage{subcaption}
\usepackage{amsfonts}
\usepackage[most]{tcolorbox}
\tcbset{
  reviewbox/.style={
    enhanced,
    colback=gray!5,
    colframe=black, 
    boxrule=0.8pt,
    arc=1.5mm,
    auto outer arc,
    boxsep=5pt,
    left=6pt,
    right=6pt,
    top=6pt,
    bottom=6pt,
    fonttitle=\bfseries,
    before skip=10pt,
    after skip=10pt,
  }
}
\usepackage{xurl}
\usepackage{float}

\title{Topic-Based Watermarks for Large Language Models}

\author{
 \textbf{Alexander Nemecek\textsuperscript{1}},
 \textbf{Yuzhou Jiang\textsuperscript{1}},
 \textbf{Erman Ayday\textsuperscript{1}}
\\
 \textsuperscript{1}Case Western Reserve University
 \\
  \texttt{\{ajn98, yxj466, exa208\}@case.edu}
}

\begin{document}
\maketitle
\begin{abstract}
The indistinguishability of large language model (LLM) output from human-authored content poses significant challenges, raising concerns about potential misuse of AI-generated text and its influence on future model training. Watermarking algorithms offer a viable solution by embedding detectable signatures into generated text. However, existing watermarking methods often involve trade-offs among attack robustness, generation quality, and additional overhead such as specialized frameworks or complex integrations. We propose a lightweight, topic-guided watermarking scheme for LLMs that partitions the vocabulary into topic-aligned token subsets. Given an input prompt, the scheme selects a relevant topic-specific token list, effectively ``green-listing'' semantically aligned tokens to embed robust marks while preserving fluency and coherence. Experimental results across multiple LLMs and state-of-the-art benchmarks demonstrate that our method achieves text quality comparable to industry-leading systems and simultaneously improves watermark robustness against paraphrasing and lexical perturbation attacks, with minimal performance overhead. Our approach avoids reliance on additional mechanisms beyond standard text generation pipelines, enabling straightforward adoption and suggesting a practical path toward globally consistent watermarking of AI-generated content. Our code is available at \url{https://github.com/ANCP2021/Topic-Based-Watermarks}.
\end{abstract}

\section{Introduction}\label{introduction}
The rapid expansion of Large Language Model (LLM) capabilities has led to unprecedented accuracy and fluency in tasks such as text generation, summarization, and dialogue. Models like OpenAI's ChatGPT~\cite{openai2022chatgpt} and Google's Gemini~\cite{deepmind2024gemini} can produce text nearly indistinguishable from human-authored content. While these advancements offer significant benefits across multiple domains, they also pose security and ethical challenges. One central issue is the misuse of LLM-generated text for malicious purposes, such as misinformation, copyright infringement, or plagiarism~\cite{chen2024combating, mueller2024llms, lee2023language}. Additionally, many large-scale language models are trained on massive corpora scraped from the web, and the prevalence of LLM-generated data raises concerns about ``model collapse,'' wherein repeatedly ingesting AI-generated text as training data leads to a gradual erosion in quality~\cite{shumailov2024ai}.

In response to these growing concerns, the research community has focused on methods for reliably attributing text to its source, specifically, to distinguish LLM-generated from human-authored text~\cite{li2021surveytextclassificationshallow, openai2023aiclassifier, mitchell2023detectgpt}. Early approaches to text attribution primarily involved training classification-based detection methods on labeled corpora of human- and machine-generated text. While such classifiers can achieve respectable accuracy under controlled conditions, they are often susceptible to adversarial paraphrasing or stylistic alterations~\cite{liang2023gpt}. Furthermore, these methods rely on maintaining large, curated training sets that reflect the rapidly evolving landscape of LLMs which in turn pose substantial scalability challenges.

Researchers have explored \textit{watermarking} as a complementary or alternative solution. Rather than detecting AI-authored text post-hoc, watermarking algorithms embed a detectable signature into text during the generation process~\cite{kirchenbauer2023watermark, aaronson2023watermarking}. By introducing controlled token-level modifications, watermarks can remain identifiable even after moderate transformations. However, existing approaches face trade-offs where methods prioritizing computational efficiency lack robustness to paraphrasing~\cite{dathathri2024scalable}, while more robust techniques require costly architectural modifications or multiple inference passes that degrade practical deployability~\cite{huo2024tokenspecificwatermarkingenhanceddetectability, kuditipudi2024robustdistortionfreewatermarkslanguage}.

\textbf{Our Contributions.}
We propose a \textit{lightweight, topic-based watermarking (TBW) scheme} that addresses these limitations by integrating semantic information into the watermarking process while preserving generation pipeline simplicity. Rather than randomly partitioning the vocabulary, we map tokens to predefined topic embeddings. During generation, we \textbf{(i)} identify relevant topics from the input prompt, \textbf{(ii)} select the corresponding semantically-aligned ``green'' token list, and \textbf{(iii)} bias generation toward these thematically appropriate tokens. This semantic alignment preserves fluency while embedding robust watermark signals that resist paraphrasing attacks. We evaluate TBW on robustness, text quality, and efficiency, demonstrating performance comparable to leading techniques with perplexity on par with current production-grade systems, all without additional inference steps beyond standard generation pipelines.

\section{Related Work}\label{related work}
Traditional AI-generated text detection relied on post-hoc classifiers trained to differentiate human and machine writing, but these often fail under adversarial transformations like paraphrasing. To address this, watermarking techniques have gained prominence as a more robust alternative, categorized into post-processing and generation methods. \textbf{Post-processing watermarking} modifies generated text to embed hidden patterns but remains vulnerable to edits that erase markers~\cite{sato2023embarrassinglysimpletextwatermarks}. \textbf{Generation watermarking} embeds signals during token generation, ensuring greater resilience against text modifications.

The KGW algorithm~\cite{kirchenbauer2023watermark} introduced model vocabulary partitioning into ``green'' and ``red'' lists, biasing sampling toward green tokens. While computationally efficient, it remains vulnerable to paraphrasing and token perturbation. The Unigram watermark~\cite{zhao2024provable} improves detection by assigning tokens based on unigram statistics rather than random partitioning. However, both methods suffer from partition recovery attacks where adversaries can approximate token subsets through repeated querying~\cite{jovanovi2024stealing}. 

Google's SynthID-Text~\cite{dathathri2024scalable} employs Tournament Sampling, ranking candidate tokens using random watermarking functions before selecting the highest-scoring option, designed for minimal generation overhead and enhanced text quality. While scalable, SynthID-Text has shown limited resistance to paraphrasing and token perturbations. Similarly, DiPmark~\cite{dipmark2024}, another lightweight biasing-based watermark, struggles under text rewriting.

To enhance robustness while preserving text quality, the SIR watermark~\cite{liu2024semanticinvariantrobustwatermark} incorporates user-provided context to guide token selection, achieving greater durability at the cost of requiring decoder modifications and prompt access, hindering adoption in large-scale commercial LLMs. While semantic-based watermarks~\cite{liu2024semanticinvariantrobustwatermark, hou-etal-2024-semstamp, lee-etal-2024-wrote} can offer greater robustness, they still  remain vulnerable to paraphrasing attacks.

In pursuit of stronger detection resilience, EXP, EXP-Edit, and ITS-Edit~\cite{kuditipudi2024robustdistortionfreewatermarkslanguage} introduce iterative decoding or re-ranking to strengthen watermark robustness. While these methods improve detection under adversarial conditions, they significantly increase computational overhead, potentially degrading fluency or inflating perplexity. Beyond these iterative approaches, some watermarking algorithms employ architectural changes that may survive more aggressive attack models (e.g., heavy paraphrasing, near-synonym substitutions), but at the cost of increased generation time and heightened implementation complexity~\cite{zhang2024remarkllmrobustefficientwatermarking, liu2024adaptivetextwatermarklarge}, making them less attractive for practical deployments in latency-sensitive applications.

\textbf{Overall gaps} persist: in-generation methods balance robustness, computational cost, and text quality. KGW-like and SynthID-Text approaches minimize perplexity shifts but struggle against adversarial edits, whereas more durable methods require costly repeated passes or model modifications. Our work occupies this middle ground by integrating semantic information without adding significant complexity, enabling efficient watermarking with improved resilience.

\section{Preliminaries}\label{prelim}
We introduce the notation and key concepts utilized in our topic-guided watermarking approach.

\subsection{Notation and Setup}
Let $V$ denote the vocabulary of an LLM with parameters $\theta$. Each token $v \in V$ is associated with an embedding $e_v \in \mathbb{R}^d$, typically obtained from the model's embedding layer or another off-the-shelf embedding source. We assume a predefined set of topics $\{t_1, t_2, \dots, t_K\}$ with  corresponding embedding $e_{t^i} \in \mathbb{R}^d$.

As a preparatory step, we assign each token $v$ to exactly one topic-aligned list based on its semantic similarity. Specifically, for each token $v$, we compute
\begin{equation}
\operatorname{sim}(v, t_i) \;=\; \frac{e_v \;\cdot\; e_{t_i}}{\|e_v\|\; \|e_{t_i}\|},
\end{equation}
and compare this value to a threshold $\tau$. If $\operatorname{sim}(v, t_i) \ge \tau$ for some topic $t_i$, then $v$ is appended to $t_i$'s list. Tokens not meeting or exceeding $\tau$ for any topic are evenly distributed among all topic lists in a round-robin fashion to ensure balanced coverage. Thus, each topic list $G_{t_i} \subseteq V$ serves as the ``green list'' for $t_i$, analogous to the subsets in prior watermarking schemes such as KGW. The number of topics $K$ determines each active green list's vocabulary fraction, analogous to KGW's $\gamma$ parameter. For example, $K=4$ yields an effective fraction of $0.25$, comparable to $\gamma=0.2$ used in KGW's reported experiments.

During text generation, given an input prompt $x^{\text{prompt}}$, an LLM predicts the next token from a probability distribution $p_{\theta}(v \mid x^{\text{prompt}})$ over $V$. To embed a watermark, we can reweight this distribution toward tokens belonging to a chosen topic list. Specifically, we extract the most relevant topics from $x^{\text{prompt}}$ using a lightweight topic extraction model, then use $k$-means clustering to map them to the closest predefined topic if no direct match is found. This yields a single ``green list'' $G_{t^*}$, which is then biased during the generation process.

\begin{figure*}[t]
\centering
\includegraphics[width=0.98\textwidth]{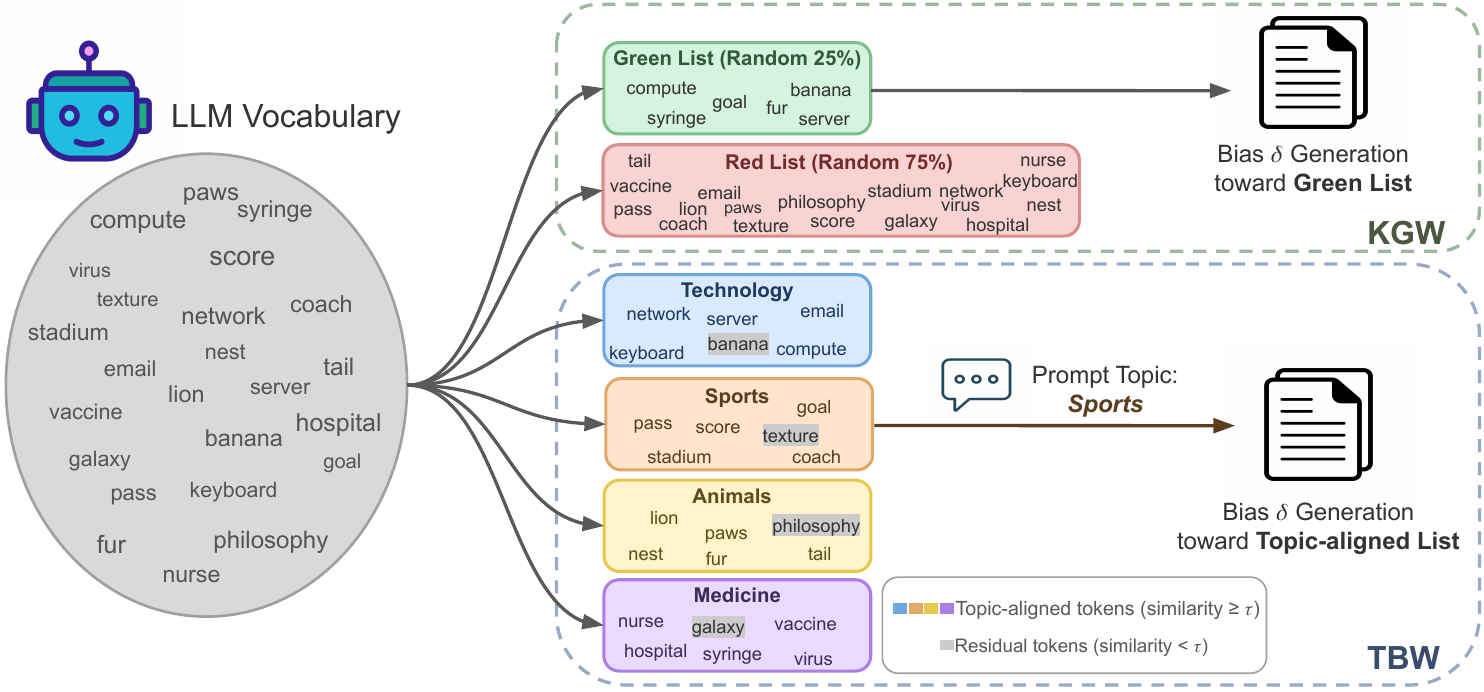}
\caption{\textbf{Comparison of KGW and TBW vocabulary partitioning.} KGW (top) randomly partitions vocabulary $V$ into green/red lists using parameter $\gamma$. TBW (bottom) creates semantically meaningful partitions by assigning tokens to predefined topic lists. Prompts are mapped to corresponding topics, making that topic list the active green list. Both methods bias generation toward green lists using parameter $\delta$ to adjust logit values.}
\label{fig:token_mapping}
\end{figure*}

\subsection{Threat Model}\label{threat_model}
We assume a standard black-box adversary with access only to LLM outputs, seeking to remove watermarks while preserving semantic fidelity. Following~\cite{kirchenbauer2023watermark, zhang2024remarkllmrobustefficientwatermarking}, the adversary cannot access model parameters, watermarking keys, or topic-partition logic. While the detection algorithm is public, specific parameters (thresholds, embeddings) remain private. We evaluate robustness against two primary attack surfaces:
\textbf{(1) text degradation} via token-level perturbations (e.g., substitutions) that disrupt the watermark signal at the cost of fluency; and \textbf{(2) semantic rephrasing} using external models for paraphrasing or summarization to rewrite content while maintaining meaning.

We additionally consider \textbf{(3) partition recovery attacks}, where an adversary attempts to reconstruct the topic-aligned green lists by exploiting the semantic structure of TBW's vocabulary partitioning. Prior work has demonstrated such attacks against KGW-style schemes through repeated positional querying~\cite{jovanovi2024stealing}. While TBW's topic-aligned partitions introduce semantic co-grouping that could appear more predictable than random assignment, the full partition remains unrecoverable without the private random seed used during round-robin distribution of residual tokens (\S\ref{proposed}). Detailed adversary assumptions, excluded attack vectors (e.g., statistical comparison), detection-based attacks, and the full partition recovery analysis are provided in~\ref{extended_threat_model}.

\section{Proposed Method}\label{proposed}
In this section, we detail our lightweight topic-based watermarking (TBW) scheme. \S\ref{semmap} explains how tokens are mapped to topic-aligned subsets, which serve as ``green lists'' for watermarking. In \S\ref{TBW}, we describe the generation procedure for embedding watermarks. \S\ref{TBD}, presents multiple watermark detection algorithms that identify watermarked text.

\subsection{Token-to-Topic Mappings}\label{semmap}
We begin by clustering tokens in the LLM vocabulary $V$ into semantically aligned lists, each associated with one of a small set of high-level ``generalized topics'' $\{t_1, \dots, t_K\}$. For illustration, one might choose topics such as \{\texttt{animals}, \texttt{technology}, \texttt{sports}, \texttt{medicine}\} to capture common themes. Using a sentence embedding model (e.g., all-MiniLM-L6-v2~\cite{reimers-gurevych-2020-making}), we encode each token $v \in V$ into an embedding $\mathbf{e}_v$ and compute its similarity to each topic embedding $\mathbf{e}_{t_i}$: $\text{sim}(v, t_i)$. If the maximum similarity across all topics exceeds a threshold $\tau$, the token is assigned to the corresponding topic's ``green list'' $G_{t_i}$. Tokens that do not exceed $\tau$ for any topic are collected into a residual set, which is subsequently distributed among $\{G_{t_1}, \dots, G_{t_K}\}$ in a round-robin fashion. This ensures comprehensive coverage of the entire vocabulary, preventing any token from being discarded.

The hyperparameter $\tau$ controls the granularity of semantic alignment and comprehensive topic coverage where a higher $\tau$ enforces stronger coherence but increases the proportion of tokens allocated via the round-robin mechanism.  Although we consider only four broad topics in our main evaluations, the same procedure naturally extends to a larger number of topics for more fine-grained coverage as the vocabulary $V$ increases. We use a general-purpose sentence model (all-MiniLM-L6-v2) for this implementation as a lightweight approach; however, any semantic embedding framework can be substituted, allowing practitioners to tailor the mapping for domain-specific or resource-constrained environments. Figure~\ref{fig:token_mapping} illustrates this token-to-topic assignment process, with the detailed implementation procedure provided in Algorithm~\ref{alg:token_mapping} (\ref{algo_implementations}).

\subsection{Topic-Based Watermarks}\label{TBW}
Building on the token-to-topic mappings, we now detail how to embed watermarks during text generation by selectively biasing tokens in a single topic list. This procedure is analogous to the KGW scheme, where a targeted subset of the vocabulary (the ``green list'') receives a higher sampling probability. However, in our approach, the specific green list is chosen via a semantic matching process that depends on the user's prompt.

Given an input prompt $x^{\text{prompt}}$, we first identify relevant keywords or topics using a lightweight extractor (KeyBERT~\cite{grootendorst2020keybert}). If one or more of these extracted topics $\mathcal{T}_{\text{detected}}$ exactly matches an entry in the predefined set $\{t_1, \dots, t_K\}$, we select the corresponding list $G_{t^*}$. Otherwise, we cluster the detected topic embeddings into a few centroids and compute their cosine similarity to each $t_i$'s embedding. We designate the topic list whose embedding is most similar to the centroid as $G_{t^*}$. This ensures that even if an exact match is unavailable, the system still picks the most semantically aligned topic.

At each generation step, the model produces logits $p_{\theta}(v \mid x^{\text{prompt}}, \mathbf{z})$ over the vocabulary $V$. We add a small bias $\delta$ to all tokens $v \in G_{t^*}$ before normalizing with a softmax function. Intuitively, this raises the selection probability of tokens in $G_{t^*}$, embedding a watermark without introducing multiple decoding passes or inflating perplexity. A larger $\delta$ yields a more robust watermark signal at the cost of potentially more noticeable shifts in text style or quality. After adjusting logits, the model samples the next token via standard methods (e.g., top-$k$ sampling, beam search). The entire generation loop is detailed in Algorithm~\ref{alg:topic-watermarking} (\ref{algo_implementations}), highlighting that topic extraction and logit biasing constitute minimal overhead compared to typical LLM pipelines.

\subsection{Topic-Based Detection}\label{TBD}
While our watermarking approach embeds topic-aligned signals during generation, practical detection faces challenges including topic ambiguity, topic drift within longer documents, and potential generation-detection misalignment. To address these challenges systematically, we propose three detection schemes with increasing robustness and decreasing reliance on perfect topic knowledge.

All detection schemes, compute a $z$-score as: 
\begin{equation}
z = \frac{g - \gamma \cdot n}{\sqrt{n \cdot \gamma \cdot (1-\gamma)}}
\end{equation}
where $g$ is the observed green token count, $n$ is the total tokens, and $\gamma$ is the expected green token fraction. Text is classified as watermarked if $z$ exceeds a threshold.

\subsubsection{Strict Topic Matching} 
Our baseline detection scheme assumes consistent topic alignment between generation and detection phases, establishing an upper bound on performance while highlighting the challenges that arise when this assumption fails in practice.

Given text $z_{test}$, we extract topics using KeyBERT. The process follows a hierarchical matching strategy: if extracted topics $T_{detected}$ directly match predefined topics $\{t_1, \ldots, t_K\}$, we select the corresponding green list $G_{t^*}$. Otherwise, we use semantic mapping to ensure consistency with the generation-time topic selection using one of two approaches. \textbf{Embedding averaging} computes the mean embedding of all detected topics and identifies the predefined topic with highest cosine similarity, where $\bar{e}_{detected}=\frac{1}{m}\sum_{i=1}^m e_{d_i}$ and selects $t^* = \operatorname{argmax}_{t_j} \operatorname{sim}(\bar{e}_{detected}, e_{t_j})$. \textbf{K-means clustering} captures topic diversity within the detected set by applying k-means to detected topic embeddings and evaluating centroids against predefined topics: $t^*=\operatorname{argmax}_{t_j}\operatorname{max}_{c_k}\operatorname{sim}(c_k, e_{t_j})$. Detection proceeds by computing the $z$-score using $G_{t^*}$.

\subsubsection{Sliding Window Detection} 
While strict matching works under ideal conditions, real text exhibits topic drift. This approach addresses local topic inconsistencies while maintaining semantic awareness through temporal aggregation.

We partition the input text into windows of size $w$. For each window, we extract topics using the hierarchical matching above, then assign the final topic via majority voting: $t^*=\operatorname{argmax}_{t_j} |\{w_i:\text{topic}(w_i)=t_j\}|$ where $\text{topic}(w_i)$ represents the assigned topic for window $w_i$. An adaptive fallback mechanism selects embedding averaging for windows with fewer than 3 detected keywords. Detection proceeds using the same $z$-score computation as strict matching.

\subsubsection{Maximum $z$-Score Detection} 
Both previous methods rely on successful topic extraction, which can fail with ambiguous or multi-topic content. Our most robust detection scheme eliminates the topic matching assumption entirely by evaluating text against each predefined green list $G_{t_i}$, computing the corresponding $z$-score $z_i$ using the same statistical framework as previous methods. The final classification uses the maximum $z$-score across all topics: $t^*=\operatorname{argmax}_{t_i}z_i$. This approach effectively allows the watermark signal itself to determine the most likely topic alignment.

This parameter-free approach provides several key advantages: (i) it requires no topic extraction or mapping steps, eliminating potential failure modes; (ii) it leverages the embedded watermark signal to guide topic selection, aligning detection with the generation process; and (iii) it provides robustness against topic ambiguity, drift, and misalignment, making it suitable for practical deployment where topic alignment cannot be guaranteed.

\section{Evaluation}\label{experiments}
We present a comprehensive evaluation of TBW against watermarking schemes across four key dimensions: text quality, robustness against both full-text paraphrasing and lexical perturbations, evaluation of watermark detection performance under our proposed schemes, and computational efficiency via average generation times. Our results demonstrate that TBW achieves an optimal balance across these metrics, offering strong robustness and high text quality while maintaining computational efficiency comparable to production-ready systems.

\subsection{Experimental Setup}\label{setup}
\textbf{Data and Models.} All experiments use sampled subsets of the C4 dataset~\cite{c4}, where we truncate the first 100 tokens as the input prompt and let each model generate 200 additional tokens (allowing a tolerance of $\pm5$ to accommodate decoding variation). We primarily evaluate two LLMs: \textsc{OPT-6.7B}~\cite{zhang2022optopenpretrainedtransformer} and \textsc{Gemma-7B}~\cite{team2024gemma}. All evaluations use NVIDIA V100 or A100 GPUs.

\textbf{Watermarking Comparisons.} We compare TBW against several baselines (see \S\ref{related work} for details): No Watermark (standard decoding), KGW~\cite{kirchenbauer2023watermark}, DiPMark (DiP)~\cite{dipmark2024}, Unigram~\cite{zhao2024provable}, SynthID-Text (SynthID)~\cite{dathathri2024scalable}, SIR~\cite{liu2024semanticinvariantrobustwatermark}, and EXP, EXP-Edit, ITS-Edit~\cite{kuditipudi2024robustdistortionfreewatermarkslanguage}. We use the \textsc{MarkLLM} library~\cite{pan-etal-2024-markllm} implementations with parameter settings from the original papers to ensure fair comparison (details in~\ref{markllm}). We additionally compare TBW against semantic watermarking schemes that operate via post-processing rather than in-generation biasing, specifically PostMark~\cite{chang2024postmark} and k-SemStamp~\cite{hou-etal-2024-k}; these results are reported separately in~\ref{semantic_comparison} due to their different generation pipelines.

\textbf{Implementation Details:} Our TBW uses KeyBERT~\cite{grootendorst2020keybert} for topic extraction alongside a partition of topic-aligned tokens (\S\ref{semmap}). We employ a predefined set of four generalized topics, \{\texttt{animals}, \texttt{technology}, \texttt{sports}, \texttt{medicine}\}, which we found to be sufficiently generic to cover large vocabulary partitions while avoiding over-specialization. In our setup, four topics correspond to an effective green-list fraction of $0.25$, which is close to the $\gamma=0.2$ used in KGW's experiments, while still maintaining broad topic coverage. Topic assignment is performed using a sentence embedding model (all-MiniLM-L6-v2) with a similarity threshold of $\tau=0.7$, though any semantic embedding framework can be substituted. We also study scalability beyond four topics by varying $K\in\{4,8,16,32\}$; as $K$ increases the watermark signal attenuates as expected yet remains strong, and text quality stays flat (see~\ref{topic_scalability} and~\ref{ablation:num_topics}).

\subsection{Text Quality}\label{text_quality}
We evaluate text quality preservation through perplexity metrics 
(\S\ref{main_perplexity_eval}) with additional human evaluation in \ref{main_human_eval} and LLM-as-a-Judge analysis in~\ref{llm-as-a-judge}.

\subsubsection{Perplexity}\label{main_perplexity_eval}
\begin{figure}[h]
\begin{center}
    \centerline{
        \includegraphics[width=0.8\columnwidth]{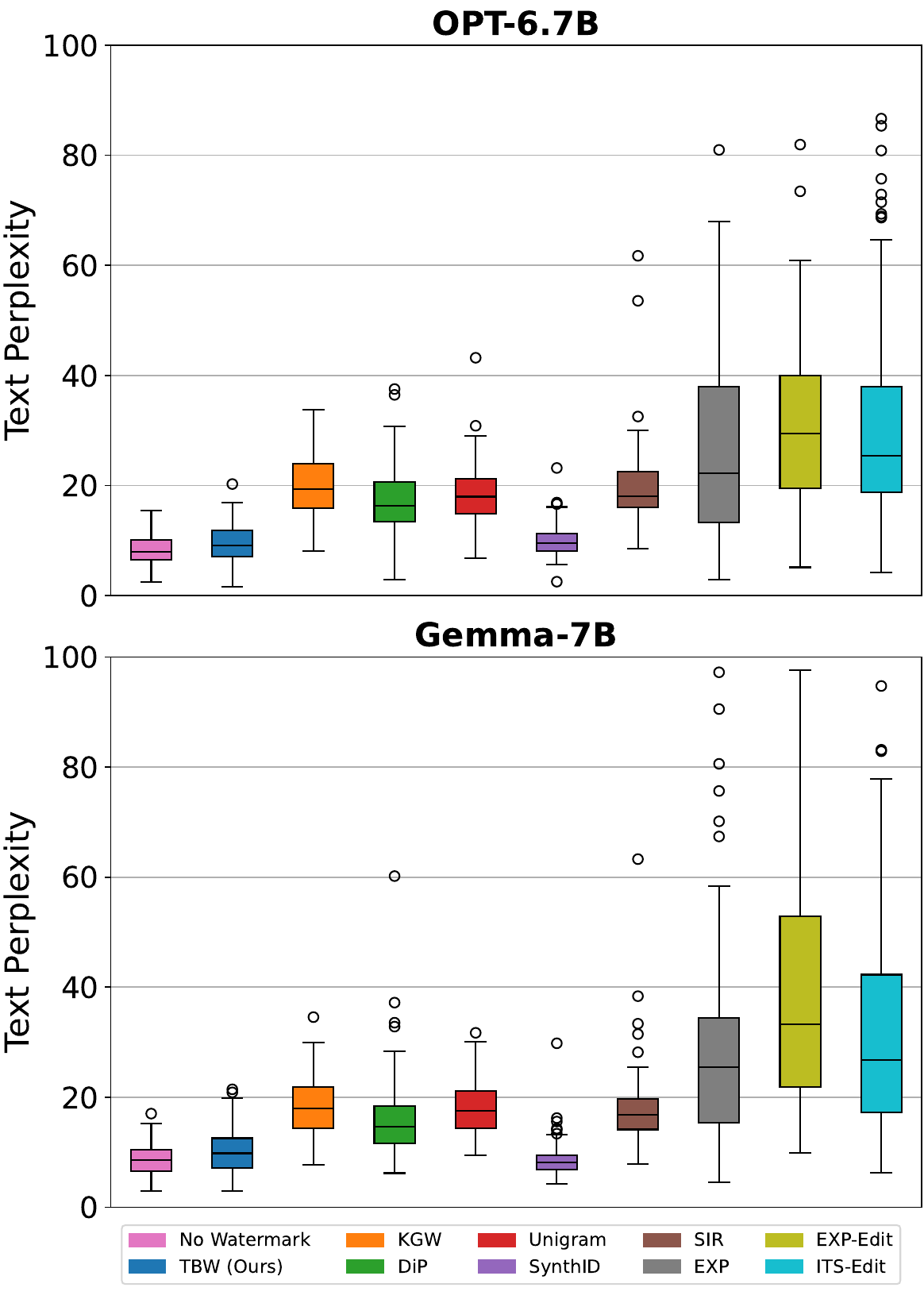}
    }
    \caption{Text perplexity comparison using baseline watermarking schemes. Boxplots show distribution of perplexity values for each watermarking method. Lower text perplexity indicates higher generated text quality.}
    \label{fig:perplexity}
\end{center}
\end{figure}

We evaluate text fluency via perplexity~\cite{niess-kern-2025-ensemble, mao2025watermarkinglowentropygenerationlarge, feng2025bimark, qu2025provablyrobustmultibitwatermarking} using a larger ``oracle'' model, \textsc{Llama-3.1-8B}~\cite{grattafiori2024llama}, which is from a different family than both \textsc{OPT} and \textsc{Gemma} to avoid bias. We fix the watermark strength at $\delta=2.0$ for direct comparability with prior work~\cite{kirchenbauer2023watermark, zhao2024provable} and optimal quality-detection trade-off (see~\ref{sensitive} for sensitivity analysis). Perplexity is measured on the generated $200\pm5$ tokens only of 100 samples. Figure~\ref{fig:perplexity} shows distributions for \textsc{OPT-6.7B} and \textsc{Gemma-7B} under all watermarking schemes, with values above 100 clipped for clarity.
We observe that TBW achieves significantly lower perplexity (higher text quality) compared to other watermarking schemes, specifically SynthID, closely matching non-watermarked outputs. TBW improves perplexity by approximately 42\% over Unigram on \textsc{OPT-6.7B} and by 48\% on \textsc{Gemma-7B}.
These results indicate that TBW embeds a robust watermark without degrading text fluency, making it unlikely that watermarked text would be perceptibly less natural to human readers. Additional results for a higher bias strength setting ($\delta=3.0$) on \textsc{OPT-6.7B} are provided in~\ref{more_perplexity}, illustrating the trade-off between increased robustness and text quality.

\subsection{Robustness}\label{robustness} 
Having established TBW's high text quality and efficiency, we next evaluate its robustness to adversarial transformations. Since EXP-based methods (e.g., ITS-Edit, EXP-Edit) exhibit substantially degraded text quality at comparable watermark strengths, we focus our main comparisons on TBW, KGW, DiP, Unigram, SynthID, and SIR. Results for ITS-Edit, the strongest EXP-based scheme, are deferred to~\ref{its_edit}. 

All robustness experiments use oracle conditions, where the generation topic is assumed to be known at detection time, as well as a higher watermark strength ($\delta=3.0$) to establish an upper bound on detection performance independent of topic extraction errors. Such conditions approximate the scenario of near-perfect topic detection achievable with our best-performing algorithm (see \S\ref{detection_schemes}) and provide a clean benchmark for evaluating the intrinsic resilience of each watermarking method. Detailed attack configurations are provided in~\ref{attack_config}.

\subsubsection{Text Degradation} 
We evaluate how each scheme's detection score deteriorates under lexical perturbations using \textsc{OPT-6.7B} as the representative model. This choice provides a conservative robustness estimate, as its smaller vocabulary makes maintaining watermark signal more challenging (\S\ref{text_quality}). Two perturbation types are considered: \textbf{random perturbations} (a mix of word insertions, deletions, and substitutions applied uniformly across the text affecting both high- and low-importance words) and \textbf{targeted perturbations} (the same operations applied selectively to semantically important words such as nouns and verbs, while avoiding low-information stop words like ``a,'' ``the,'' and ``to,'' thereby maximizing disruption to watermark signal).

\begin{figure}[t]
\begin{center}
\centerline{\includegraphics[width=0.95\columnwidth]{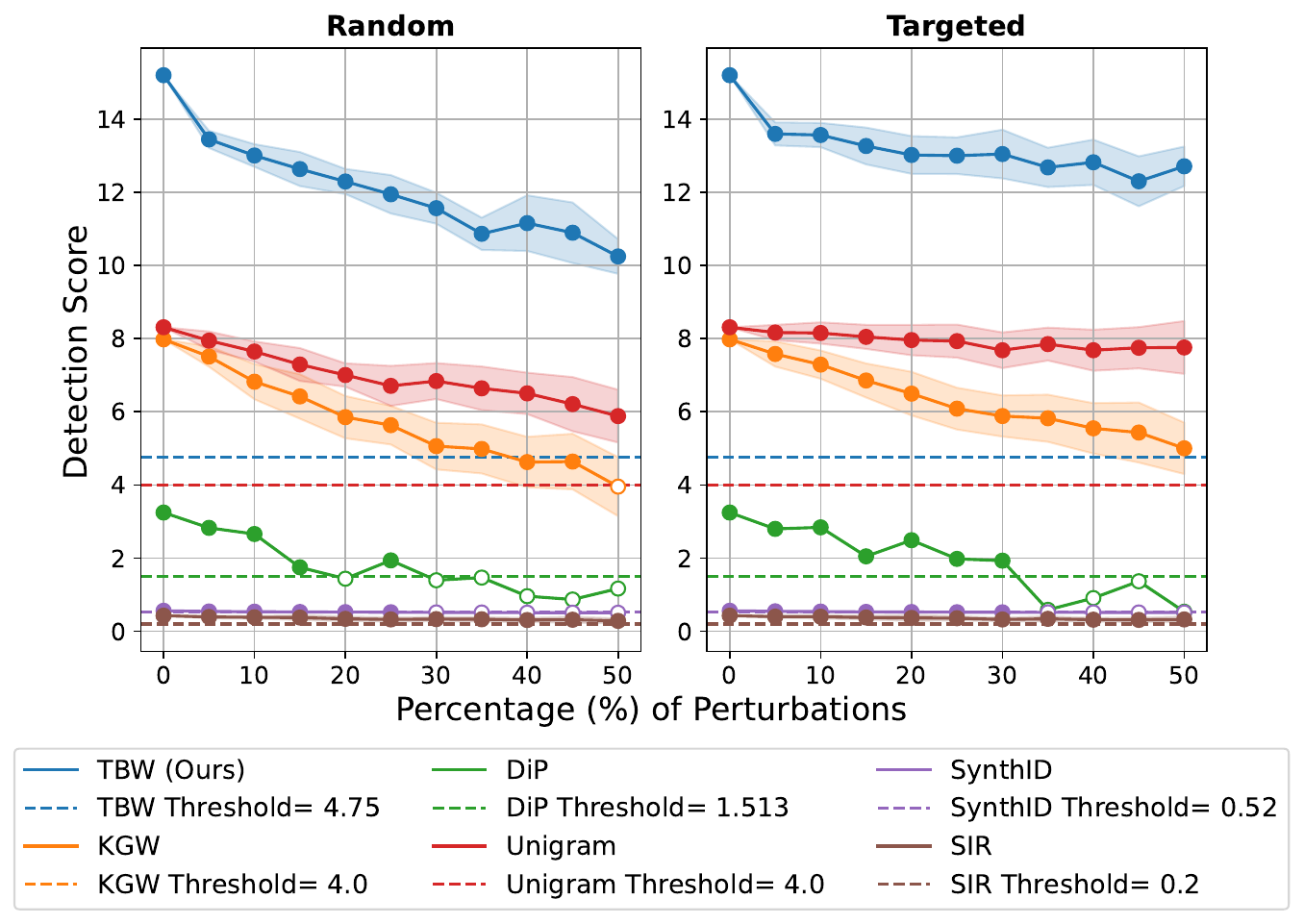}}
\caption{Detection scores under random (left) and targeted (right) word perturbations. Solid ticks indicate scores above threshold; white ticks indicate scores below. Higher scores indicate higher robustness to attacks.}
\label{zscore-degradation}
\end{center}
\end{figure}

For each scheme, we generate 100 watermarked texts, then apply perturbations in 5\% increments up to 50\%. At each level, we average the binary detection outcome (``watermarked'' vs. ``not watermarked'') across 20 independent trials. Figure~\ref{zscore-degradation} presents the resulting score trajectories.

All watermarking schemes show a gradual decline in classification scores as perturbation levels increase, except for DiP, which does not rely on a $z$-statistic for detection. Unigram, despite its robustness to paraphrasing, deteriorates under even modest perturbations, crossing its classification threshold earlier than TBW. This highlights a key vulnerability that simple, low-effort edits can bypass detection despite strong paraphrasing resistance. In contrast, TBW maintains higher detection rates across all perturbation levels for both random and targeted attacks, demonstrating resilience to lexical perturbations.

\subsubsection{Semantic Paraphrasing} 
We evaluate detection robustness against strong paraphrasing by generating 500 watermarked and 500 non-watermarked samples from both \textsc{OPT-6.7B} and \textsc{Gemma-7B}. Each 200-token completion is transformed using two paraphrasers widely adopted in watermarking research, PEGASUS~\cite{pegasus} and DIPPER~\cite{dipper}, consistent with prior evaluation~\cite{liu2024adaptivetextwatermarklarge, hou-etal-2024-semstamp, hou-etal-2024-k}. 

We first report ROC-AUC and Best F1 scores in Table~\ref{metrics}. Without attack, all methods reliably separate watermarked from non-watermarked text. Under paraphrasing, however, DIPPER and PEGASUS substantially degrade the performance of SynthID and DiP. TBW and Unigram retain the highest ROC-AUC and Best F1 scores across both attack scenarios, with TBW often matching Unigram and outperforming all other baselines. Full ROC-AUC curves are included in~\ref{auc_roc}.

In practical applications, maintaining a high True Positive Rate (TPR) at consistently low False Positive Rates (FPR) ensure non-watermarked texts are not misclassified as watermarked. We therefore report TPR@1\%FPR and TPR@10\%FPR for \textsc{OPT-6.7B} and \textsc{Gemma-7B} in Table~\ref{tpr_fpr_results}, with additional human-text FPR analysis in~\ref{fpr_analysis}.

Across both paraphrasers, TBW achieves the highest TPR@1\%FPR of all watermarking schemes, indicating superior detection in conservative operating setting. At TPR@10\%FPR, TBW remains competitive with Unigram while outperforming other baselines, demonstrating that it maintains strong paraphrasing robustness without compromising the practical constraints of real-world watermark detection.

\begin{table*}[t]
\centering
\scalebox{0.98}{
\scriptsize
\begin{tabular}{clcccccccccccc}
\toprule
 & & \multicolumn{6}{c}{\textbf{ROC-AUC}} & \multicolumn{6}{c}{\textbf{Best F1 Score}} \\
\cmidrule(lr){3-8}\cmidrule(lr){9-14}
\textbf{Model} & \textbf{Attacks} & \textbf{TBW} & KGW & DiP & Unigram & SynthID & SIR & \textbf{TBW} & KGW & DiP & Unigram & SynthID & SIR  \\
\midrule
\multirow{3}{*}{OPT-6.7B}
  & No Attack & \textbf{1.000} & \textbf{1.000} & \underline{0.999} & \textbf{1.000} & \underline{0.999} & 0.995 & \underline{0.995} & \textbf{0.998} & 0.994 & 0.994 & \underline{0.995} & 0.978 \\
  & Pegasus  & \textbf{0.990} & 0.975 & 0.824 & \underline{0.987} & 0.910 & 0.971 & \underline{0.960} & 0.933 & 0.756 & \textbf{0.970} & 0.837 & 0.920 \\
  & DIPPER   & \underline{0.945} & 0.826 & 0.576 & \textbf{0.955} & 0.650 & 0.891 &  \underline{0.888} & 0.770 & 0.667 & \textbf{0.893} & 0.675 & 0.829 \\
\midrule
\multirow{3}{*}{Gemma-7B}
  & No Attack & \underline{0.998} & 0.995 & \textbf{1.000} & \underline{0.998} & \textbf{1.000} & 0.990 & \textbf{0.999} & \underline{0.997} & \textbf{0.999} & 0.996 & \underline{0.997} & 0.973 \\
  & Pegasus   & 0.981 & \underline{0.983} & 0.836 & \textbf{0.985} & 0.912 & 0.952 & 0.951 & \textbf{0.962} & 0.759 & \underline{0.959} & 0.842 & 0.903 \\
  & DIPPER    & \underline{0.871} & 0.825 & 0.546 & \textbf{0.911} & 0.656 & 0.822 & \underline{0.811} & 0.773 & 0.668 & \textbf{0.851} & 0.676 & 0.775 \\
\bottomrule
\end{tabular}
}
\caption{ROC-AUC and Best F1 scores for watermarking approaches under no attack, PEGASUS paraphrasing, and DIPPER paraphrasing. Best results are in \textbf{bold}; second-best results are \underline{underlined}.}
\label{metrics}
\end{table*}

\begin{table*}[t]
\centering
\scalebox{0.98}{
\scriptsize
\begin{tabular}{clcccccccccccc}
\toprule
 & & \multicolumn{6}{c}{\textbf{TPR@1\% FPR}} & \multicolumn{6}{c}{\textbf{TPR@10\% FPR}} \\
\cmidrule(lr){3-8}\cmidrule(lr){9-14}
\textbf{Model} & \textbf{Attacks} & \textbf{TBW} & KGW & DiP & Unigram & SynthID & SIR & \textbf{TBW} & KGW & DiP & Unigram & SynthID & SIR  \\
\midrule
\multirow{3}{*}{OPT-6.7B}
  & No Attack & \underline{0.994} & \textbf{0.996} & 0.992 & \textbf{0.996} & 0.992 & 0.964 & \textbf{1.000} & \textbf{1.000} & \underline{0.996} & \underline{0.996} & \underline{0.996} & 0.986 \\
  & Pegasus  & \textbf{0.910} & 0.578 & 0.228 & \underline{0.900} & 0.446 & 0.726 & \underline{0.980} & 0.948 & 0.552 & \textbf{0.986} & 0.768 & 0.930 \\
  & DIPPER   & \textbf{0.536} & 0.124 & 0.028 & \underline{0.516} & 0.058 & 0.248 &  \underline{0.866} & 0.534 & 0.170 & \textbf{0.872} & 0.258 & 0.702 \\
\midrule
\multirow{3}{*}{Gemma-7B}
  & No Attack & \textbf{1.000} & \textbf{1.000} & \textbf{1.000} & \underline{0.998} & \textbf{1.000} & 0.890 & \textbf{1.000} & \textbf{1.000} & \textbf{1.000} & \textbf{1.000} & \textbf{1.000} & \underline{0.998} \\
  & Pegasus   & \textbf{0.842} & 0.246 & 0.282 & \underline{0.598} & 0.484 & 0.476 & 0.960 & \underline{0.974} & 0.614 & \textbf{0.980} & 0.750 & 0.896 \\
  & DIPPER    & \textbf{0.196} & 0.052 & 0.022 & 0.034 & 0.024 & \underline{0.190} & \underline{0.612} & 0.568 & 0.164 & \textbf{0.766} & 0.288 & 0.524 \\
\bottomrule
\end{tabular}
}
\caption{True Positive Rate (TPR) at fixed False Positive Rate (FPR) for watermarking approaches under no attack, PEGASUS paraphrasing, and DIPPER paraphrasing. Best results are in \textbf{bold}; second-best results are \underline{underlined}.}
\label{tpr_fpr_results}
\end{table*}

\subsection{Detection Analysis}\label{detection_schemes}
While our robustness experiments in \S\ref{robustness} established watermark resilience under oracle conditions, practical deployment requires detection schemes that operate without prior topic knowledge. We evaluate the detection approaches introduced in \S\ref{TBD}, moving from theoretical validation to real-world applicability, using 500 watermarked samples under our standard experimental configuration. We further provide discussion in~\ref{detection_discussion} on practical deployment contexts of detection schemes.

Table~\ref{tab:detection_comparison} compares all detection schemes for both \textsc{OPT-6.7B} and \textsc{GEMMA-7B}. Strict topic matching slightly outperforms sliding window detection despite the latter's intent to mitigate topic drift. Embedding averaging yields higher detection rates than k-means in both contexts. We attribute sliding window's weaker performance to the 50-word window size: while offering local topic resolution, it limits available context compared to processing the full text. Exploring alternative window sizes is considered out of scope but presents a possible avenue for future work.

The maximum $z$-score detection method delivers near-perfect results, achieving 99.6\% and 100\% detection rates for \textsc{OPT-6.7B} and \textsc{GEMMA-7B}, respectively, with the highest mean $z$-scores and minimal variance. This performance stems from its independence from topic extraction by directly evaluating all topic partitions and selecting the highest $z$-score, it avoids the alignment errors that constrain topic-dependent approaches. We confirm this with 500 watermarked and 500 non-watermarked samples with detailed performance metrics provided in~\ref{max_z_detailed}. These results connect our oracle-condition evaluations in \S\ref{robustness} with practical, topic-agnostic detection, showing that maximum $z$-score preserves robustness while providing tunable separation adaptable to different operational constraints.

\begin{figure}[t]
\begin{center}
\centerline{\includegraphics[width=0.95\columnwidth]{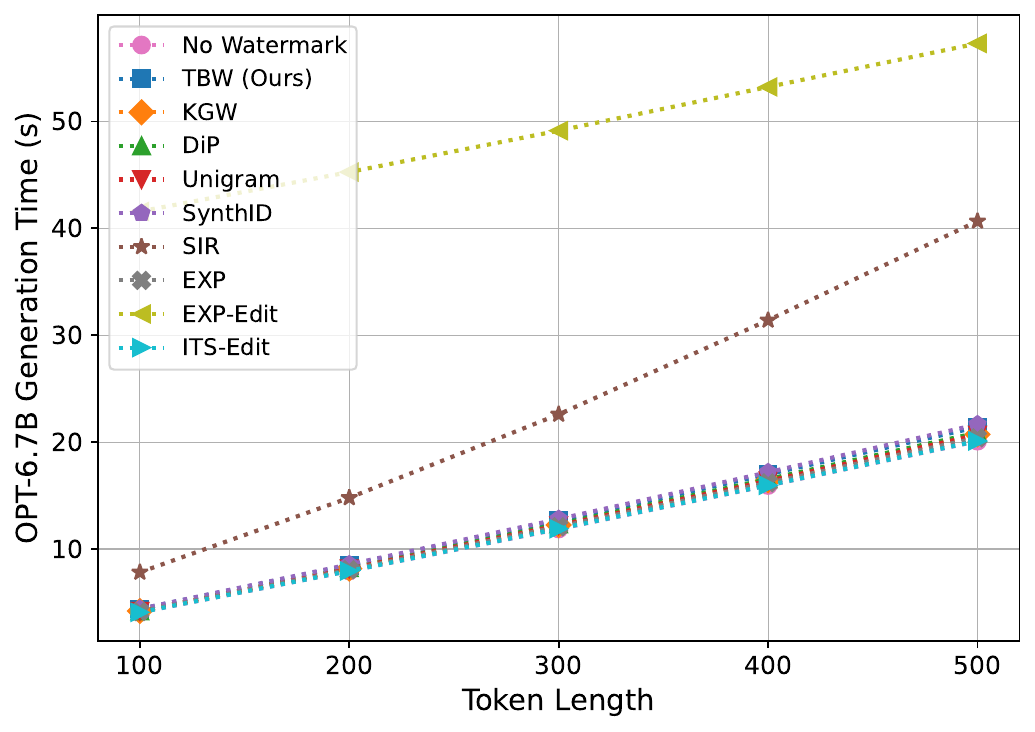}}
\caption{Comparison of average generation time (seconds) over various output token lengths from multiple watermarking schemes on \textsc{OPT-6.7B}.}
\label{efficiency-opt6p7b}
\end{center}
\end{figure}

\begin{table*}[t]
\centering
\scalebox{0.97}{
\scriptsize
\begin{tabular}{lccccccc}
\toprule
 & & \multicolumn{3}{c}{\textbf{OPT-6.7B}} & \multicolumn{3}{c}{\textbf{GEMMA-7B}} \\
\cmidrule(lr){3-5}\cmidrule(lr){6-8}
\textbf{Detection Scheme} & & \textbf{Detection Rate} & \textbf{$z$-Score (Mean$\pm$Sd)} & \textbf{Topic Accuracy} & \textbf{Detection Rate} & \textbf{$z$-Score (Mean$\pm$Sd)} & \textbf{Topic Accuracy} \\
\midrule
Strict K-means & & 0.540 & $6.32\pm10.80$ & 0.542 & 0.570 & $6.71\pm10.58$ & 0.570 \\
Strict Embedding & & 0.574 & $7.05\pm10.68$ & 0.624 & 0.584 & $6.99\pm10.53$ & 0.584 \\
Sliding K-means & & 0.516 & $5.80\pm10.77$ & 0.520 & 0.532 & $5.95\pm10.69$ & 0.532 \\
Sliding Embedding & & 0.566 & $6.91\pm10.67$ & 0.602 & 0.516 & $5.63\pm10.70$ & 0.524 \\
Sliding K-means+Emb & & 0.526 &$ 6.02\pm10.79 $& 0.530 & 0.528 & $5.85\pm10.68$ & 0.528 \\
\textbf{Max. $z$-Score} & & \textbf{0.996} & $\mathbf{15.88\pm3.03}$ & \textbf{1.000} & \textbf{1.000} & $\mathbf{15.92\pm1.39}$ & \textbf{1.000} \\
\bottomrule
\end{tabular}
}
\caption{Comparison of watermark detection schemes on watermarked text under realistic (non-oracle) detection conditions. \textbf{Bold} indicates the highest performing detection scheme in each metric.}
\label{tab:detection_comparison}
\end{table*}

\subsection{Efficiency}\label{efficiency} 
We measure the computational overhead imposed by each watermarking method on \textsc{OPT-6.7B}, generating sequences of lengths \{100, 200, 300, 400, 500\}. For each token-length, we record the average generation time over 10 samples from \textsc{C4}. Figure~\ref{efficiency-opt6p7b} shows that TBW introduces negligible overhead compared to non-watermarked generation, matching lightweight methods such as KGW and SynthID. In contrast, EXP-Edit requires multiple re-ranking passes and SIR incurs additional complexity, resulting in noticeably higher generation times. We observe consistent trends on \textsc{OPT-2.7B} (see~\ref{small_eff}), confirming that TBW's efficiency advantage holds across model scales.

\subsection{Topic Scalability}\label{topic_scalability}
While our main experiments use $K=4$ topics, practitioners may require finer-grained coverage. We evaluate TBW with $K \in \{4, 8, 16, 32\}$ on \textsc{Gemma-7B}, whose larger vocabulary is more representative of deployed LLMs. We apply $\delta=2.0$ and $\tau=0.5$, relaxed relative to the main experiments to compensate for smaller per-topic coverage at larger $K$. Detection uses the maximum $z$-score scheme. Full experimental details, topic inventories, and additional visualizations are provided in~\ref{ablation:num_topics}.

\begin{figure}[h]
\begin{center}
\centerline{\includegraphics[width=0.95\columnwidth]{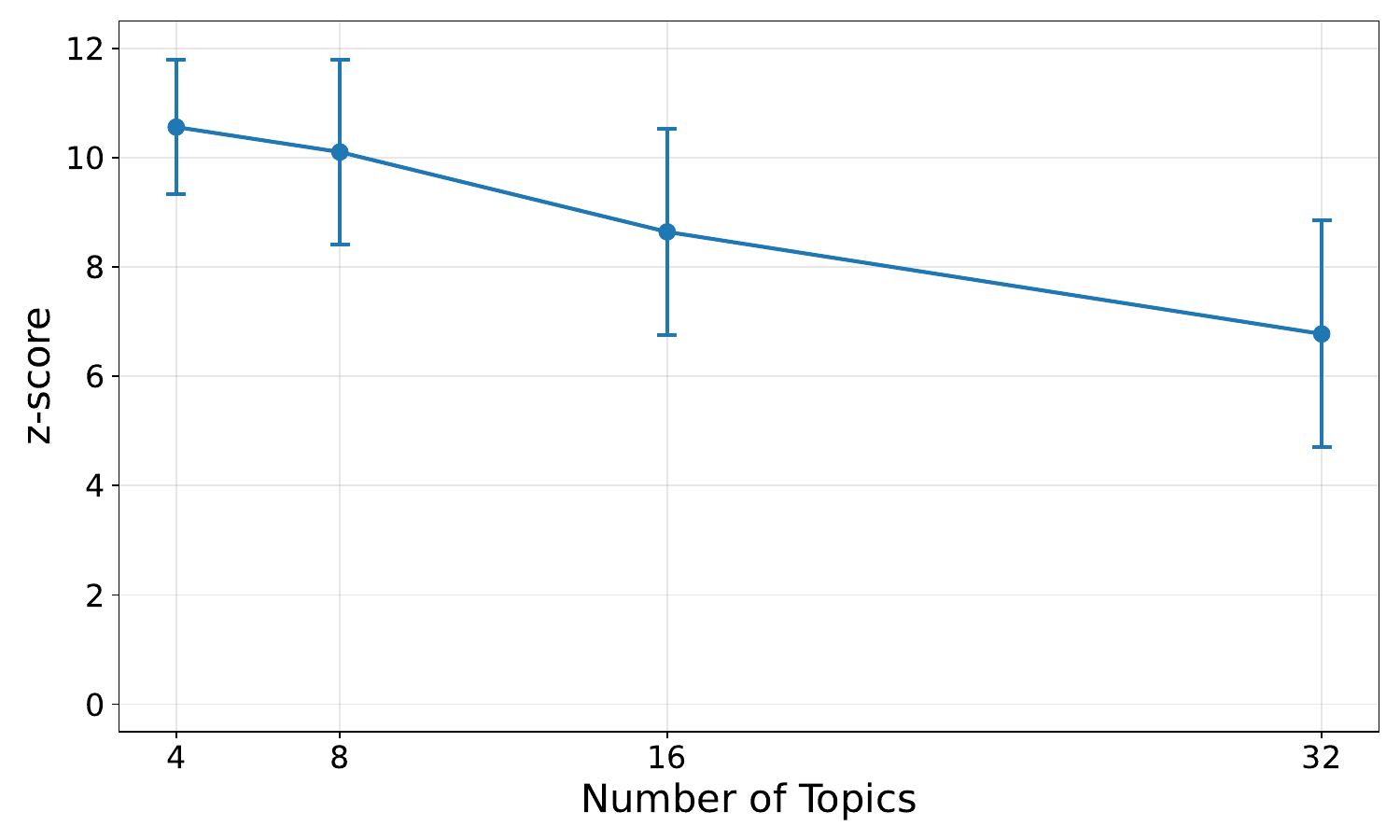}}
\caption{Detection strength vs.\ number of topics. Mean detector $z$-score (max-$z$) as we scale $K\in\{4,8,16,32\}$ on \textsc{Gemma-7B} with $\delta=2.0$, $\tau=0.5$ over 100 prompts. Error bars denote $\pm$s.d.}
\label{zcore_lists}
\end{center}
\end{figure}

Figure~\ref{zcore_lists} reports detection strength as a function of $K$. The $z$-score decreases gracefully from ${\sim}11$ at $K=4$ to ${\sim}7$ at $K=32$, yet remains comparable to KGW and Unigram baselines at equivalent watermark strength. Text quality (BERTScore F1) remains flat across all $K$ values (see~\ref{ablation:num_topics}), confirming that increased topic granularity does not degrade fluency or semantics. Table~\ref{tab:detection_times} reports detection runtime, which scales approximately linearly with $K$ but remains tractable; because detection is performed offline by the model owner, this overhead does not impact user-facing latency.

\begin{table}[t]
\centering
\small
\begin{tabular}{cc}
\toprule
\textbf{Number of Topics ($K$)} & \textbf{Detection Time (s)} \\
\midrule
4 & $2.002 \pm 0.041$ \\
8 & $2.910 \pm 0.047$ \\
16 & $3.833 \pm 0.046$ \\
32 & $6.565 \pm 0.066$ \\
\bottomrule
\end{tabular}
\caption{Detector runtime vs.\ number of topics. Mean $\pm$ s.d.\ per-sample time in seconds.}
\label{tab:detection_times}
\end{table}

\section{Conclusion}
Watermarking has emerged as a critical tool for mitigating risks in large language models, yet existing schemes face a persistent trade-off where lightweight methods such as SynthID-Text or KGW offer efficiency and text quality but collapse under paraphrasing, while heavyweight multi-pass methods deliver robustness at the cost of fluency and practicality. We presented Topic-Based Watermarking (TBW), which addresses this divide by embedding semantically aligned signals through topic-driven token partitioning. TBW preserves the speed and fluency of lightweight approaches while achieving robustness closer to heavyweight schemes. Our evaluation confirms that TBW provides strong detection accuracy, resilience to lexical and semantic attacks, and negligible runtime overhead. By overcoming this trade-off, TBW offers a practical path toward deployable watermarking.

\clearpage

\section*{Limitations}
Our results indicate that models with smaller vocabularies may exhibit slightly reduced performance compared to larger vocabulary models. For instance, \textsc{Opt-6.7B} (50,272 tokens) shows marginally lower detection scores and higher false positive rates than \textsc{GEMMA-7B} (256,000 tokens) (\ref{max_z_detailed}). This disparity arises from how vocabulary size interacts with topic partitioning. When the vocabulary is divided among $K$ predefined topics, larger vocabularies yield denser topic-aligned green lists, providing more semantically coherent tokens to bias during generation. Consequently, if practitioners require finer-grained topic coverage (e.g., $K=32$ topics), smaller vocabulary models may experience more pronounced performance degradation. We empirically examine this scaling behavior in~\ref{ablation:num_topics}, demonstrating that while detection strength ($z$-score) decreases as $K$ increases, text quality remains stable and the watermark signal persists even at $K=32$ topics.

The interaction between vocabulary size and topic granularity suggests an avenue for domain-specific adaptation. Rather than increasing $K$ to achieve broader coverage, practitioners deploying TBW on smaller models may instead tailor the predefined topic lists to their specific use case, maintaining a modest number of topics while ensuring semantic alignment with the target domain. Prior work has demonstrated the feasibility of such domain-specific topic-based watermarking in specialized contexts such as AI-assisted peer review~\cite{nemecek-etal-2025-feasibility}, suggesting that thoughtful topic selection can preserve watermark effectiveness without requiring large vocabularies or extensive topic sets.

If TBW were to be deployed with a large number of topics, we would expect larger vocabulary models to yield better results. Our maximum $z$-score detection scheme, which achieves near-perfect separation between watermarked and non-watermarked text, requires evaluating all $K$ topic partitions to identify the strongest watermark signal. As the number of topics increases, detection time scales approximately linearly with $K$. We report these timing results in~\ref{ablation:num_topics}, observing an increase from approximately $2$ seconds at $K=4$ to $6.5$ seconds at $K=32$. While this overhead remains tractable, it represents a consideration for deployments requiring rapid detection at scale. Importantly, detection is performed offline by model owners rather than during user-facing generation, mitigating latency concerns in practice.

Beyond TBW's specific implementation, a limitation it shares with other list-based watermarking schemes, including KGW~\cite{kirchenbauer2023watermark} and SynthID-Text~\cite{dathathri2024scalable}, is reduced detection reliability on shorter text sequences. Our primary evaluations use 200-token generations, which we consider relatively short for watermark detection. The statistical nature of $z$-score-based detection requires sufficient tokens to distinguish the watermark signal from random variation; as text length decreases, the gap between watermarked and non-watermarked $z$-score distributions narrows, increasing the likelihood of misclassification. We hypothesize that detection performance would degrade on shorter sequences, a constraint inherent to the underlying framework.

\section*{Ethical Considerations}
Watermarking techniques inherently carry dual-use risks. While watermarking enables provenance tracking to combat misuse, it could potentially be exploited for punitive enforcement based on probabilistic signals. Adversarial actors may further weaponize watermarking to falsely attribute content to specific organizations or models. We additionally acknowledge residual concerns regarding false positives, which could lead to wrongful accusations. We recommend deployment with conservative thresholds, transparent documentation of error rates, and appeals processes for affected users, calibrated to the specific deployment context.

Our evaluation used a public dataset (\textsc{C4}) and locally-run models without probing live services or violating terms of service. The human evaluation study (\ref{main_human_eval}) involved only non-sensitive, model-generated text; participants provided informed consent, and no personally identifiable information was collected.

\section*{Acknowledgments}
This work was supported in part by the National Science Foundation under Grant No.\ OAC-2112606. ChatGPT and Claude were used in the preparation of this work for editing and grammar checking. No passages were copied without full author review, and all substantive ideas, analysis, and conclusions are the product and responsibility of the authors. The AI tools were also utilized for code development and early manuscript reviews.

\clearpage

\bibliography{custom}

@misc{openai2022chatgpt,
  title = {Introducing ChatGPT},
  author = {{OpenAI}},
  year = {2022},
  howpublished = {\url{https://openai.com/index/chatgpt/}},
  note = {Accessed: 2025-07-29}
}

@misc{dipperCode,
  title = {ai-detection-paraphrases},
  author = {Kalpesh Krishna},
  year = {2023},
  howpublished = {\url{https://github.com/martiansideofthemoon/ai-detection-paraphrases
}},
  note = {Accessed: 2025-08-01}
}

@misc{deepmind2024gemini,
  title        = {Introducing Gemini 2.0: Our New AI Model for the Agentic Era},
  author       = {{Google}},
  year         = {2024},
  howpublished = {\url{https://blog.google/technology/google-deepmind/google-gemini-ai-update-december-2024/}},
  note         = {Accessed: 2025-07-29}
}

@article{chen2024combating,
  title={Combating misinformation in the age of llms: Opportunities and challenges},
  author={Chen, Canyu and Shu, Kai},
  journal={AI Magazine},
  volume={45},
  number={3},
  pages={354--368},
  year={2024},
  publisher={Wiley Online Library}
}

@inproceedings{mueller2024llms,
  title={LLMs and memorization: On quality and specificity of copyright compliance},
  author={Mueller, Felix B and G{\"o}rge, Rebekka and Bernzen, Anna K and Pirk, Janna C and Poretschkin, Maximilian},
  booktitle={Proceedings of the AAAI/ACM Conference on AI, Ethics, and Society},
  volume={7},
  pages={984--996},
  year={2024}
}

@inproceedings{lee2023language,
  title={Do language models plagiarize?},
  author={Lee, Jooyoung and Le, Thai and Chen, Jinghui and Lee, Dongwon},
  booktitle={Proceedings of the ACM Web Conference 2023},
  pages={3637--3647},
  year={2023}
}

@article{shumailov2024ai,
  title={AI models collapse when trained on recursively generated data},
  author={Shumailov, Ilia and Shumaylov, Zakhar and Zhao, Yiren and Papernot, Nicolas and Anderson, Ross and Gal, Yarin},
  journal={Nature},
  volume={631},
  number={8022},
  pages={755--759},
  year={2024},
  publisher={Nature Publishing Group UK London}
}

@misc{li2021surveytextclassificationshallow,
      title={A Survey on Text Classification: From Shallow to Deep Learning}, 
      author={Qian Li and Hao Peng and Jianxin Li and Congying Xia and Renyu Yang and Lichao Sun and Philip S. Yu and Lifang He},
      year={2021},
      eprint={2008.00364},
      archivePrefix={arXiv},
      primaryClass={cs.CL},
      url={https://arxiv.org/abs/2008.00364}, 
}

@misc{openai2023aiclassifier,
  title        = {New AI classifier for indicating AI‑written text},
  author       = {{OpenAI}},
  year         = {2023},
  howpublished = {\url{https://openai.com/index/new-ai-classifier-for-indicating-ai-written-text}},
  note         = {Accessed: 2025-07-29}
}

@inproceedings{mitchell2023detectgpt,
  title={Detectgpt: Zero-shot machine-generated text detection using probability curvature},
  author={Mitchell, Eric and Lee, Yoonho and Khazatsky, Alexander and Manning, Christopher D and Finn, Chelsea},
  booktitle={International conference on machine learning},
  pages={24950--24962},
  year={2023},
  organization={PMLR}
}

@article{liang2023gpt,
  title={GPT detectors are biased against non-native English writers},
  author={Liang, Weixin and Yuksekgonul, Mert and Mao, Yining and Wu, Eric and Zou, James},
  journal={Patterns},
  volume={4},
  number={7},
  year={2023},
  publisher={Elsevier}
}

@inproceedings{kirchenbauer2023watermark,
  title={A watermark for large language models},
  author={Kirchenbauer, John and Geiping, Jonas and Wen, Yuxin and Katz, Jonathan and Miers, Ian and Goldstein, Tom},
  booktitle={International Conference on Machine Learning},
  pages={17061--17084},
  year={2023},
  organization={PMLR}
}

@misc{aaronson2023watermarking,
  title        = {Watermarking of Large Language Models},
  author       = {{Scott Aaronson}},
  year         = {2023},
  howpublished = {\url{https://simons.berkeley.edu/talks/scott-aaronson-ut-austin-openai-2023-08-17}},
  note         = {Accessed: 2025‑07‑29}
}

@misc{kuditipudi2024robustdistortionfreewatermarkslanguage,
      title={Robust Distortion-free Watermarks for Language Models}, 
      author={Rohith Kuditipudi and John Thickstun and Tatsunori Hashimoto and Percy Liang},
      year={2024},
      eprint={2307.15593},
      archivePrefix={arXiv},
      primaryClass={cs.LG},
      url={https://arxiv.org/abs/2307.15593}, 
}

@inproceedings{
zhao2024provable,
title={Provable Robust Watermarking for {AI}-Generated Text},
author={Xuandong Zhao and Prabhanjan Vijendra Ananth and Lei Li and Yu-Xiang Wang},
booktitle={The Twelfth International Conference on Learning Representations},
year={2024},
url={https://openreview.net/forum?id=SsmT8aO45L}
}

@article{dathathri2024scalable,
  title={Scalable watermarking for identifying large language model outputs},
  author={Dathathri, Sumanth and See, Abigail and Ghaisas, Sumedh and Huang, Po-Sen and McAdam, Rob and Welbl, Johannes and Bachani, Vandana and Kaskasoli, Alex and Stanforth, Robert and Matejovicova, Tatiana and others},
  journal={Nature},
  volume={634},
  number={8035},
  pages={818--823},
  year={2024},
  publisher={Nature Publishing Group UK London}
}

@misc{liu2024semanticinvariantrobustwatermark,
      title={A Semantic Invariant Robust Watermark for Large Language Models}, 
      author={Aiwei Liu and Leyi Pan and Xuming Hu and Shiao Meng and Lijie Wen},
      year={2024},
      eprint={2310.06356},
      archivePrefix={arXiv},
      primaryClass={cs.CR},
      url={https://arxiv.org/abs/2310.06356}, 
}

@inproceedings{hou-etal-2024-semstamp,
    title = "{S}em{S}tamp: A Semantic Watermark with Paraphrastic Robustness for Text Generation",
    author = "Hou, Abe  and
      Zhang, Jingyu  and
      He, Tianxing  and
      Wang, Yichen  and
      Chuang, Yung-Sung  and
      Wang, Hongwei  and
      Shen, Lingfeng  and
      Van Durme, Benjamin  and
      Khashabi, Daniel  and
      Tsvetkov, Yulia",
    editor = "Duh, Kevin  and
      Gomez, Helena  and
      Bethard, Steven",
    booktitle = "Proceedings of the 2024 Conference of the North American Chapter of the Association for Computational Linguistics: Human Language Technologies (Volume 1: Long Papers)",
    month = jun,
    year = "2024",
    address = "Mexico City, Mexico",
    publisher = "Association for Computational Linguistics",
    url = "https://aclanthology.org/2024.naacl-long.226/",
    doi = "10.18653/v1/2024.naacl-long.226",
    pages = "4067--4082"
}

@inproceedings{hou-etal-2024-k,
    title = "k-{S}em{S}tamp: A Clustering-Based Semantic Watermark for Detection of Machine-Generated Text",
    author = "Hou, Abe  and
      Zhang, Jingyu  and
      Wang, Yichen  and
      Khashabi, Daniel  and
      He, Tianxing",
    editor = "Ku, Lun-Wei  and
      Martins, Andre  and
      Srikumar, Vivek",
    booktitle = "Findings of the Association for Computational Linguistics: ACL 2024",
    month = aug,
    year = "2024",
    address = "Bangkok, Thailand",
    publisher = "Association for Computational Linguistics",
    url = "https://aclanthology.org/2024.findings-acl.98/",
    doi = "10.18653/v1/2024.findings-acl.98",
    pages = "1706--1715"
}

@inproceedings{huo2024tokenspecificwatermarkingenhanceddetectability,
author = {Huo, Mingjia and Somayajula, Sai Ashish and Liang, Youwei and Zhang, Ruisi and Koushanfar, Farinaz and Xie, Pengtao},
title = {Token-specific watermarking with enhanced detectability and semantic coherence for large language models},
year = {2024},
publisher = {JMLR.org},
booktitle = {Proceedings of the 41st International Conference on Machine Learning},
articleno = {833},
numpages = {22},
location = {Vienna, Austria},
series = {ICML'24}
}

@inproceedings{liu2024adaptivetextwatermarklarge, author = {Liu, Yepeng and Bu, Yuheng}, title = {Adaptive text watermark for large language models}, year = {2024}, publisher = {JMLR.org}, booktitle = {Proceedings of the 41st International Conference on Machine Learning}, articleno = {1238}, numpages = {20}, location = {Vienna, Austria}, series = {ICML'24} }

@inproceedings{zhang2024remarkllmrobustefficientwatermarking,
author = {Zhang, Ruisi and Hussain, Shehzeen Samarah and Neekhara, Paarth and Koushanfar, Farinaz},
title = {REMARK-LLM: a robust and efficient watermarking framework for generative large language models},
year = {2024},
isbn = {978-1-939133-44-1},
publisher = {USENIX Association},
address = {USA},
booktitle = {Proceedings of the 33rd USENIX Conference on Security Symposium},
articleno = {102},
numpages = {18},
location = {Philadelphia, PA, USA},
series = {SEC '24}
}

@misc{sato2023embarrassinglysimpletextwatermarks,
      title={Embarrassingly Simple Text Watermarks}, 
      author={Ryoma Sato and Yuki Takezawa and Han Bao and Kenta Niwa and Makoto Yamada},
      year={2023},
      eprint={2310.08920},
      archivePrefix={arXiv},
      primaryClass={cs.LG},
      url={https://arxiv.org/abs/2310.08920}, 
}

@inproceedings{jovanovi2024stealing,
author = {Jovanovi\'{c}, Nikola and Staab, Robin and Vechev, Martin},
title = {Watermark stealing in large language models},
year = {2024},
publisher = {JMLR.org},
booktitle = {Proceedings of the 41st International Conference on Machine Learning},
articleno = {907},
numpages = {24},
location = {Vienna, Austria},
series = {ICML'24}
}

@inproceedings{dipmark2024,
author = {Wu, Yihan and Hu, Zhengmian and Guo, Junfeng and Zhang, Hongyang and Huang, Heng},
title = {A resilient and accessible distribution-preserving watermark for large language models},
year = {2024},
publisher = {JMLR.org},
booktitle = {Proceedings of the 41st International Conference on Machine Learning},
articleno = {2190},
numpages = {28},
location = {Vienna, Austria},
series = {ICML'24}
}

@inproceedings{lee-etal-2024-wrote,
    title = "Who Wrote this Code? Watermarking for Code Generation",
    author = "Lee, Taehyun  and
      Hong, Seokhee  and
      Ahn, Jaewoo  and
      Hong, Ilgee  and
      Lee, Hwaran  and
      Yun, Sangdoo  and
      Shin, Jamin  and
      Kim, Gunhee",
    editor = "Ku, Lun-Wei  and
      Martins, Andre  and
      Srikumar, Vivek",
    booktitle = "Proceedings of the 62nd Annual Meeting of the Association for Computational Linguistics (Volume 1: Long Papers)",
    month = aug,
    year = "2024",
    address = "Bangkok, Thailand",
    publisher = "Association for Computational Linguistics",
    url = "https://aclanthology.org/2024.acl-long.268/",
    doi = "10.18653/v1/2024.acl-long.268",
    pages = "4890--4911",
}

@inproceedings{reimers-gurevych-2020-making,
    title = "Making Monolingual Sentence Embeddings Multilingual using Knowledge Distillation",
    author = "Reimers, Nils  and
      Gurevych, Iryna",
    editor = "Webber, Bonnie  and
      Cohn, Trevor  and
      He, Yulan  and
      Liu, Yang",
    booktitle = "Proceedings of the 2020 Conference on Empirical Methods in Natural Language Processing (EMNLP)",
    month = nov,
    year = "2020",
    address = "Online",
    publisher = "Association for Computational Linguistics",
    url = "https://aclanthology.org/2020.emnlp-main.365/",
    doi = "10.18653/v1/2020.emnlp-main.365",
    pages = "4512--4525",
}

@misc{grootendorst2020keybert,
  title        = {KeyBERT: Minimal keyword extraction with BERT},
  author       = {Maarten Grootendorst},
  year         = {2020},
  howpublished = {\url{https://maartengr.github.io/KeyBERT/}},
  note         = {Accessed: 2025‑07‑29}
}

@article{c4,
author = {Raffel, Colin and Shazeer, Noam and Roberts, Adam and Lee, Katherine and Narang, Sharan and Matena, Michael and Zhou, Yanqi and Li, Wei and Liu, Peter J.},
title = {Exploring the limits of transfer learning with a unified text-to-text transformer},
year = {2020},
issue_date = {January 2020},
publisher = {JMLR.org},
volume = {21},
number = {1},
issn = {1532-4435},
journal = {J. Mach. Learn. Res.},
month = jan,
articleno = {140},
numpages = {67}
}

@misc{zhang2022optopenpretrainedtransformer,
      title={OPT: Open Pre-trained Transformer Language Models}, 
      author={Susan Zhang and Stephen Roller and Naman Goyal and Mikel Artetxe and Moya Chen and Shuohui Chen and Christopher Dewan and Mona Diab and Xian Li and Xi Victoria Lin and Todor Mihaylov and Myle Ott and Sam Shleifer and Kurt Shuster and Daniel Simig and Punit Singh Koura and Anjali Sridhar and Tianlu Wang and Luke Zettlemoyer},
      year={2022},
      eprint={2205.01068},
      archivePrefix={arXiv},
      primaryClass={cs.CL},
      url={https://arxiv.org/abs/2205.01068}, 
}

@article{team2024gemma,
  title={Gemma: Open models based on gemini research and technology},
  author={Team, Gemma and Mesnard, Thomas and Hardin, Cassidy and Dadashi, Robert and Bhupatiraju, Surya and Pathak, Shreya and Sifre, Laurent and Rivi{\`e}re, Morgane and Kale, Mihir Sanjay and Love, Juliette and others},
  journal={arXiv preprint arXiv:2403.08295},
  year={2024}
}

@inproceedings{pan-etal-2024-markllm,
    title = "{M}ark{LLM}: An Open-Source Toolkit for {LLM} Watermarking",
    author = "Pan, Leyi  and
      Liu, Aiwei  and
      He, Zhiwei  and
      Gao, Zitian  and
      Zhao, Xuandong  and
      Lu, Yijian  and
      Zhou, Binglin  and
      Liu, Shuliang  and
      Hu, Xuming  and
      Wen, Lijie  and
      King, Irwin  and
      Yu, Philip S.",
    editor = "Hernandez Farias, Delia Irazu  and
      Hope, Tom  and
      Li, Manling",
    booktitle = "Proceedings of the 2024 Conference on Empirical Methods in Natural Language Processing: System Demonstrations",
    month = nov,
    year = "2024",
    address = "Miami, Florida, USA",
    publisher = "Association for Computational Linguistics",
    url = "https://aclanthology.org/2024.emnlp-demo.7/",
    doi = "10.18653/v1/2024.emnlp-demo.7",
    pages = "61--71",
}

@article{grattafiori2024llama,
  title={The llama 3 herd of models},
  author={Grattafiori, Aaron and Dubey, Abhimanyu and Jauhri, Abhinav and Pandey, Abhinav and Kadian, Abhishek and Al-Dahle, Ahmad and Letman, Aiesha and Mathur, Akhil and Schelten, Alan and Vaughan, Alex and others},
  journal={arXiv preprint arXiv:2407.21783},
  year={2024}
}

@inproceedings{pegasus,
author = {Zhang, Jingqing and Zhao, Yao and Saleh, Mohammad and Liu, Peter J.},
title = {PEGASUS: pre-training with extracted gap-sentences for abstractive summarization},
year = {2020},
publisher = {JMLR.org},
booktitle = {Proceedings of the 37th International Conference on Machine Learning},
articleno = {1051},
numpages = {12},
series = {ICML'20}
}

@inproceedings{dipper,
author = {Krishna, Kalpesh and Song, Yixiao and Karpinska, Marzena and Wieting, John and Iyyer, Mohit},
title = {Paraphrasing evades detectors of AI-generated text, but retrieval is an effective defense},
year = {2023},
publisher = {Curran Associates Inc.},
address = {Red Hook, NY, USA},
booktitle = {Proceedings of the 37th International Conference on Neural Information Processing Systems},
articleno = {1195},
numpages = {32},
location = {New Orleans, LA, USA},
series = {NIPS '23}
}

@inproceedings{cachola-etal-2020-tldr,
    title = "{TLDR}: Extreme Summarization of Scientific Documents",
    author = "Cachola, Isabel  and
      Lo, Kyle  and
      Cohan, Arman  and
      Weld, Daniel",
    editor = "Cohn, Trevor  and
      He, Yulan  and
      Liu, Yang",
    booktitle = "Findings of the Association for Computational Linguistics: EMNLP 2020",
    month = nov,
    year = "2020",
    address = "Online",
    publisher = "Association for Computational Linguistics",
    url = "https://aclanthology.org/2020.findings-emnlp.428/",
    doi = "10.18653/v1/2020.findings-emnlp.428",
    pages = "4766--4777",
}

@inproceedings{niess-kern-2025-ensemble,
    title = "Ensemble Watermarks for Large Language Models",
    author = "Niess, Georg  and
      Kern, Roman",
    editor = "Che, Wanxiang  and
      Nabende, Joyce  and
      Shutova, Ekaterina  and
      Pilehvar, Mohammad Taher",
    booktitle = "Proceedings of the 63rd Annual Meeting of the Association for Computational Linguistics (Volume 1: Long Papers)",
    month = jul,
    year = "2025",
    address = "Vienna, Austria",
    publisher = "Association for Computational Linguistics",
    url = "https://aclanthology.org/2025.acl-long.145/",
    pages = "2903--2916",
    ISBN = "979-8-89176-251-0",
}

@misc{mao2025watermarkinglowentropygenerationlarge,
      title={Watermarking Low-entropy Generation for Large Language Models: An Unbiased and Low-risk Method}, 
      author={Minjia Mao and Dongjun Wei and Zeyu Chen and Xiao Fang and Michael Chau},
      year={2025},
      eprint={2405.14604},
      archivePrefix={arXiv},
      primaryClass={cs.CL},
      url={https://arxiv.org/abs/2405.14604}, 
}

@inproceedings{
feng2025bimark,
title={BiMark: Unbiased Multilayer Watermarking for Large Language Models},
author={Xiaoyan Feng and He Zhang and Yanjun Zhang and Leo Yu Zhang and Shirui Pan},
booktitle={Forty-second International Conference on Machine Learning},
year={2025},
url={https://openreview.net/forum?id=Zvyb3WAg03}
}

@article{lewis2004rcv1,
  title={Rcv1: A new benchmark collection for text categorization research},
  author={Lewis, David D and Yang, Yiming and Rose, Tony G and Li, Fan},
  journal={Journal of machine learning research},
  volume={5},
  number={Apr},
  pages={361--397},
  year={2004}
}

@book{fellbaum1998wordnet,
  title={WordNet: An electronic lexical database},
  author={Fellbaum, Christiane},
  year={1998},
  publisher={MIT press}
}

@misc{qu2025provablyrobustmultibitwatermarking,
      title={Provably Robust Multi-bit Watermarking for AI-generated Text}, 
      author={Wenjie Qu and Wengrui Zheng and Tianyang Tao and Dong Yin and Yanze Jiang and Zhihua Tian and Wei Zou and Jinyuan Jia and Jiaheng Zhang},
      year={2025},
      eprint={2401.16820},
      archivePrefix={arXiv},
      primaryClass={cs.CR},
      url={https://arxiv.org/abs/2401.16820}, 
}

@inproceedings{tu-etal-2024-waterbench,
    title = "{W}ater{B}ench: Towards Holistic Evaluation of Watermarks for Large Language Models",
    author = "Tu, Shangqing  and
      Sun, Yuliang  and
      Bai, Yushi  and
      Yu, Jifan  and
      Hou, Lei  and
      Li, Juanzi",
    editor = "Ku, Lun-Wei  and
      Martins, Andre  and
      Srikumar, Vivek",
    booktitle = "Proceedings of the 62nd Annual Meeting of the Association for Computational Linguistics (Volume 1: Long Papers)",
    month = aug,
    year = "2024",
    address = "Bangkok, Thailand",
    publisher = "Association for Computational Linguistics",
    url = "https://aclanthology.org/2024.acl-long.83/",
    doi = "10.18653/v1/2024.acl-long.83",
    pages = "1517--1542",
}

@misc{zhang2020bertscoreevaluatingtextgeneration,
      title={BERTScore: Evaluating Text Generation with BERT}, 
      author={Tianyi Zhang and Varsha Kishore and Felix Wu and Kilian Q. Weinberger and Yoav Artzi},
      year={2020},
      eprint={1904.09675},
      archivePrefix={arXiv},
      primaryClass={cs.CL},
      url={https://arxiv.org/abs/1904.09675}, 
}

@misc{newsweek2025_best_hospitals,
  author       = {{Newsweek}},
  title        = {{World's Best Hospitals 2025 - United States of America}},
  howpublished = {\url{https://rankings.newsweek.com/worlds-best-hospitals-2025/united-states-america}},
  year         = {2025},
  note         = {Accessed: 2025-08-25}
}

@misc{clevelandclinic2025_ambienceAI,
  author       = {{Cleveland Clinic}},
  title        = {{Cleveland Clinic Announces Rollout of Ambience Healthcare's AI Platform}},
  howpublished = {\url{https://newsroom.clevelandclinic.org/2025/02/19/cleveland-clinic-announces-the-rollout-of-ambience-healthcares-ai-platform}},
  year         = {2025},
  month        = feb,
  day          = {19},
  note         = {Accessed: 2025-08-25}
}

@misc{aaai2025_peer_review_assessment,
  author       = {{Association for the Advancement of Artificial Intelligence (AAAI)}},
  title        = {{AAAI Launches AI-Powered Peer Review Assessment System}},
  howpublished = {\url{https://aaai.org/aaai-launches-ai-powered-peer-review-assessment-system/}},
  year         = {2025},
  month        = may,
  day          = {16},
  note         = {Accessed: 2025-08-25}
}

@online{deepmind_synthid_2024,
  author       = {{SynthID-Team}},
  title        = {Watermarking AI-generated text and video with SynthID},
  year         = {2024},
  url          = {https://deepmind.google/discover/blog/watermarking-ai-generated-text-and-video-with-synthid/},
  note         = {Accessed: 2025-08-26}
}

@misc{devlin2019bertpretrainingdeepbidirectional,
      title={BERT: Pre-training of Deep Bidirectional Transformers for Language Understanding}, 
      author={Jacob Devlin and Ming-Wei Chang and Kenton Lee and Kristina Toutanova},
      year={2019},
      eprint={1810.04805},
      archivePrefix={arXiv},
      primaryClass={cs.CL},
      url={https://arxiv.org/abs/1810.04805}, 
}

@misc{bafna2024mastkalandarsemeval2024task,
      title={Mast Kalandar at SemEval-2024 Task 8: On the Trail of Textual Origins: RoBERTa-BiLSTM Approach to Detect AI-Generated Text}, 
      author={Jainit Sushil Bafna and Hardik Mittal and Suyash Sethia and Manish Shrivastava and Radhika Mamidi},
      year={2024},
      eprint={2407.02978},
      archivePrefix={arXiv},
      primaryClass={cs.CL},
      url={https://arxiv.org/abs/2407.02978}, 
}

@misc{islam2023distinguishinghumangeneratedtext,
      title={Distinguishing Human Generated Text From ChatGPT Generated Text Using Machine Learning}, 
      author={Niful Islam and Debopom Sutradhar and Humaira Noor and Jarin Tasnim Raya and Monowara Tabassum Maisha and Dewan Md Farid},
      year={2023},
      eprint={2306.01761},
      archivePrefix={arXiv},
      primaryClass={cs.CL},
      url={https://arxiv.org/abs/2306.01761}, 
}

@misc{huggingface2025,
  author       = {{Hugging Face}},
  title        = {Hugging Face - The AI community building the future},
  year         = {2025},
  howpublished = {\url{https://huggingface.co}},
  note         = {Accessed: 2025-08-26}
}

@inproceedings{nemecek-etal-2025-feasibility,
    title = "The Feasibility of Topic-Based Watermarking on Academic Peer Reviews",
    author = "Nemecek, Alexander  and
      Jiang, Yuzhou  and
      Ayday, Erman",
    editor = "Inui, Kentaro  and
      Sakti, Sakriani  and
      Wang, Haofen  and
      Wong, Derek F.  and
      Bhattacharyya, Pushpak  and
      Banerjee, Biplab  and
      Ekbal, Asif  and
      Chakraborty, Tanmoy  and
      Singh, Dhirendra Pratap",
    booktitle = "Proceedings of the 14th International Joint Conference on Natural Language Processing and the 4th Conference of the Asia-Pacific Chapter of the Association for Computational Linguistics",
    month = dec,
    year = "2025",
    address = "Mumbai, India",
    publisher = "The Asian Federation of Natural Language Processing and The Association for Computational Linguistics",
    url = "https://aclanthology.org/2025.findings-ijcnlp.36/",
    doi = "10.18653/v1/2025.findings-ijcnlp.36",
    pages = "616--634",
    ISBN = "979-8-89176-303-6"
}

@inproceedings{chang2024postmark,
  title={Postmark: A robust blackbox watermark for large language models},
  author={Chang, Yapei and Krishna, Kalpesh and Houmansadr, Amir and Wieting, John Frederick and Iyyer, Mohit},
  booktitle={Proceedings of the 2024 Conference on Empirical Methods in Natural Language Processing},
  pages={8969--8987},
  year={2024}
}

\clearpage

\appendix

\section*{Appendix Contents}
\addcontentsline{toc}{section}{Appendix Contents}
\begin{itemize}
    \item[\ref{algo_implementations}] \nameref{algo_implementations} \dotfill \pageref{algo_implementations}
    
    \item[\ref{extended_threat_model}] \nameref{extended_threat_model} \dotfill \pageref{extended_threat_model}

    \item[\ref{detection_discussion}] \nameref{detection_discussion} \dotfill \pageref{detection_discussion}
    \begin{itemize} 
        \item[\ref{strict_sliding_detection}] \nameref{strict_sliding_detection} \dotfill \pageref{strict_sliding_detection}
        \item[\ref{max_z_detection}] \nameref{max_z_detection} \dotfill \pageref{max_z_detection}
    \end{itemize}

    \item[\ref{markllm}] \nameref{markllm} \dotfill \pageref{markllm}

    \item[\ref{additional_eval_results_from_main_text}] \nameref{additional_eval_results_from_main_text} \dotfill \pageref{additional_eval_results_from_main_text}
    \begin{itemize} 
        \item[\ref{more_perplexity}] \nameref{more_perplexity} \dotfill \pageref{more_perplexity}
        \item[\ref{main_human_eval}] \nameref{main_human_eval} \dotfill \pageref{main_human_eval}
        \item[\ref{llm-as-a-judge}] \nameref{llm-as-a-judge} \dotfill \pageref{llm-as-a-judge}
        \item[\ref{small_eff}] \nameref{small_eff} \dotfill \pageref{small_eff}
        \item[\ref{its_edit}] \nameref{its_edit} \dotfill \pageref{its_edit}
        \item[\ref{semantic_comparison}] \nameref{semantic_comparison} \dotfill \pageref{semantic_comparison}
        \item[\ref{auc_roc}] \nameref{auc_roc} \dotfill \pageref{auc_roc}
        \item[\ref{fpr_analysis}] \nameref{fpr_analysis} \dotfill \pageref{fpr_analysis}
        \item[\ref{detection_attacks}] \nameref{detection_attacks} \dotfill \pageref{detection_attacks}
        \item[\ref{max_z_detailed}] \nameref{max_z_detailed} \dotfill \pageref{max_z_detailed}
    \end{itemize}

    \item[\ref{ablation_studies_from_main_text}] \nameref{ablation_studies_from_main_text} \dotfill \pageref{ablation_studies_from_main_text}
    \begin{itemize} 
        \item[\ref{sensitive}] \nameref{sensitive} \dotfill \pageref{sensitive}
        \item[\ref{parameter_space}] \nameref{parameter_space} \dotfill \pageref{parameter_space}
        \item[\ref{task_specific_ablation}] \nameref{task_specific_ablation} \dotfill \pageref{task_specific_ablation}
        \item[\ref{ablation:num_topics}] \nameref{ablation:num_topics} \dotfill \pageref{ablation:num_topics}
    \end{itemize}

    \item[\ref{attack_config}] \nameref{attack_config} \dotfill \pageref{attack_config}
    \begin{itemize} 
        \item[\ref{details_on_paraphrasing_attacks}] \nameref{details_on_paraphrasing_attacks} \dotfill \pageref{details_on_paraphrasing_attacks}
        \item[\ref{details_on_perturbation_attacks}] \nameref{details_on_perturbation_attacks} \dotfill \pageref{details_on_perturbation_attacks}
    \end{itemize}

    \item[\ref{examples}] \nameref{examples} \dotfill \pageref{examples}
    
\end{itemize}

\section{Algorithm Implementations}\label{algo_implementations}
We provide implementation details for our topic-based watermarking approach. Algorithm~\ref{alg:token_mapping} presents the token-to-topic mapping procedure, while Algorithm~\ref{alg:topic-watermarking} details the watermark generation process. 
\begin{algorithm*}[h!]
   \caption{Token-to-Topic Mapping}
   \label{alg:token_mapping}
\begin{algorithmic}
   \STATE \textbf{Input:} Vocabulary $V$, predefined topic set $\{t_1, \dots, t_K\}$, embedding function $\text{Enc}(\cdot)$, similarity threshold $\tau$.
   \STATE Compute topic embeddings: $E_T = \{\mathbf{e}_{t_i} \mid t_i \in \{t_1, \dots, t_K\}\}$
   \STATE Compute token embeddings: $E_V = \{\mathbf{e}_v \mid v \in V\}$
   \STATE Initialize topic-aligned lists: $G_{t_i} = \emptyset, \forall i \in \{1, \dots, K\}$
   \STATE Initialize residual set: $\mathcal{B} = \emptyset$
   
   \FOR{each token $v \in V$}
      \STATE Compute similarity scores: $\textit{sim}(v, t_i) = \frac{\mathbf{e}_v \cdot \mathbf{e}_{t_i}}{\|\mathbf{e}_v\| \|\mathbf{e}_{t_i}\|}, \forall i$
      \STATE $m, i^* \gets \max(\textit{sim}(v, t_i)), \arg\max(\textit{sim}(v, t_i))$
      \IF{$m \geq \tau$}
         \STATE Assign $v$ to topic $t_{i^*}$: $G_{t_{i^*}} \gets G_{t_{i^*}} \cup \{v\}$
      \ELSE
         \STATE Add $v$ to residual set: $\mathcal{B} \gets \mathcal{B} \cup \{v\}$
      \ENDIF
   \ENDFOR
   
   \STATE Distribute remaining tokens:
   \STATE Initialize counter: $i \gets 1$
   \FOR{each token $v \in \mathcal{B}$}
      \STATE Assign $v$ to $t_{\text{target}} = t_{(i \bmod K) + 1}$
      \STATE $G_{t_{\text{target}}} \gets G_{t_{\text{target}}} \cup \{v\}$
      \STATE $i \gets i + 1$
   \ENDFOR

   \STATE \textbf{Return} $\{G_{t_1}, \dots, G_{t_K}\}$ \COMMENT{Final topic-aligned token lists}
\end{algorithmic}
\end{algorithm*}

\begin{algorithm*}[h!]
   \caption{Topic-Based Watermark Generation}
   \label{alg:topic-watermarking}
\begin{algorithmic}
   \STATE \textbf{Input:} Prompt $\mathbf{x}^{\text{prompt}}$, topic set $\{t_1, \dots, t_K\}$, topic-aligned lists $\{G_{t_1}, \dots, G_{t_K}\}$, logit bias $\delta$.
   
   \STATE Extract topics: $\mathcal{T}_{\text{detected}} \gets \text{KeyBERT}(\mathbf{x}^{\text{prompt}})$
   
   \IF{$\exists t_i \in \{t_1, \dots, t_K\}$ such that $t_i \in \mathcal{T}_{\text{detected}}$}
      \STATE Select direct match: $t^* \gets t_i$ 
   \ELSE
      \STATE Assign via clustering: \\ $t^* \gets \text{KMeans}(\mathcal{T}_{\text{detected}}, \{t_1, \dots, t_K\})$
   \ENDIF

   \STATE Retrieve topic-aligned list: $G_{t^*} \gets G_{t^*}$
   \STATE Initialize output sequence: $\mathbf{z} \gets \emptyset$
   
   \WHILE{not end-of-sequence}
      \STATE Compute logits: $\mathbf{logits} \gets p_{\theta}(\cdot \mid \mathbf{x}^{\text{prompt}}, \mathbf{z})$ 
      \FOR{each token $v \in V$}
         \IF{$v \in G_{t^*}$}
            \STATE Adjust logit: $\mathbf{logits}[v] \gets \mathbf{logits}[v] + \delta$
         \ENDIF
      \ENDFOR
      \STATE Compute probabilities: $\mathbf{p} \gets \text{Softmax}(\mathbf{logits})$
      \STATE Sample next token: $v_{\text{next}} \gets \text{SampleToken}(\mathbf{p})$
      \STATE Append token to sequence: $\mathbf{z} \gets \mathbf{z} \cup \{v_{\text{next}}\}$
   \ENDWHILE

   \STATE \textbf{Return} $\mathbf{z}$ \COMMENT{Watermarked output text}
\end{algorithmic}
\end{algorithm*}

\section{Extended Threat Model}\label{extended_threat_model}
This section provides additional formal constraints and justifications for the adversary assumptions introduced in \S\ref{threat_model}.

\textbf{Adversary Assumptions and Constraints.} 
We assume an adversary consistent with standard threat models adopted across the watermarking literature~\cite{kirchenbauer2023watermark, zhao2024provable, zhang2024remarkllmrobustefficientwatermarking}. In particular, the adversary cannot fine-tune or retrain the model, nor inspect its parameters or internal mechanisms. Interaction is strictly via black-box queries to the generator; the adversary has no oracle access to the watermark detection algorithm and receives neither per-text decisions nor detection scores from the system. While the detection algorithmic form may be public, the operating keys, topic partitions, and thresholds remain private, and detection is performed offline by a trusted party. This scope reflects realistic end-user capabilities and explicitly excludes insider or white-box attacks where model internals are visible.

\textbf{Excluded and Impractical Attacks.} We further assume the adversary cannot mount impractical strategies that rely on excessive querying or access to paired datasets (watermarked vs. non-watermarked versions of the same prompt). Statistical comparison or supervised classification attacks are excluded from our primary model as they require assumptions that are rarely feasible for typical end-users. We provide an empirical evaluation of these ``detection-based'' attacks in~\ref{detection_attacks} for completeness, though they fall outside our primary security focus.

\textbf{Partition Recovery Attacks.} Prior work has demonstrated partition recovery attacks against KGW-style watermarks, where adversaries approximate green/red list assignments through repeated positional querying~\cite{jovanovi2024stealing}. Because TBW uses a single fixed green list per generation rather than per-token hashing, this specific attack vector does not directly transfer. However, TBW's semantic partitioning introduces a different concern: an adversary who knows the algorithm, topic list, and embedding approach might attempt to reconstruct the topic-aligned green lists through structural inference.

To evaluate this, we consider a strong attacker who knows TBW's partitioning algorithm, the predefined topic set, and the general embedding approach, but not the private parameters: the specific embedding model weights, similarity threshold $\tau$, or the random seed used for round-robin distribution of residual tokens. We measure partition recovery across 12 attacker configurations, varying the embedding model (3 models, including an oracle configuration using the identical architecture as the watermark deployer) and similarity threshold ($\tau \in \{0.3, 0.5, 0.7, 0.9\}$). For each configuration, we compute Jaccard similarity, precision, recall, and F1 between the attacker's reconstructed partition and the true partition, averaged across all four topic lists. Tables~\ref{tab:partition_recovery_opt} and~\ref{tab:partition_recovery_gemma} report results for \textsc{OPT-6.7B} and \textsc{Gemma-7B}, respectively.

\begin{table}[h]
\centering
\scriptsize
\begin{tabular}{llcccc}
\toprule
\textbf{Model} & $\boldsymbol{\tau}$ & \textbf{Jaccard} & \textbf{Precision} & \textbf{Recall} & \textbf{F1} \\
\midrule
all-MiniLM & 0.3 & 0.1433 & 0.2512 & 0.2514 & 0.2505 \\
all-MiniLM & 0.5 & 0.1448 & 0.2529 & 0.2529 & 0.2529 \\
all-MiniLM & 0.7 & 0.1419 & 0.2486 & 0.2486 & 0.2486 \\
all-MiniLM & 0.9 & 0.1415 & 0.2478 & 0.2478 & 0.2478 \\
\midrule
all-mpnet & 0.3 & 0.1423 & 0.2491 & 0.2492 & 0.2491 \\
all-mpnet & 0.5 & 0.1414 & 0.2478 & 0.2478 & 0.2478 \\
all-mpnet & 0.7 & 0.1413 & 0.2476 & 0.2476 & 0.2476 \\
all-mpnet & 0.9 & 0.1415 & 0.2478 & 0.2478 & 0.2478 \\
\midrule
bge-small & 0.3 & 0.1299 & 0.2546 & 0.2532 & 0.2256 \\
bge-small & 0.5 & 0.1301 & 0.2542 & 0.2532 & 0.2259 \\
bge-small & 0.7 & 0.1423 & 0.2491 & 0.2491 & 0.2491 \\
bge-small & 0.9 & 0.1438 & 0.2515 & 0.2515 & 0.2515 \\
\bottomrule
\end{tabular}
\caption{Partition recovery attack results on \textsc{OPT-6.7B} across 12 attacker configurations. The all-MiniLM rows represent an oracle attacker using the identical embedding model as the watermark deployer.}
\label{tab:partition_recovery_opt}
\end{table}

\begin{table}[h]
\centering
\scriptsize
\begin{tabular}{llcccc}
\toprule
\textbf{Model} & $\boldsymbol{\tau}$ & \textbf{Jaccard} & \textbf{Precision} & \textbf{Recall} & \textbf{F1} \\
\midrule
all-MiniLM & 0.3 & 0.1427 & 0.2498 & 0.2498 & 0.2497 \\
all-MiniLM & 0.5 & 0.1422 & 0.2490 & 0.2490 & 0.2490 \\
all-MiniLM & 0.7 & 0.1425 & 0.2494 & 0.2494 & 0.2494 \\
all-MiniLM & 0.9 & 0.1434 & 0.2509 & 0.2509 & 0.2509 \\
\midrule
all-mpnet & 0.3 & 0.1181 & 0.2489 & 0.2491 & 0.2064 \\
all-mpnet & 0.5 & 0.1428 & 0.2499 & 0.2499 & 0.2499 \\
all-mpnet & 0.7 & 0.1445 & 0.2525 & 0.2525 & 0.2525 \\
all-mpnet & 0.9 & 0.1434 & 0.2509 & 0.2509 & 0.2509 \\
\midrule
bge-small & 0.3 & 0.0940 & 0.2502 & 0.2498 & 0.1617 \\
bge-small & 0.5 & 0.0940 & 0.2497 & 0.2497 & 0.1618 \\
bge-small & 0.7 & 0.1432 & 0.2506 & 0.2506 & 0.2506 \\
bge-small & 0.9 & 0.1431 & 0.2503 & 0.2503 & 0.2503 \\
\bottomrule
\end{tabular}
\caption{Partition recovery attack results on \textsc{Gemma-7B} across 12 attacker configurations. Results are consistent with \textsc{OPT-6.7B}, confirming chance-level recovery across both models.}
\label{tab:partition_recovery_gemma}
\end{table}

Even the oracle attacker using the identical embedding model (all-MiniLM) and matching $\tau$ achieves only chance-level recovery across both models. With $K=4$ topics, random assignment would yield Jaccard $= 0.25$ and F1 $= 0.25$ in expectation; the observed values fall at or below this baseline across all configurations. Some non-oracle configurations (e.g., bge-small at $\tau=0.3$) perform below chance, reflecting topic-specific misalignment that further underscores the difficulty of partition inference. This result stems from TBW's dual assignment mechanism: tokens exceeding the similarity threshold $\tau$ are assigned semantically, while residual tokens are distributed via a randomly shuffled round-robin process (\S\ref{proposed}). Since the random seed governing the round-robin assignment is private, the full partition is unrecoverable regardless of the attacker's knowledge of the embedding model or threshold. Semantic co-grouping alone is insufficient without the private round-robin seed. As complementary evidence, our word cloud analysis (Figure~\ref{fig:wordclouds_combined}) confirms that watermarked outputs show no systematic elevation of topic-specific tokens, preventing observational inference of the active partition from generated text.

\section{Practical Use of Detection Schemes}\label{detection_discussion}
\subsection{Strict and Sliding Window Detection}\label{strict_sliding_detection}
As discussed in \S\ref{detection_schemes}, both the strict and sliding window detection schemes exhibit degraded performance due to imperfect topic matching between input prompts and output text. This limitation arises because the detection pipeline assumes alignment between topics extracted from the input and the generated output, an assumption that often fails in general-purpose, black-box deployment. However, there are domains where such access to the input text is natural and the trade-off becomes favorable.

For example, \S\ref{ablation:num_topics} shows that increasing the number of topics from $K=4$ to $K=32$ increases detection runtime under maximum $z$-score detection (from approximately 2 seconds to 6.5 seconds). In latency-sensitive applications, strict or sliding detection may therefore be more practical when input prompts are accessible, since they avoid evaluating all topic partitions. Healthcare settings provide a clear illustration: Cleveland Clinic, ranked among the world's top hospitals~\cite{newsweek2025_best_hospitals}, recently announced deployment of an ambient AI platform for clinical documentation~\cite{clevelandclinic2025_ambienceAI}. In such environments, LLM-generated text is deeply integrated into provider workflows, and provenance guarantees are paramount to avoid blind reliance on automatically produced notes. With multiple campuses and a high volume of patients, runtime efficiency is critical, and strict/sliding detection can provide timely watermark checks while still leveraging topic-aware alignment. The natural pairing of spoken prompts (provider voice memos) with textual outputs further supports this alignment assumption.

Similarly, the Association for the Advancement of Artificial Intelligence (AAAI) initiative on AI-assisted peer review for 2026~\cite{aaai2025_peer_review_assessment} offers another case where the input, the submitted paper, is always available alongside generated reviewing text. Here, strict and sliding window detection align directly with the review pipeline, enabling topic-aware watermark verification without the cost of iterating across all topic partitions. Looking forward, these use cases also suggest an avenue for future work allowing for hybrid ``sliding-chunk'' watermarking that adapts to multiple topical segments within long-form input documents, such as scientific papers, while retaining prompt-to-output alignment.

\subsection{Maximum $z$-Score Detection}\label{max_z_detection}
In contrast, the maximum $z$-score detection scheme (\S\ref{detection_schemes}) is best suited for general deployment where access to input prompts cannot be assumed. In such settings, strict or sliding detection would fail due to topic misalignment, whereas maximum $z$-score avoids this assumption by evaluating all topic partitions and selecting the strongest watermark signal. While detection incurs additional cost as the number of topics increases (e.g., 6.5 seconds at $K=32$), our evaluation shows that this overhead remains tractable, particularly relative to the robustness gains. Maximum z-score detection delivers near-perfect separation of watermarked and non-watermarked text even under strong adversarial attacks, making it the preferred scheme for open-world scenarios where robustness is the overriding priority.

\section{Watermarking Evaluation Parameters}\label{markllm}
To conduct our evaluations, we utilize MarkLLM~\cite{pan-etal-2024-markllm}, an open-source framework designed to facilitate the implementation and evaluation of LLM watermarking methods. MarkLLM provides an approach to watermarking by integrating different watermarking schemes within a unified framework. Its modular structure supports both the KGW-based family, which modifies token selection probabilities through logit adjustments, and the EXP-based family, which introduces pseudo-random guided sampling to embed watermarks.

We apply MarkLLM to evaluate our diverse set of watermarkings we compared to our proposed topic-based watermark (TBW). We use this framework exclusively for watermark generation and detection in alignment with the respective watermarking approach. Other utilities within the framework, such as robustness evaluation or text quality analysis, are not utilized in our study. The framework ensures that the configurations used in our study remain consistent with the original parameter choices presented in the respective papers, enabling a rigorous and reproducible assessment of each method.

\section{Additional Evaluation Results}\label{additional_eval_results_from_main_text}

\subsection{Perplexity at Higher Bias Strength}\label{more_perplexity}
We report perplexity results for \textsc{OPT-6.7B} when applying a stronger watermark bias parameter of $\delta = 3.0$ \textit{for TBW only}. This matches the strength used in our robustness evaluation (\S\ref{robustness}) and allows a direct quality-robustness trade-off comparison. The experimental setup follows \S\ref{text_quality}: we measure perplexity on the generated $200 \pm 5$ tokens using the \textsc{Llama-3.1-8B} oracle, clipping values above 100 for visualization. All baseline watermarking methods are reported with their default hyperparameters from the open-source \textsc{MARKLLM} library; only TBW uses the increased $\delta$.
\begin{figure}[h]
\begin{center}
\centerline{\includegraphics[width=\columnwidth]{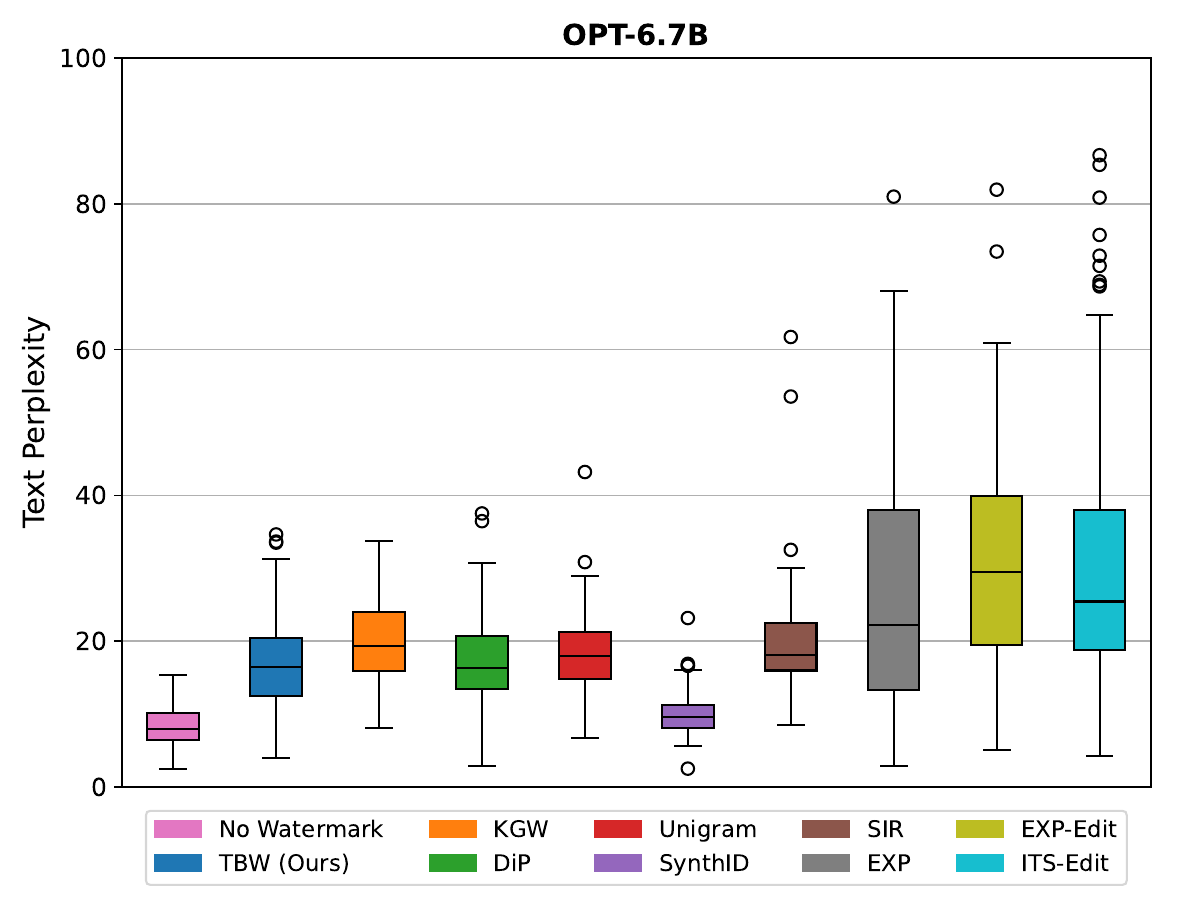}}
\caption{Perplexity comparison for \textsc{OPT-6.7B} with TBW at higher watermark strength ($\delta = 3.0$) and all other schemes at their standard settings. Compared to TBW's $\delta = 2.0$ results (see \S\ref{text_quality}), the increased bias leads to moderate quality degradation while retaining the lowest perplexity among watermarking methods. Lower values indicate higher text quality.}
\label{fig:perplexity_3}
\end{center}
\end{figure}

\begin{table*}[t]
\centering
\small
\begin{tabular}{llccccc}
\toprule
\textbf{Model} & \textbf{Method} & \textbf{Fluency} & \textbf{Coherence} & \textbf{Grammar} & \textbf{Lexical Variety} & \textbf{Overall Quality} \\
\midrule
\multirow{2}{*}{{OPT-6.7B}} & TBW & $3.23 \pm 1.16$ & $3.33 \pm 1.16$ & $3.63 \pm 0.80$ & $3.20 \pm 1.02$ & $3.17 \pm 1.14$ \\
& No Watermark & $\mathbf{3.47 \pm 0.95}$ & $\mathbf{3.45 \pm 1.11}$ & $\mathbf{3.90 \pm 0.90}$ & $\mathbf{3.37 \pm 0.97}$ & $\mathbf{3.45 \pm 1.00}$ \\
\midrule
\multirow{2}{*}{{GEMMA-7B}} & TBW & $3.53 \pm 1.05$ & $\mathbf{3.48 \pm 1.13}$ & $\mathbf{3.77 \pm 0.96}$ & $\mathbf{3.63 \pm 0.86}$ & $\mathbf{3.50 \pm 1.02}$ \\
& No Watermark & $\mathbf{3.55 \pm 1.05}$ & $3.42 \pm 1.15$ & $3.75 \pm 1.05$ & $3.50 \pm 0.93$ & $3.40 \pm 1.14$ \\
\bottomrule
\end{tabular}
\caption{Human evaluation results for single-text quality assessment. Values represent average ratings ($\pm$standard deviation) on a 5-point Likert scale from 12 evaluators.}
\label{tab:human_single_eval}
\end{table*}

Results indicate that increasing TBW's $\delta$ from 2.0 to 3.0 boosts robustness (\S\ref{robustness}) at the cost of a moderate rise in perplexity. Despite this stronger bias, TBW still achieves lower perplexity than other baselines, suggesting that semantic alignment mitigates some of the quality loss typically associated with high bias values.

\subsection{Human Evaluation}\label{main_human_eval}
To complement our automated text quality metrics, we conduct a human evaluation to assess the perceptual quality of TBW text. Human evaluation provides crucial insights into whether watermarking artifacts are detectable to end users, which is essential for practical deployment.

We conduct a single-text evaluation where human evaluators assess individual text samples across multiple quality dimensions. Evaluators are presented with 20 generated text samples, comprising 10 samples from \textsc{OPT-6.7B} and 10 samples from \textsc{GEMMA-7B}. Within each model's samples, 5 are watermarked with TBW and 5 are non-watermarked. Again, we fix the watermark strength at $\delta=2.0$ to mimic our perplexity as a metric study.

Evaluators rate each text sample on five quality dimensions using a 5-point Likert scale (1 = Very Poor, 5 = Excellent): \textbf{fluency} (how smooth and natural the writing flows), \textbf{coherence} (logical consistency and ease of following), \textbf{grammar} (correctness of grammar, spelling, and punctuation), \textbf{lexical variety} (appropriateness and diversity of word choices), and \textbf{overall quality} (holistic judgment). We collect responses from 12 evaluators, ensuring sufficient inter-rater reliability for statistical analysis. Table~\ref{tab:human_single_eval} summarizes the results across all quality dimensions.

The results demonstrate that TBW maintains competitive text quality across all evaluated dimensions, with performance variations that align with our automated perplexity assessments. \textsc{OPT-6.7B} exhibits modest quality decrements when watermarked, with performance gaps ranging from 0.12 to 0.28 points across the five evaluation criteria. However, these differences remain within acceptable bounds, as no metric, whether watermarked or non-watermarked, exceeds a score of 4.0, indicating that all generated text falls within a respectable quality range. Conversely, TBW demonstrates particularly stronger performance on \textsc{GEMMA-7B}, where watermarked text achieves comparable or better quality relative to non-watermarked baselines. Notably, TBW underperforms non-watermarked text only in fluency, with a negligible difference of 0.02 points, while achieving higher scores in lexical variety and overall quality metrics.

\begin{table*}[t]
\centering
\begin{tabular}{lcccc}
\toprule
\textbf{Method} & \textbf{Fluency} & \textbf{Coherence} & \textbf{Grammar} & \textbf{Lexical Variety} \\
\midrule
No Watermark & $\mathbf{3.20 \pm 0.62}$ & $\mathbf{3.15 \pm 0.59}$ & \underline{$3.85 \pm 0.59$} & $\mathbf{3.00 \pm 0.32}$ \\
TBW                  & \underline{$3.10 \pm 0.72$} & \underline{$3.05 \pm 0.69$} & $\mathbf{3.90 \pm 0.72}$ & $\mathbf{3.00 \pm 0.46}$ \\
Unigram              & $2.65 \pm 0.67$ & $2.70 \pm 0.73$ & $3.35 \pm 0.88$ & \underline{$2.90 \pm 0.55$} \\
SynthID         & $3.00 \pm 0.73$ & \underline{$3.05 \pm 0.76$} & $3.70 \pm 0.73$ & \underline{$2.90 \pm 0.55$} \\
\bottomrule
\end{tabular}
\caption{Automated GPT-4o-based text quality evaluation across watermarking methods. Values represent average scores ($\pm$standard deviation) on a 5-point scale over 20 prompts. \textbf{Bold} shows best values while \underline{underline} indicates second best results.}
\label{tab:gpt4o_eval}
\end{table*}

We attribute this difference to the disparity in vocabulary sizes between the two models. \textsc{OPT-6.7B} has a vocabulary of 50,272 tokens, whereas \textsc{GEMMA-7B} uses 256,000 tokens. A smaller vocabulary limits the number of semantically coherent tokens that exceed our similarity threshold $\tau$ during topic partitioning (Algorithm~\ref{alg:token_mapping}), slightly reducing lexical flexibility in watermarked outputs. In contrast, \textsc{GEMMA-7B}'s larger vocabulary yields denser topic-aligned partitions, enabling TBW to preserve or even improve perceived quality. This pattern is consistent with modern production-grade LLMs, which typically adopt vocabularies exceeding 50k tokens, suggesting that TBW's benefits are most fully realized in contemporary large-scale deployments. Automated LLM-as-a-Judge evaluation (\ref{llm-as-a-judge}) confirms these findings, showing TBW outperforms Unigram and SynthID-Text across all quality dimensions while matching non-watermarked text.

\subsection{LLM-as-a-Judge}\label{llm-as-a-judge}
While human evaluation provides valuable qualitative insight, its scalability is inherently limited. To extend our comparative analysis across a wider range of watermarking schemes, we employ an LLM-as-a-Judge setup using the chat functionality of GPT-4o. At the time of our study, GPT-4o was the strongest publicly available general-purpose model, offering high accuracy, consistent reasoning, and strong natural language understanding. This approach enables systematic, large-scale evaluation of TBW's performance not only against non-watermarked baselines, but also relative to other state-of-the-art watermarking methods, including SynthID-Text. Similar LLM-judge protocols have been successfully adopted in prior watermarking research~\cite{pan-etal-2024-markllm, tu-etal-2024-waterbench}, demonstrating the validity and growing acceptance of this methodology in the field.

We conduct this evaluation using \textsc{OPT-6.7B} to provide a conservative assessment of TBW's quality advantages. Given that our human evaluation demonstrated TBW's stronger performance on models with larger vocabularies, using \textsc{OPT-6.7B} (with its smaller 50,272-token vocabulary) represents a more challenging test case for our method. 

We evaluate text quality across four of the five dimensions used in our human evaluation using 20 C4 prompts and the same watermark strength ($\delta=2.0$). For each prompt, we generate watermarked text using TBW, Unigram, and SynthID-Text, alongside non-watermarked baselines. Each evaluation uses a fresh GPT-4o conversation to avoid memory effects that could bias assessments. We use the following evaluation prompt to ensure consistent and impartial assessment:
\begin{tcolorbox}[reviewbox]
You are an impartial evaluator judging the fluency and naturalness of a text sample. Please rate the sample without knowing how it was produced. Based on the readability and quality, answer the following criteria using a 1--5 scale: 
\begin{itemize}
    \item Fluency (1 = disfluent/unnatural, 5 = perfectly smooth and human-like)
    \item Grammatical correctness (1 = frequent grammar issues, 5 = grammatically perfect)
    \item Coherence (1 = ideas are disjointed or confusing, 5 = ideas flow logically)
    \item Lexical variety (1 = repetitive/basic word use, 5 = rich, diverse vocabulary)
\end{itemize}
Sample: 

\textit{\{Sample Text\}}
\end{tcolorbox}

Table~\ref{tab:gpt4o_eval} presents the automated evaluation results. The findings demonstrate that TBW maintains text quality nearly identical to non-watermarked output across all evaluated dimensions. Most notably, TBW achieves comparable fluency (3.10 vs. 3.20), equivalent lexical variety (3.00 vs 3.00), and even slightly superior grammatical correctness (3.90 vs 3.85) compared to non-watermarked text.

In comparison to competing watermarking methods, TBW demonstrates clear advantages. Against Unigram, which showed similar robustness in our adversarial evaluations (\S\ref{robustness}), TBW achieves quality improvements: $+$0.45 in fluency, $+$0.55 in grammatical correctness, and $+$0.35 in coherence. Similarly, TBW outperforms the industry-standard SynthID-Text across all dimensions. 

Overall, TBW consistently outperforms other watermarking schemes, matching or exceeding the text quality of non-watermarked outputs across most dimensions. These findings are consistent with our human evaluation, where TBW maintained comparable quality scores to non-watermarked text, confirming that its robustness advantages do not come at the expense of fluency, coherence, or grammatical correctness.

\subsection{Efficiency Results}\label{small_eff}
\begin{figure}[h]
\begin{center}
\centerline{\includegraphics[width=\columnwidth]{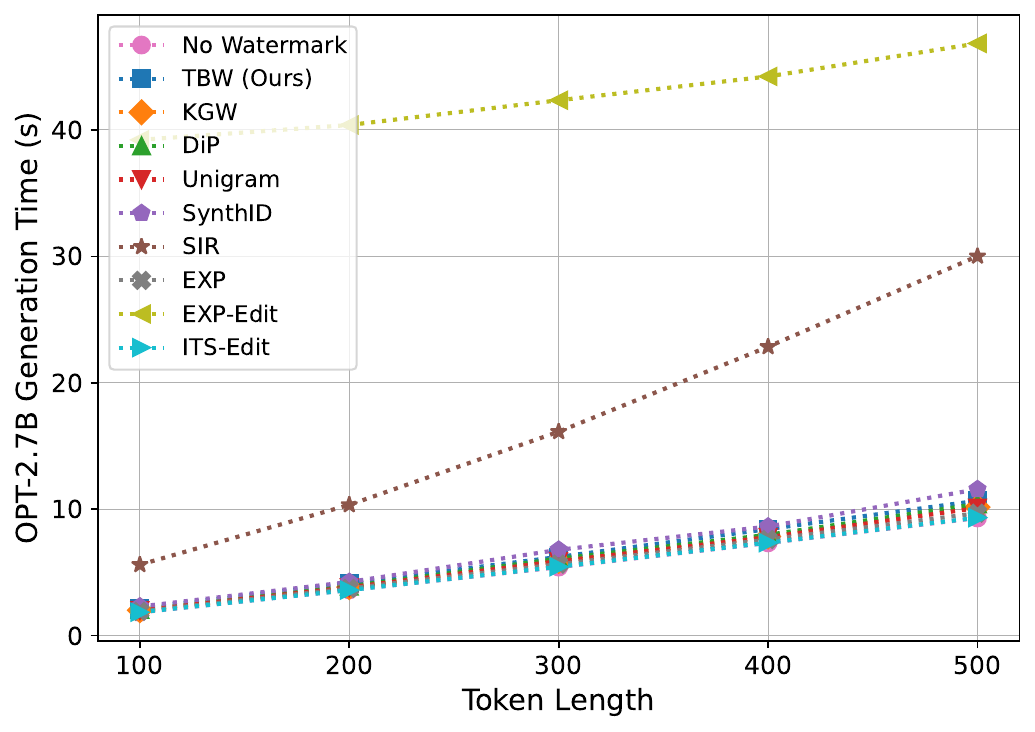}}
\caption{Comparison of average generation time (seconds) over various output token lengths from multiple watermarking schemes on \textsc{OPT-2.7B}.}
\label{small_27b}
\end{center}
\end{figure}

To verify that TBW's efficiency advantage holds across model scales, we repeat the generation time evaluation from \S\ref{efficiency} on \textsc{OPT-2.7B}. Figure~\ref{small_27b} confirms that the same trends observed on \textsc{OPT-6.7B} hold at smaller scale since TBW introduces negligible overhead relative to non-watermarked generation, matching lightweight methods across all sequence lengths.

\subsection{Comparison to ITS-Edit}\label{its_edit}
In the main evaluation, we exclude EXP-Edit and ITS-Edit due to their poor perplexity values, which make them impractical for real-world use. However, for completeness, we assess the robustness of ITS-Edit in terms of ROC-AUC score, Best F1 score, TPR@1\%FPR, and TPR@10\%FPR, comparing it to TBW.

We follow the same evaluation procedure used in the main paper under a smaller sample size, generating 50 watermarked and 50 non-watermarked (baseline) samples using \textsc{OPT-6.7B}. We then apply two paraphrasers, PEGASUS and DIPPER, to the 200-token completions. ITS-Edit performs 500 detection runs per sample during the detection phase to better estimate the presence of the watermark.

As shown in Table~\ref{metrics_its}, TBW consistently outperforms ITS-Edit across all evaluated robustness metrics, demonstrating superior detection performance while maintaining practical efficiency.

\begin{table*}[!htbp]
    \centering
    \begin{tabular}{llcccc}
        \toprule
        \textbf{Attack} & \textbf{Method} & \textbf{ROC-AUC} & \textbf{Best F1 Score} & \textbf{TPR@1\% FPR} & \textbf{TPR@10\% FPR} \\
        \midrule
        \multirow{2}{*}{No Attack} & TBW & \textbf{0.999} & \textbf{0.990} & \textbf{0.980} & \textbf{1.000} \\
        & ITS-EDIT & 0.043 & 0.667 & 0.000 & 0.000 \\
        \midrule
        \multirow{2}{*}{PEGASUS} & TBW & \textbf{0.959} & \textbf{0.939} & \textbf{0.800} & \textbf{0.920} \\
        & ITS-EDIT & 0.417 & 0.667 & 0.000 & 0.100 \\
        \midrule
        \multirow{2}{*}{DIPPER} & TBW & \textbf{0.929} & \textbf{0.875} & \textbf{0.575} & \textbf{0.840} \\
        & ITS-EDIT & 0.519 & 0.667 & 0.020 & 0.040  \\
        \bottomrule
    \end{tabular}
    \caption{Comparison of robustness metrics between TBW and ITS-Edit on \textsc{OPT-6.7B}, evaluated using PEGASUS and DIPPER paraphrasers where \textbf{bold} indicates better scores.}
    \label{metrics_its}
\end{table*}

\begin{table*}[t]
\centering
\begin{tabular}{lcccccc}
\toprule
\multirow{2}{*}{\textbf{Method}} & \textbf{Time} & \textbf{PPL} & \multicolumn{2}{c}{\textbf{No Attack}} & \multicolumn{2}{c}{\textbf{DIPPER}} \\
\cmidrule(lr){4-5} \cmidrule(lr){6-7}
& (s/sample) & (mean $\pm$ std) & ROC-AUC & TPR@1\% & ROC-AUC & TPR@1\% \\
\midrule
TBW ($\delta$=2.0) & 5.23 $\pm$ 1.93 & 12.36 $\pm$ 3.64 & 0.971 & 0.879 & 0.694 & 0.222 \\
TBW ($\delta$=3.0) & 5.08 $\pm$ 2.01 & 12.46 $\pm$ 4.80 & \textbf{0.986} & \textbf{0.949} & \textbf{0.802} & \textbf{0.357} \\
k-SemStamp & 65.93 $\pm$ 75.03 & \textbf{9.16 $\pm$ 2.85} & 0.960 & 0.750 & 0.743 & 0.150 \\
PostMark & 29.61 $\pm$ 17.56 & 12.43 $\pm$ 2.90 & 0.968 & 0.920 & 0.780 & 0.210 \\
\bottomrule
\end{tabular}
\caption{Comparison of TBW against semantic black-box watermarking methods on OPT-6.7B over 100 C4 prompts. \textbf{Bold} indicates best results. TPR@1\% denotes TPR at 1\% FPR.}
\label{tab:semantic_comparison}
\end{table*}

\begin{figure*}[!htbp]
\begin{center}
\begin{minipage}[b]{\columnwidth}
\centering
\includegraphics[width=\columnwidth]{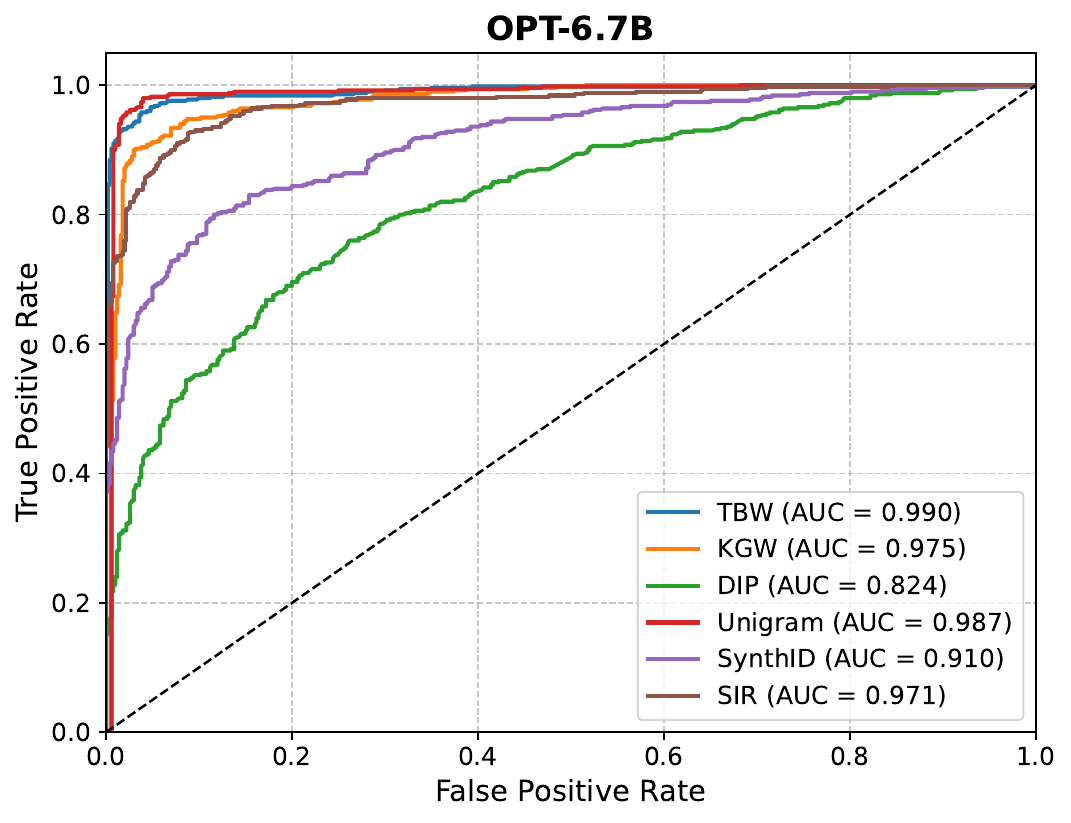}
\subcaption{OPT-6.7B vs. PEGASUS}
\end{minipage}
\hfill
\begin{minipage}[b]{\columnwidth}
\centering
\includegraphics[width=\columnwidth]{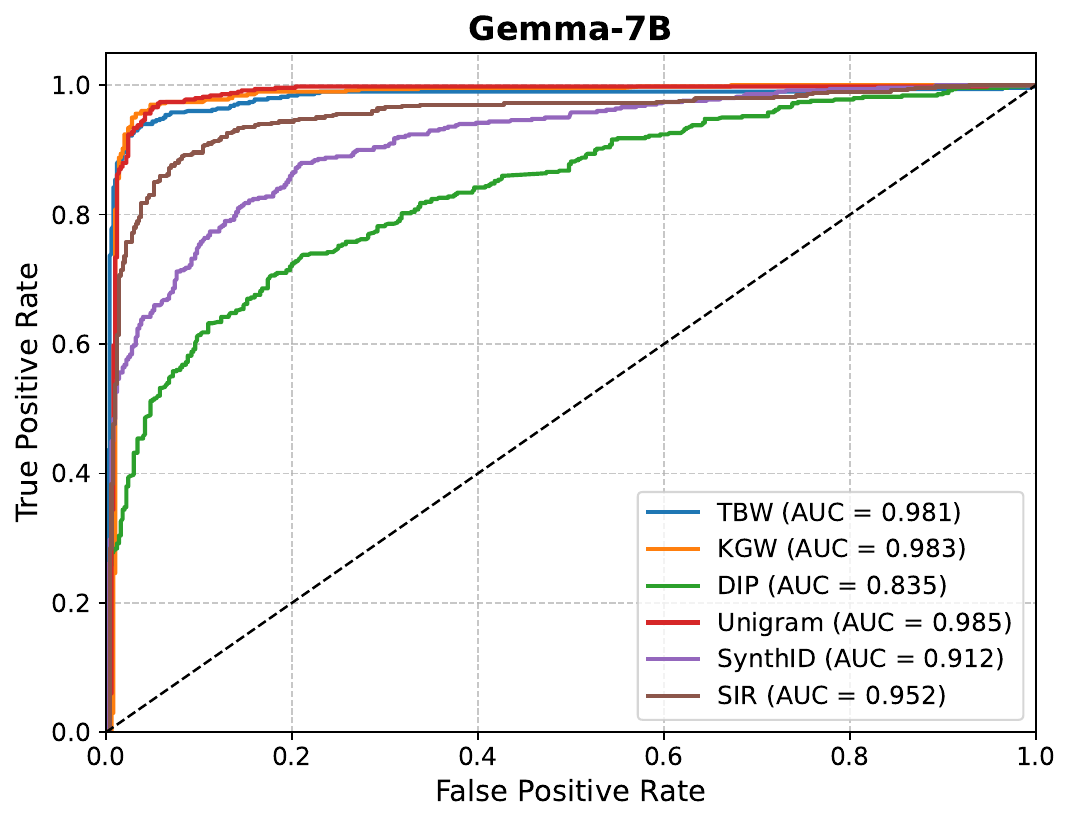}
\subcaption{Gemma-7B vs. PEGASUS}
\end{minipage}
\begin{minipage}[b]{\columnwidth}
\centering
\includegraphics[width=\columnwidth]{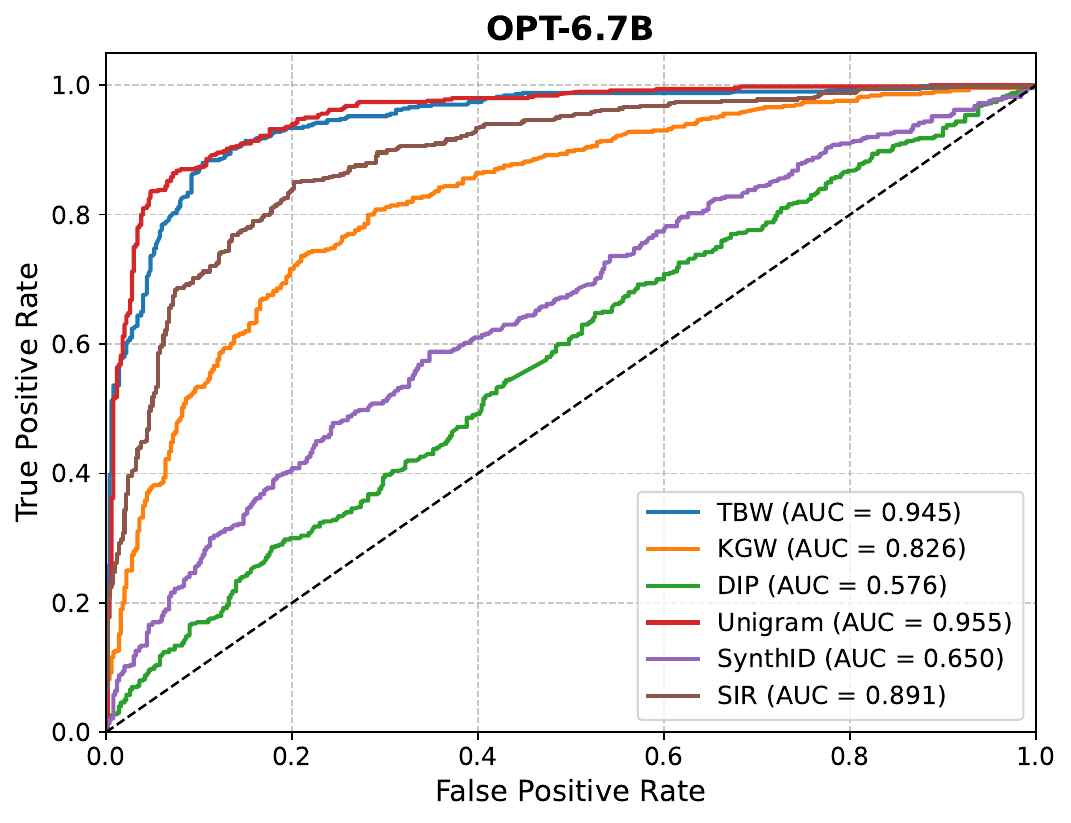}
\subcaption{OPT-6.7B vs. DIPPER}
\end{minipage}
\hfill
\begin{minipage}[b]{\columnwidth}
\centering
\includegraphics[width=\columnwidth]{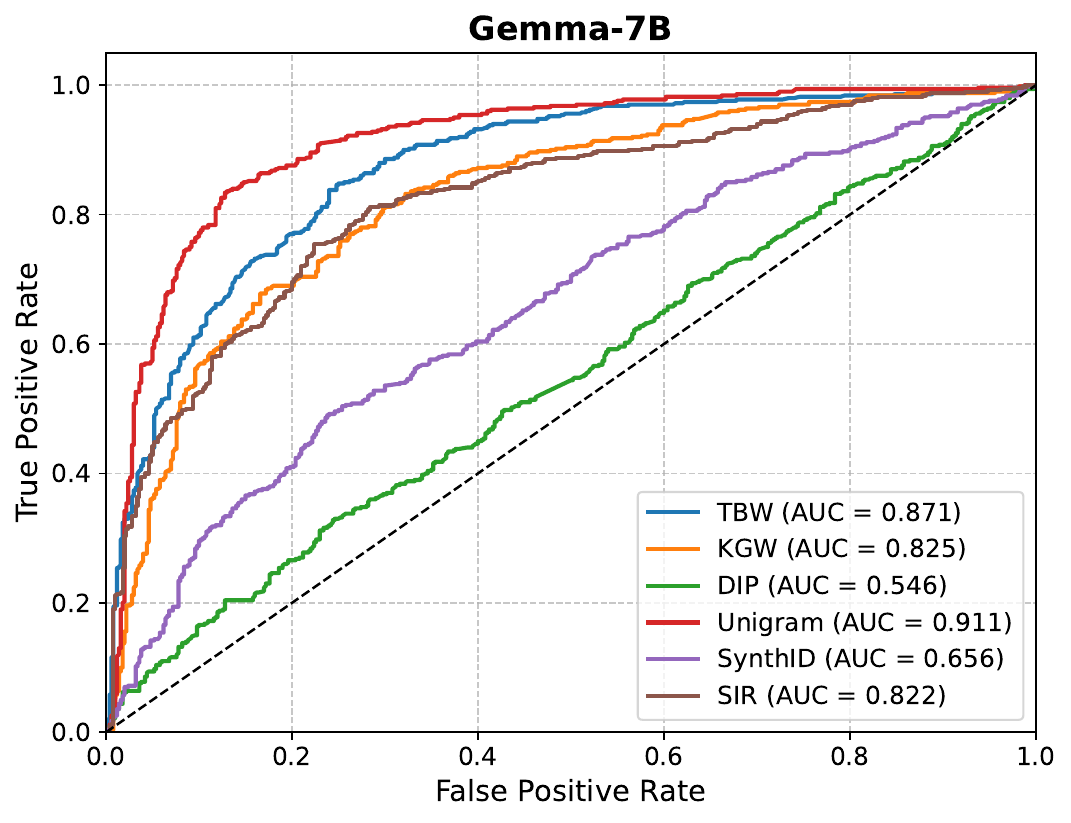}
\subcaption{Gemma-7B vs. DIPPER}
\end{minipage}
\end{center}
\caption{ROC curves comparing watermark methods on OPT-6.7B and Gemma-7B against PEGASUS (top) and DIPPER (bottom) paraphrasing attacks.}
\label{roc_comparison}
\end{figure*}

\subsection{Comparison to Semantic Watermarks}
\label{semantic_comparison}
To situate TBW relative to semantic watermarking approaches that operate via post-processing, we compare against PostMark~\cite{chang2024postmark} and k-SemStamp~\cite{hou-etal-2024-k}. A key methodological distinction is that both PostMark and k-SemStamp embed watermarks by rewriting already-generated text through additional LLM inference passes, whereas TBW embeds the watermark during a single standard decoding pass. This difference has direct implications for generation cost and deployment complexity.

We evaluate all methods over 100 C4 prompts on \textsc{OPT-6.7B}, measuring perplexity, generation time, ROC-AUC, and TPR@1\%FPR under both no-attack and DIPPER paraphrasing conditions. TBW is evaluated at two watermark strengths ($\delta = 2.0$ and $\delta = 3.0$) to illustrate the quality-robustness trade-off. Results are reported in Table~\ref{tab:semantic_comparison}.

TBW at $\delta = 3.0$ achieves the highest ROC-AUC and TPR@1\%FPR under both no-attack and DIPPER paraphrasing conditions, while being 6-13$\times$ faster than the semantic baselines. Perplexity remains comparable to PostMark and only moderately higher than k-SemStamp, whose lower perplexity comes at the cost of substantially higher generation time and lower detection performance. These results confirm that TBW's in-generation approach provides a fundamentally different and practically advantageous trade-off compared to post-processing semantic watermarks as it avoids the additional LLM inference passes required by rewriting-based methods while achieving stronger robustness and detection at a fraction of the computational cost.

\subsection{TBW Robustness ROC Curves}\label{auc_roc}
Figure~\ref{roc_comparison} presents the ROC curves and corresponding AUC values for the evaluated watermarking methods. For \textsc{OPT-6.7B}, our method achieves comparable robustness to Unigram, demonstrating strong detection performance. For \textsc{Gemma-7B}, we observe a slight reduction in robustness; however, this trade-off comes with improved text quality. The difference in AUC between our method and Unigram is minimal, approximately 4\%, highlighting the balance between robustness and text quality.

\begin{table*}[t]
\centering
\begin{tabular}{llccrr}
\toprule
\textbf{Model} & \textbf{Method} & \textbf{FPR (\%)} & \textbf{False Positives} & \textbf{Detection Score} & \textbf{Threshold} \\
\midrule
\multirow{5}{*}{OPT-6.7B} 
    & \textbf{TBW} & 0.70 & 7/1000 & $0.437 \pm 1.664$ & 4.75 \\
    & KGW & 0.00 & 0/1000 & $-0.112 \pm 1.033$ & 4.0 \\
    & DiP & 0.20 & 2/1000 & $-0.016 \pm 0.504$ & 1.513 \\
    & Unigram & 0.00 & 0/1000 & $-0.205 \pm 1.268$ & 4.0 \\
    & SynthID & 0.40 & 4/1000 & $0.500 \pm 0.007$ & 0.52 \\
\midrule
\multirow{5}{*}{GEMMA-7B} 
    & \textbf{TBW} & 0.20 & 2/1000 & $-0.200 \pm 1.405$ & 4.75 \\
    & KGW & 0.00 & 0/1000 & $-0.194 \pm 1.103$ & 4.0 \\
    & DiP & 0.10 & 1/1000 & $-0.019 \pm 0.502$ & 1.513 \\
    & Unigram & 0.00 & 0/1000 & $-0.459 \pm 1.252$ & 4.0 \\
    & SynthID & 0.60 & 6/1000 & $0.500 \pm 0.007$ & 0.52 \\
\bottomrule
\end{tabular}
\caption{False positive rate (FPR) analysis on 1,000 human-written C4 samples (200 tokens) across different watermarking schemes and language models. Detection thresholds follow the defaults from respective original implementations.}
\label{tab:fpr_analysis}
\end{table*}

\subsection{False Positive Rate (FPR) on Human-Written Text}\label{fpr_analysis}

\subsubsection{Watermarking Methods FPR}
To evaluate the practical viability of watermarking schemes, we assess false positive rates (FPR) on human-written content using the same experimental parameters as described in \S\ref{setup}. We evaluated 1,000 human-written C4 samples (200 tokens) using TBW, KGW, DiP, Unigram, and SynthID detection schemes on both \textsc{OPT-6.7B} (vocabulary size: 50,272) and \textsc{GEMMA-7B} (vocabulary size: 256,000).

The results from Table~\ref{tab:fpr_analysis} demonstrate that TBW maintains practical false positive rates across both models, with notably better performance on \textsc{GEMMA-7B} (0.20\% FPR) compared to \textsc{OPT-6.7B} (0.70\% FPR). This vocabulary-dependent behavior aligns with our hypothesis that larger vocabularies enable more precise semantic token partitioning, reducing false classifications of human-written content. While TBW shows slightly higher FPR on smaller vocabulary models, this trade-off is offset by the significant robustness advantages demonstrated in our main evaluation.

\subsubsection{TBW Threshold FPR Sensitivity} To understand how the similarity threshold $\tau$ (used in Algorithm~\ref{alg:token_mapping} for token-to-topic mapping) affects false positive rates, we conducted a threshold sensitivity analysis on \textsc{GEMMA-7B} using the same 1,000 human-written C4 samples (200 tokens). As the threshold increases, fewer tokens meet the semantic similarity criteria for topic assignment, resulting in more tokens being distributed via the round-robin mechanism.

\begin{table}[t]
\centering
\scriptsize
\begin{tabular}{cccr}
\toprule
\textbf{Threshold ($\tau$)} & \textbf{FPR} & \textbf{False Positives} & \textbf{Detection Score} \\
\midrule
0.1 & 33.2 & 332/1000 & $0.346 \pm 4.830$ \\
0.2 & 33.2 & 332/1000 & $0.365 \pm 4.876$ \\
0.3 & 1.60 & 16/1000 & $0.543 \pm 2.012$ \\
0.4 & 0.10 & 1/1000 & $-0.217 \pm 1.591$ \\
0.5 & 0.20 & 2/1000 & $0.318 \pm 1.467$ \\
0.6 & 0.30 & 3/1000 & $-0.375 \pm 1.521$ \\
\textbf{0.7} & \textbf{0.60} & \textbf{6/1000} & $\mathbf{0.261 \pm 1.795}$ \\
0.8 & 0.40 & 4/1000 & $0.232 \pm 1.455$ \\
0.9 & 0.90 & 9/1000 & $0.216 \pm 1.912$ \\
\bottomrule
\end{tabular}
\caption{Threshold sensitivity analysis for TBW on \textsc{GEMMA-7B} using 1,000 human-written C4 samples. Lower thresholds result in less semantically coherent topic partitions.}
\label{tab:threshold_sensitivity}
\end{table}

Table~\ref{tab:threshold_sensitivity} results demonstrate a clear threshold effect with very low thresholds ($\tau \leq 0.2$) lead to unacceptably high false positive rates ($>$30\%), as most semantically aligned tokens meet the similarity criteria among topic lists, resulting in overly broad topic partitions that reduce watermark specificity. The optimal range appears to be $\tau \in [0.3, 0.9]$, where FPR remains below 2\%, with $\tau \in [0.4, 0.9]$ achieving FPR below 1\%. Our chosen threshold of $\tau = 0.7$ (highlighted in bold) provides a reasonable balance between semantic coherence and false positive control.

\subsection{Detection Attacks}\label{detection_attacks}
Building on our threat model in \S\ref{threat_model}, we also consider attacks that attempt to \emph{detect} the presence of a watermark directly, rather than removing or obfuscating it through text manipulation. In this setting, the adversary does not seek to degrade or rephrase the text but instead aims to distinguish between watermarked and non-watermarked outputs based on statistical signals. One can argue that this corresponds to testing the ``stealthiness'' of the watermark. However, given that our topic-based watermarking scheme preserves text quality comparable to non-watermarked outputs, the watermark remains effectively stealthy in practice.

\begin{figure*}[t]
\centering
% First row (OPT)
\begin{subfigure}{\textwidth}
    \centering
    \includegraphics[width=0.3\textwidth]{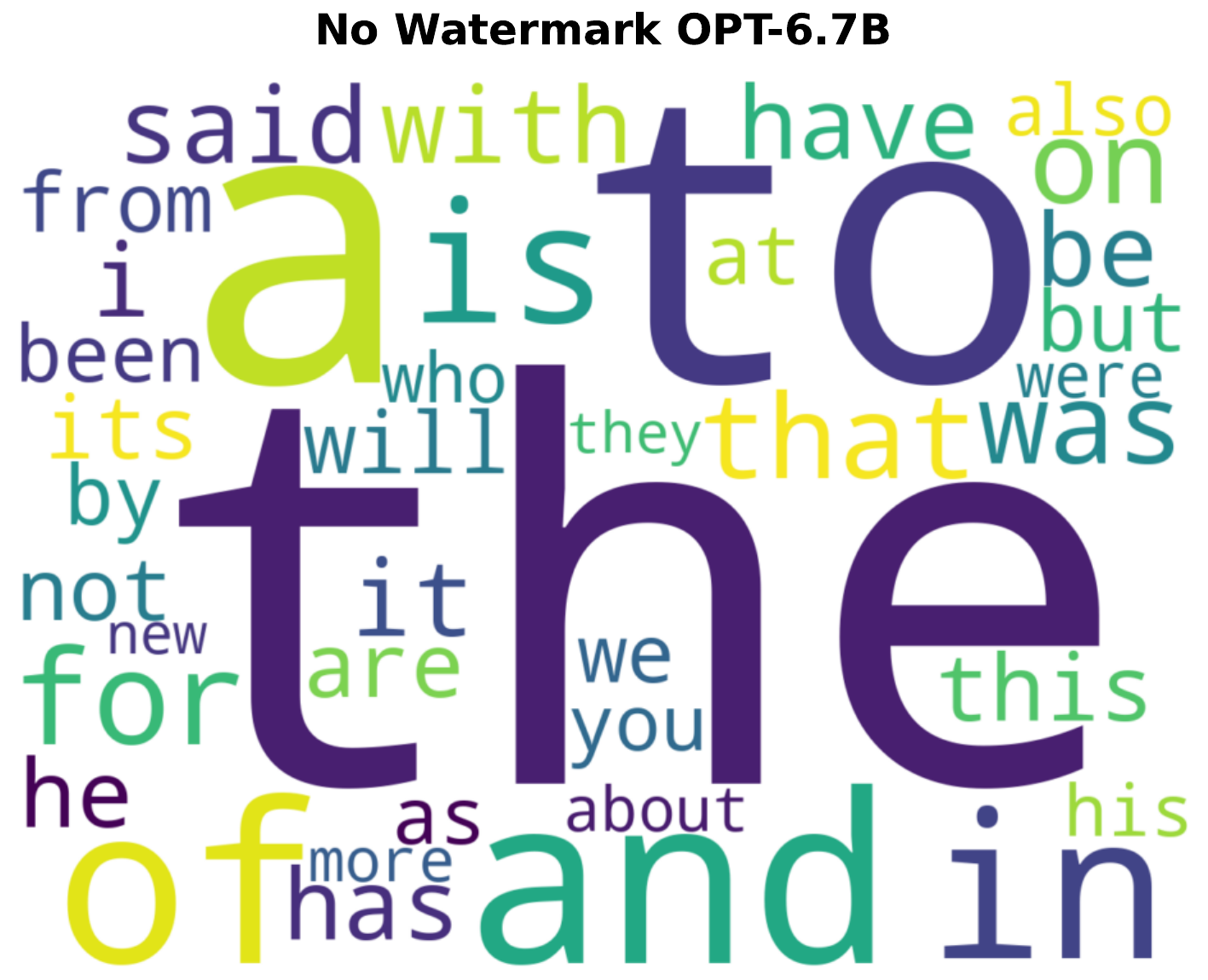}
    \hfill
    \includegraphics[width=0.3\textwidth]{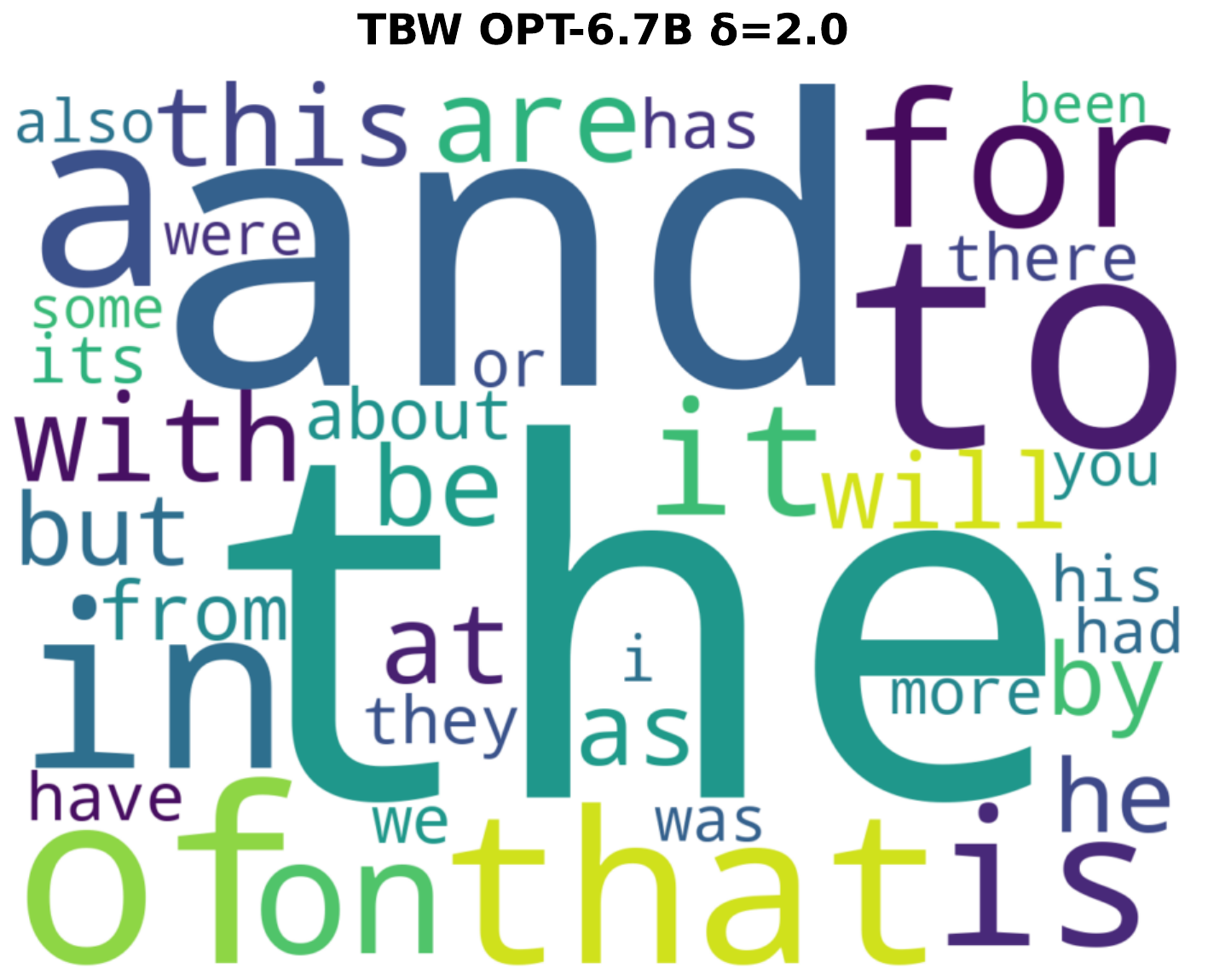}
    \hfill
    \includegraphics[width=0.3\textwidth]{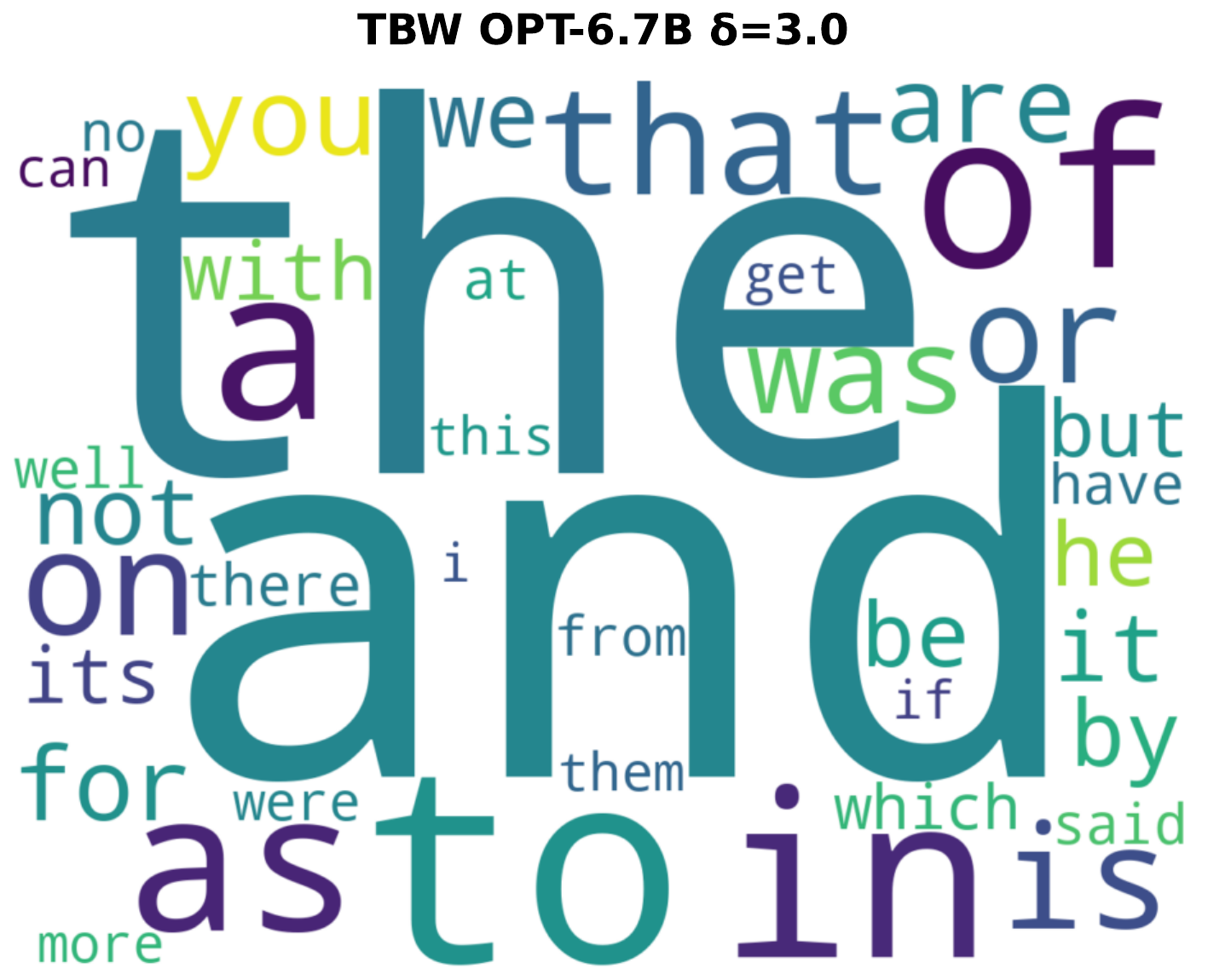}
    \caption{\textsc{OPT-6.7B}}
\end{subfigure}
% Second row (Gemma)
\begin{subfigure}{\textwidth}
    \centering
    \includegraphics[width=0.3\textwidth]{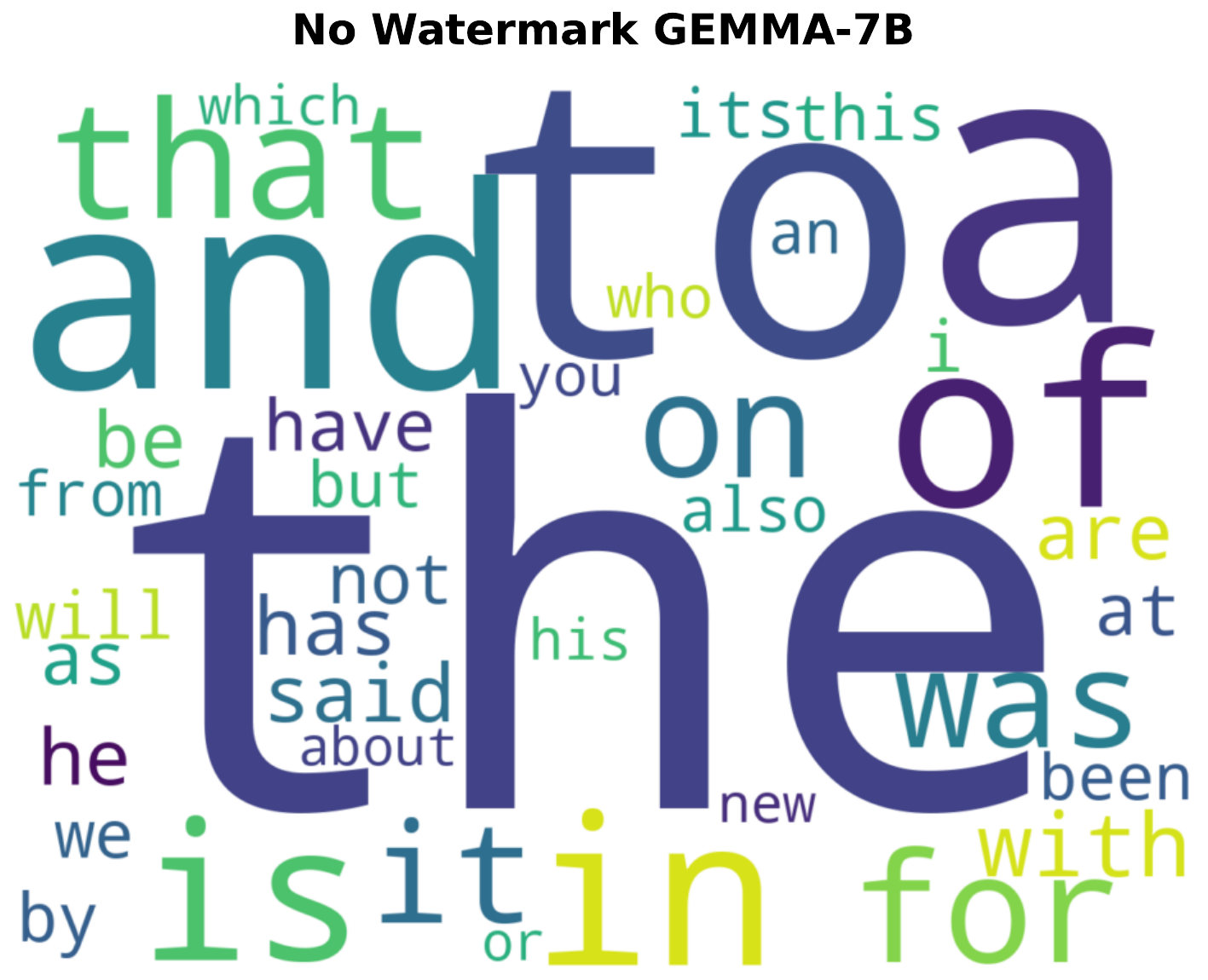}
    \hfill
    \includegraphics[width=0.3\textwidth]{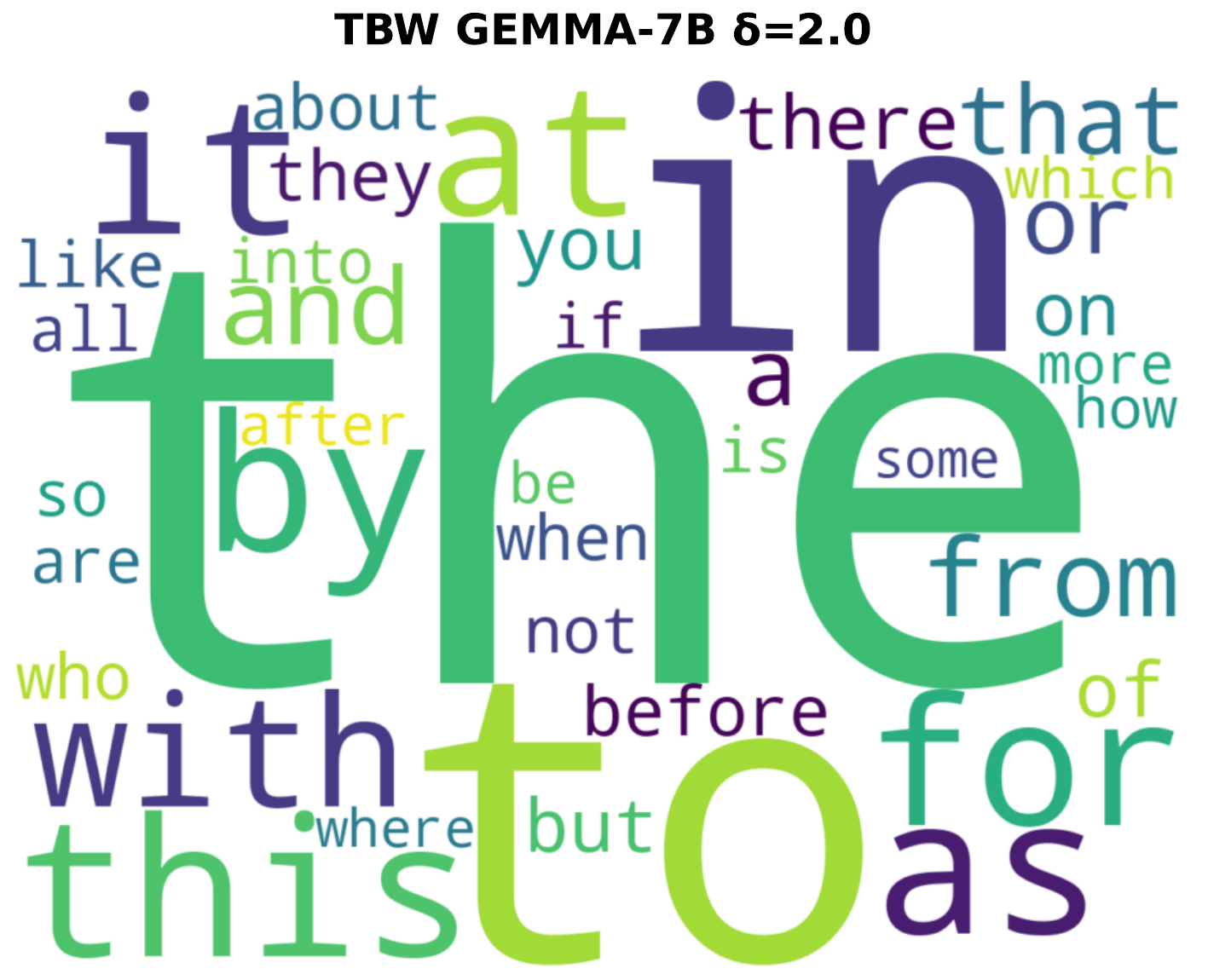}
    \hfill
    \includegraphics[width=0.3\textwidth]{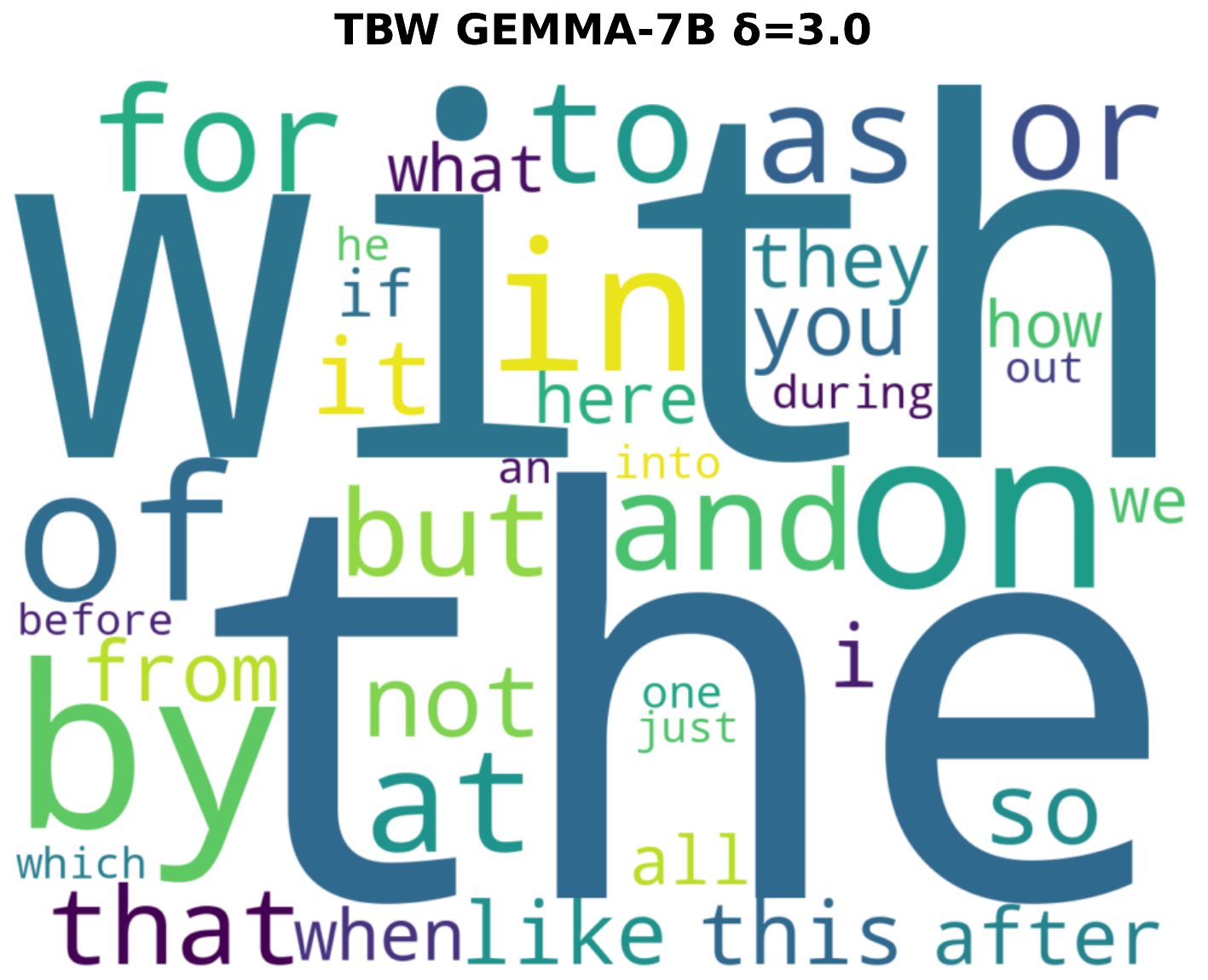}
    \caption{\textsc{GEMMA-7B}}
\end{subfigure}
\caption{Word clouds of the top-$40$ normalized token frequencies for \textsc{OPT-6.7B} (top) and \textsc{Gemma-7B} (bottom). From left to right, each row shows outputs from: (i) non-watermarked generations, (ii) TBW with bias $\delta=2.0$, and (iii) TBW with bias $\delta=3.0$. Across both models and bias strengths, the distributions are dominated by common function words (e.g., ``the,'' ``and,'' ``to''), with no systematic elevation of topic-specific tokens.}
\label{fig:wordclouds_combined}
\end{figure*}

For completeness, we evaluate our method against non-watermarked text using two classes of detection strategies: (i) statistical hypothesis tests over generated output and (ii) supervised machine learning classification trained to separate watermarked and non-watermarked corpora. We conduct experiments across our two representative models (\textsc{Gemma-7B} and \textsc{OPT-6.7B}) and evaluate at bias strengths $\delta = 2.0$ and $\delta = 3.0$. These values match those used in our main evaluations of text quality and robustness, allowing us to study detection performance under the same operating points and to compare trade-offs across quality, robustness, and detectability. While these experiments provide additional perspective, they rely on assumptions such as access to large paired corpora of watermarked and non-watermarked outputs that rarely hold in deployed systems, and thus fall outside the scope of practical adversaries targeted by our primary threat model.

\subsubsection{Statistical Tests}
We probe whether simple distributional cues can separate watermarked from non-watermarked outputs by comparing bag-of-words statistics over large corpora. From an attacker's perspective, the goal would be to identify watermark presence based on observable shifts in word frequencies. In particular, because our scheme biases topic-specific tokens, one might expect that watermarked outputs exhibit disproportionately higher rates of such tokens compared to non-watermarked outputs. 

To test this, for each model and bias strength ($\delta\in\{2.0,3.0\}$) we generate 1,000 watermarked and 1,000 non-watermarked completions using the same prompts as in our main experiments: the first 100 tokens of each C4 sample serve as the prompt, and the model produces $200 \pm 5$ tokens for analysis. To remain tokenizer-agnostic and focus on surface lexical signals, we lowercase, strip punctuation, split on whitespace, and compute corpus-level word counts. From these counts we derive the top-$N$ normalized frequencies and visualize the relative mass using word clouds in Figure~\ref{fig:wordclouds_combined}.

Across all models and bias strengths, the most frequent tokens are dominated by common function words such as ``the,'' ``and,'' ``to,'' and ``that.'' While certain settings show minor variation (e.g., the word ``with'' appears slightly more often in \textsc{Gemma-7B} TBW at $\delta=3.0$), these tokens are also prevalent in the corresponding non-watermarked corpora, and no systematic elevation of topic-specific tokens is observed. This outcome is precisely what we expect from a stealthy watermark where the lexical distribution remains consistent with non-watermarked text even under naive frequency analysis. 

We note that for Gemma, the watermarked outputs exhibit slightly flatter token distributions compared to the corresponding non-watermarked outputs. While this effect is subtle and not readily visible without corpus-level aggregation, it is consistent across bias strengths and may indicate a mild redistribution of lexical mass. Nonetheless, in realistic deployments, attackers would not have access to paired corpora of watermarked and non-watermarked text from the same model, and thus could not easily identify subtle differences even  if they existed. The closest analogue is Google's evaluation of SynthID-Text, where generations from the same model included both watermarked and non-watermarked outputs without disclosure to users~\cite{deepmind_synthid_2024, dathathri2024scalable}. However, since TBW preserves the same dominant distribution of function words across conditions, such comparisons likewise fail to yield clear evidence of watermark presence. In short, statistical tests do not provide a practical detection vector against TBW.

\subsubsection{Machine Learning Classification}
We next test whether a supervised classifier can distinguish between watermarked and non-watermarked text. For each model and bias strength ($\delta \in \{2.0, 3.0\}$), we collect 1,000 non-watermarked completions from unique prompts and 1,000 watermarked completions from a disjoint set of prompts. As in our main experiments, the first $100$ tokens of each C4 sample are used as input, and the model generates $200 \pm 5$ tokens for analysis. These 2,000 samples form the training set, where we assign a label of 0 for non-watermarked text and 1 for watermarked text. For evaluation, we use 500 unseen prompts from the same pool to generate 500 non-watermarked and 500 watermarked samples, forming a balanced test set. We train a \textsc{BERT-base}~\cite{devlin2019bertpretrainingdeepbidirectional} classifier on the training data using the hyperparameters reported in Table~\ref{tab:ml_hyperparams}, and evaluate on the held-out test set. 

\begin{table}[h]
\centering
\begin{tabular}{lc}
\toprule
\textbf{Hyperparameter} & \textbf{Value} \\
\midrule
Epochs          & 3 \\
Batch size (per device) & 16 \\
Learning rate   & 5e-5 \\
Optimizer       & AdamW \\
\bottomrule
\end{tabular}
\caption{Fine-tuning hyperparameters for \textsc{BERT-base} used in the classification experiments.}
\label{tab:ml_hyperparams}
\end{table}

Table~\ref{tab:ml_classification} reports accuracy and F1 scores for \textsc{OPT-6.7B} and \textsc{Gemma-7B} under $\delta \in \{2.0,3.0\}$. For \textsc{OPT}, classification accuracy reaches $0.80$-$0.83$ with F1 scores of $0.75$-$0.80$, indicating that supervised models can learn a moderately reliable separation between watermarked and non-watermarked corpora. In contrast, \textsc{Gemma} achieves near-perfect accuracy and F1 ($\approx 0.99$) across both bias strengths. We hypothesize this stems from the slightly flattened token distributions observed in the statistical analysis, which \textsc{BERT-base} is able to exploit effectively.

\begin{table}[h]
\centering
\begin{tabular}{cccc}
\toprule
\textbf{Language Model} & \textbf{Bias $\delta$} & \textbf{Accuracy} & \textbf{F1} \\
\midrule
\textsc{OPT-6.7B}    & 2.0 & 0.80 & 0.75 \\
\textsc{OPT-6.7B}    & 3.0 & 0.83 & 0.80 \\
\textsc{Gemma-7B}  & 2.0 & 0.99 & 0.99 \\
\textsc{Gemma-7B}  & 3.0 & 0.99 & 0.99 \\
\bottomrule
\end{tabular}
\caption{Classification performance of \textsc{BERT-base} trained to distinguish watermarked from non-watermarked text. Results are reported as Accuracy and F1 on a held-out test set of 1,000 samples (500 watermarked, 500 non-watermarked).}
\label{tab:ml_classification}
\end{table}

Despite these results, the assumptions behind this evaluation remain unrealistic for real-world adversaries. Training the classifier requires thousands of labeled watermarked and non-watermarked samples, whereas deployed systems consistently apply watermarking and never reveal non-watermarked outputs. In principle, an attacker could attempt to scrape known non-watermarked text online and compare it against API outputs, but such a classifier would primarily be distinguishing across models or domains rather than detecting watermark presence. This approach has been attempted in prior work~\cite{islam2023distinguishinghumangeneratedtext, bafna2024mastkalandarsemeval2024task}, but it misses the purpose of watermarking to provide a lightweight provenance signal that avoids constant retraining of detectors as model outputs evolve. We further discuss the practicality and assumptions of these detection-based attacks, and why they remain outside our primary threat model, in the following section.

\begin{table*}[t]
\centering
\begin{tabular}{lcccccc}
\toprule
\textbf{Language Model} & \textbf{Accuracy} & \textbf{Precision} & \textbf{Recall} & \textbf{F1 score} & \textbf{ROC--AUC} & \textbf{FPR} \\
\midrule
\textsc{OPT-6.7B} & 0.921 & 0.866 & 0.996 & 0.927 & 0.996 & 0.154 \\
\textsc{GEMMA-7B} & 0.994 & 0.988 & 1.000 & 0.994 & 1.000 & 0.012 \\
\bottomrule
\end{tabular}
\caption{Classification performance of the maximum $z$-score detection method on 500 watermarked and 500 non-watermarked samples. The method achieves consistently high accuracy, precision, recall, and ROC-AUC, with low false positive rates (FPR).}
\label{tab:max_z_performance}
\end{table*}

\subsubsection{Discussion and Implications}
This section considers the broader implications of our detection-based evaluations. While our statistical and machine-learning tests provide insight into the distributional behavior of TBW, they do not translate into practical adversarial strategies under real deployment conditions. 

As discussed previously, an attacker would almost never have access to both watermarked and non-watermarked outputs from the same model. The most feasible alternative is that an attacker could continuously query a deployed API and examine lexical distributions for anomalies. However, our word cloud analysis (Figure~\ref{fig:wordclouds_combined}) shows that the most frequent tokens remain dominated by common function words (e.g., ``the,'' ``and,'' ``to''), with no systematic inflation of topic-specific tokens. This suggests that distributional probing would not expose the watermark in practice. 

Similarly, supervised classification attacks (Table~\ref{tab:ml_classification}) assume the availability of thousands of labeled samples of both watermarked and non-watermarked text. In deployed settings, such paired corpora are inaccessible, since watermarking is applied uniformly and non-watermarked outputs are never exposed to end users. At best, an adversary could attempt to scrape known non-watermarked text online and compare it against API outputs. Yet in this case, the classifier would be distinguishing differences across models or domains, not detecting the presence of a watermark. The distinction is important as watermarking aims to provide a provenance signal robust to model drift, not to enable adversaries to retrain detectors whenever models evolve linguistically. 

One might argue that detection attacks could become more feasible in open-source ecosystems where both watermarked and non-watermarked versions of a model are released (e.g., through repositories like Hugging Face~\cite{huggingface2025}). While this is a theoretical concern, watermarking is most critical in contexts of malicious misuse, academic dishonesty, or provenance tracking which are cases that overwhelmingly involve flagship proprietary systems rather than community-maintained open-source checkpoints. In practice, deployed systems do not expose both versions of a model to end users, and thus the adversarial assumptions behind detection-based attacks are rarely satisfied. 

In summary, our analysis reinforces that detection attacks are neither representative of realistic adversarial capabilities nor an effective threat to TBW. By contrast, the practical adversaries captured in our primary threat model by those performing text degradation or semantic rephrasing are explicitly addressed by our scheme, and our evaluation demonstrates that TBW maintains robustness under these conditions. 

\subsection{Maximum $z$-Score Detection: Detailed Analysis}\label{max_z_detailed}
Given the superior performance of the maximum $z$-score detection method in Table~\ref{tab:detection_comparison}, we provide detailed performance metrics and distribution analysis. This evaluation uses 500 watermarked and 500 non-watermarked samples generated under our standard experimental configuration.

\subsubsection{Performance Metrics}
\begin{figure}[ht]
\begin{center}
\centerline{\includegraphics[width=\columnwidth]{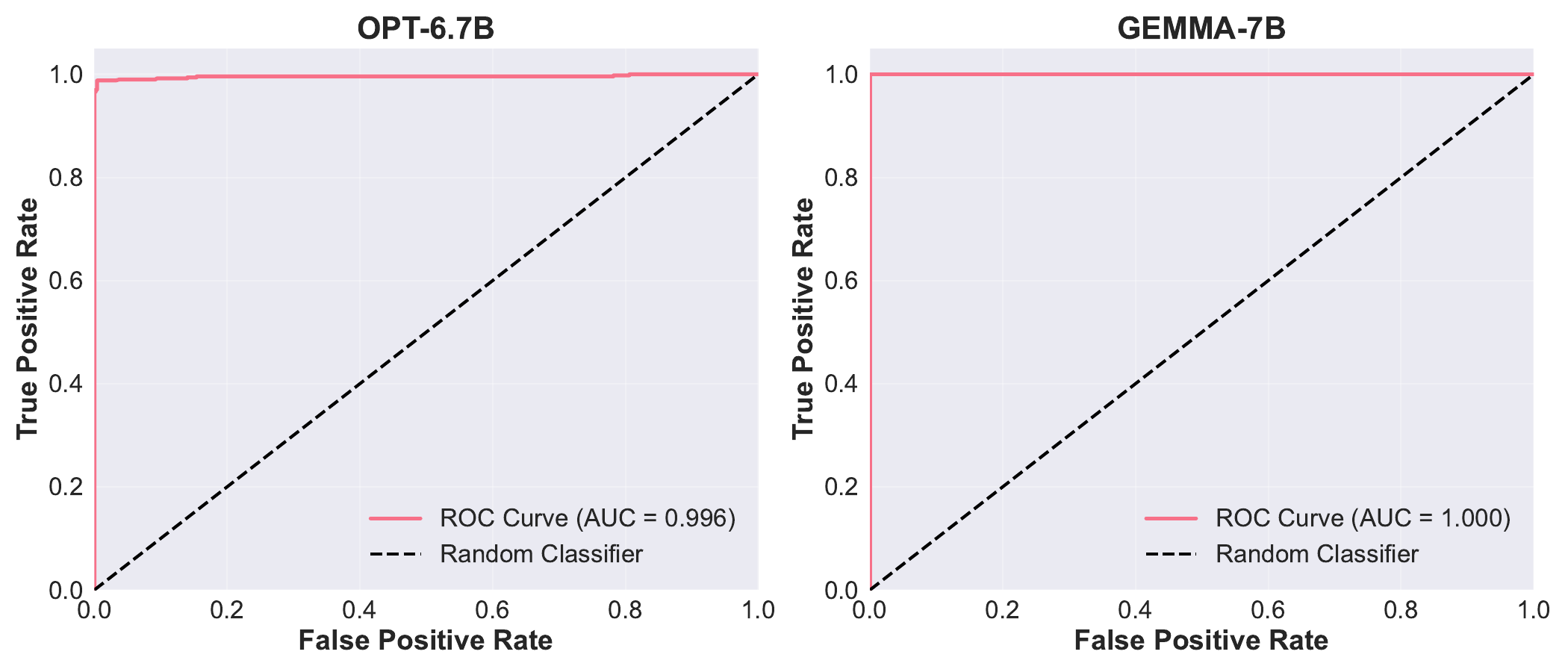}}
\caption{ROC curves for maximum $z$-score detection on \textsc{OPT-6.7B} and \textsc{GEMMA-7B}. Both models achieve AUC values of 0.996 and 1.000, indicating near-perfect separation between watermarked and non-watermarked content.}
\label{fig:max_z_roc}
\end{center}
\end{figure}

Table~\ref{tab:max_z_performance} reports standard classification metrics at the default operating threshold for maximum $z$-score detection. Results confirm exceptional performance across accuracy, precision, recall, F1-score, and ROC-AUC, with \textsc{GEMMA-7B} achieving the highest scores in all categories. The method maintains extremely low false positive rates (0.154 for \textsc{OPT-6.7B} and 0.012 for \textsc{GEMMA-7B}) while sustaining near-perfect recall, making it well-suited for deployment scenarios where false accusations of AI generation must be minimized. Figure~\ref{fig:max_z_roc} presents the corresponding ROC curves, illustrating the near-perfect separation between watermarked and non-watermarked content.

\subsubsection{$z$-Score Distribution Analysis} 
\begin{figure}[t]
\begin{center}
\centerline{\includegraphics[width=\columnwidth]{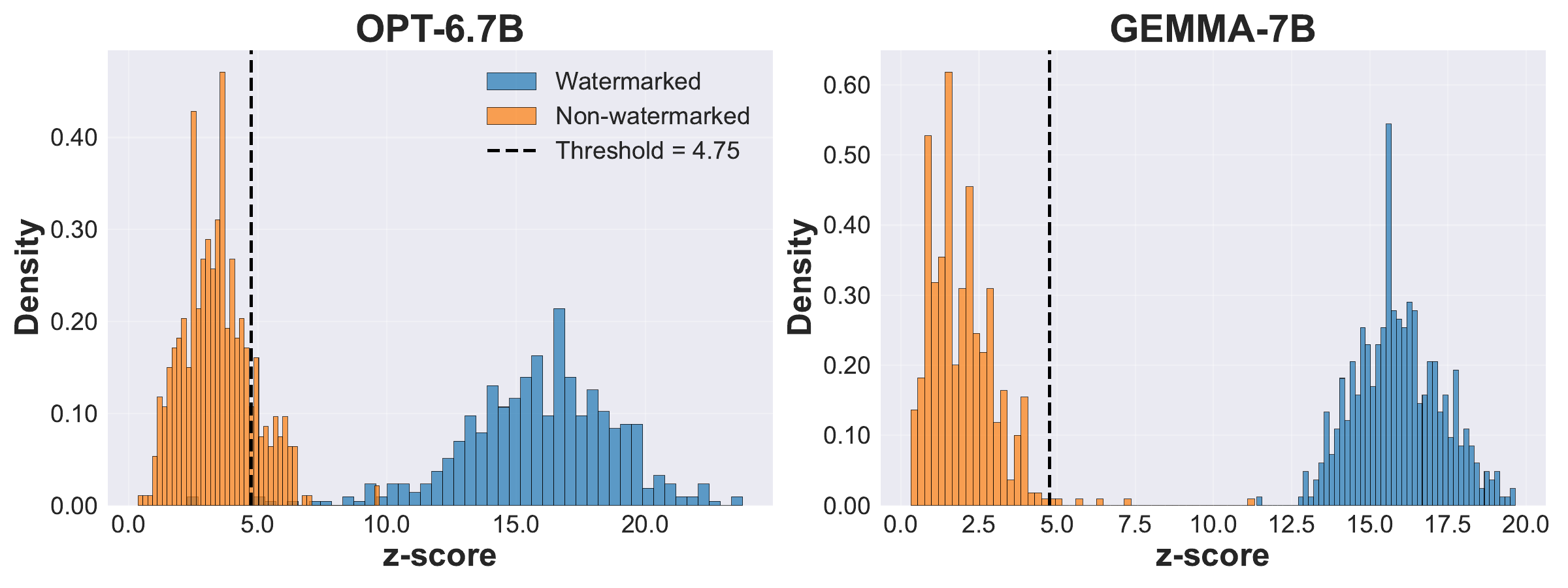}}
\caption{$z$-score distributions for the maximum $z$-score detection method on \textsc{OPT-6.7B} and \textsc{GEMMA-7B}. Both show clear separation between watermarked and non-watermarked text, with \textsc{OPT-6.7B}'s closer overlap explaining its higher false positive rate.}
\label{distribution}
\end{center}
\end{figure}

As discussed in Table \ref{tab:max_z_performance}, the false positive rate (FPR) for \textsc{OPT-6.7B} is higher than for \textsc{GEMMA-7B}, despite both achieving near-perfect recall. To better understand this behavior, we examine the $z$-score distributions for the maximum $z$-score detection method. Figure~\ref{distribution} shows that while both models exhibit clear bimodal separation between watermarked and non-watermarked text, the non-watermarked distribution for \textsc{OPT-6.7B} lies slightly closer to the detection threshold than for \textsc{GEMMA-7B}. This overlap explains the higher FPR observed for \textsc{OPT-6.7B}.

Importantly, this analysis indicates that adjusting the decision threshold upward would reduce false positives for \textsc{OPT-6.7B} without significantly impacting recall, as the watermarked distribution remains well-separated from the non-watermarked one. For \textsc{GEMMA-7B}, the separation is even more pronounced, producing extremely low FPR values under the same threshold. These findings highlight the flexibility of the maximum $z$-score method with strong separation allowing operators to tune detection thresholds to meet application-specific trade-offs between false positives and false negatives. Our sensitivity analysis (\ref{sensitive}) further demonstrates that this separation is maintained across watermark strengths.

\section{Ablation Studies}\label{ablation_studies_from_main_text}

\subsection{TBW Strength Sensitivity}\label{sensitive}
To provide deeper insights into the trade-offs between watermark detectability and text quality, we analyze how varying the bias parameter $\delta$ affects both detection performance and lexical diversity. This analysis addresses key questions about parameter selection and the practical implications of different watermark strengths for deployment scenarios.

\textbf{Experimental Setup.} We conduct this evaluation using \textsc{OPT-6.7B}, which serves as a \textit{worst-case evaluation setting} due to its relatively small vocabulary size (50,272 tokens) compared to \textsc{GEMMA-7B} (256,000 tokens). As demonstrated throughout our evaluation, TBW's performance scales positively with vocabulary size, so any robustness and diversity preservation observed here represents a conservative lower bound for larger models.

We evaluate 100 samples from the C4 dataset with 100-token prompts, generating $200 \pm 5$ tokens. We test three similarity thresholds $\tau \in \{0.3, 0.5, 0.7\}$ across five bias strengths $\delta \in \{0.0, 1.0, 2.0, 5.0, 10.0\}$. Multiple $\tau$ values are included to validate that TBW's semantic partitioning remains stable under both looser and stricter similarity constraints. To quantify lexical diversity, we report Distinct-N metrics ($N = 1, 2, 3, 4$), which measure the ratio of unique n-grams to total n-grams. Higher Distinct-N values indicate greater lexical variety, while lower values suggest increased repetition and potential quality degradation.

\begin{figure}[h]
\begin{center}
\begin{minipage}[b]{\columnwidth}
\centerline{\includegraphics[width=0.9\columnwidth]{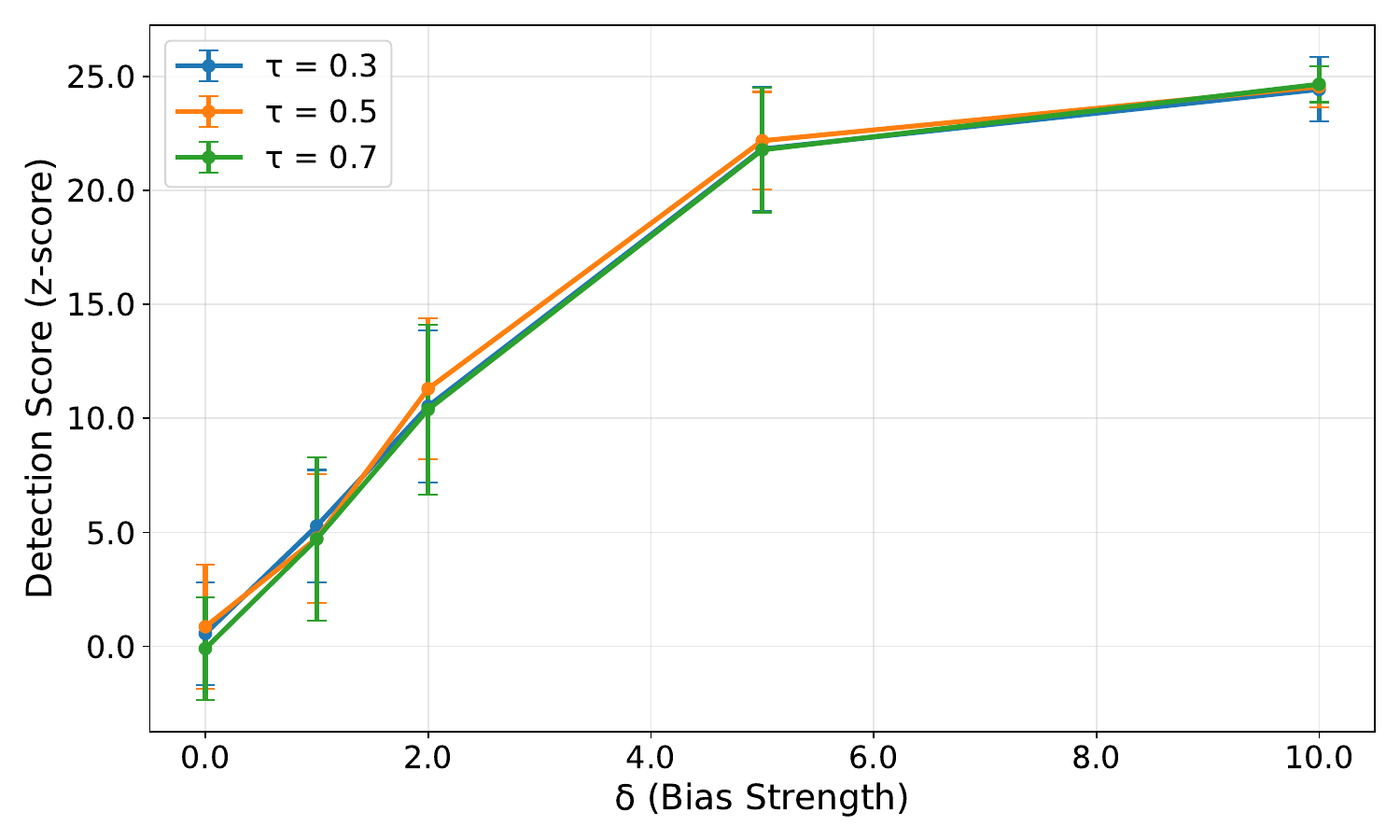}}
\caption{Detection score ($z$-score) vs. bias strength $\delta$ across similarity thresholds. Higher $\delta$ yields stronger watermark signals, with detection saturating around $\delta = 5.0$.}
\label{detectionVdelta}
\end{minipage}
\hfill
\begin{minipage}[b]{\columnwidth}
\centerline{\includegraphics[width=\columnwidth]{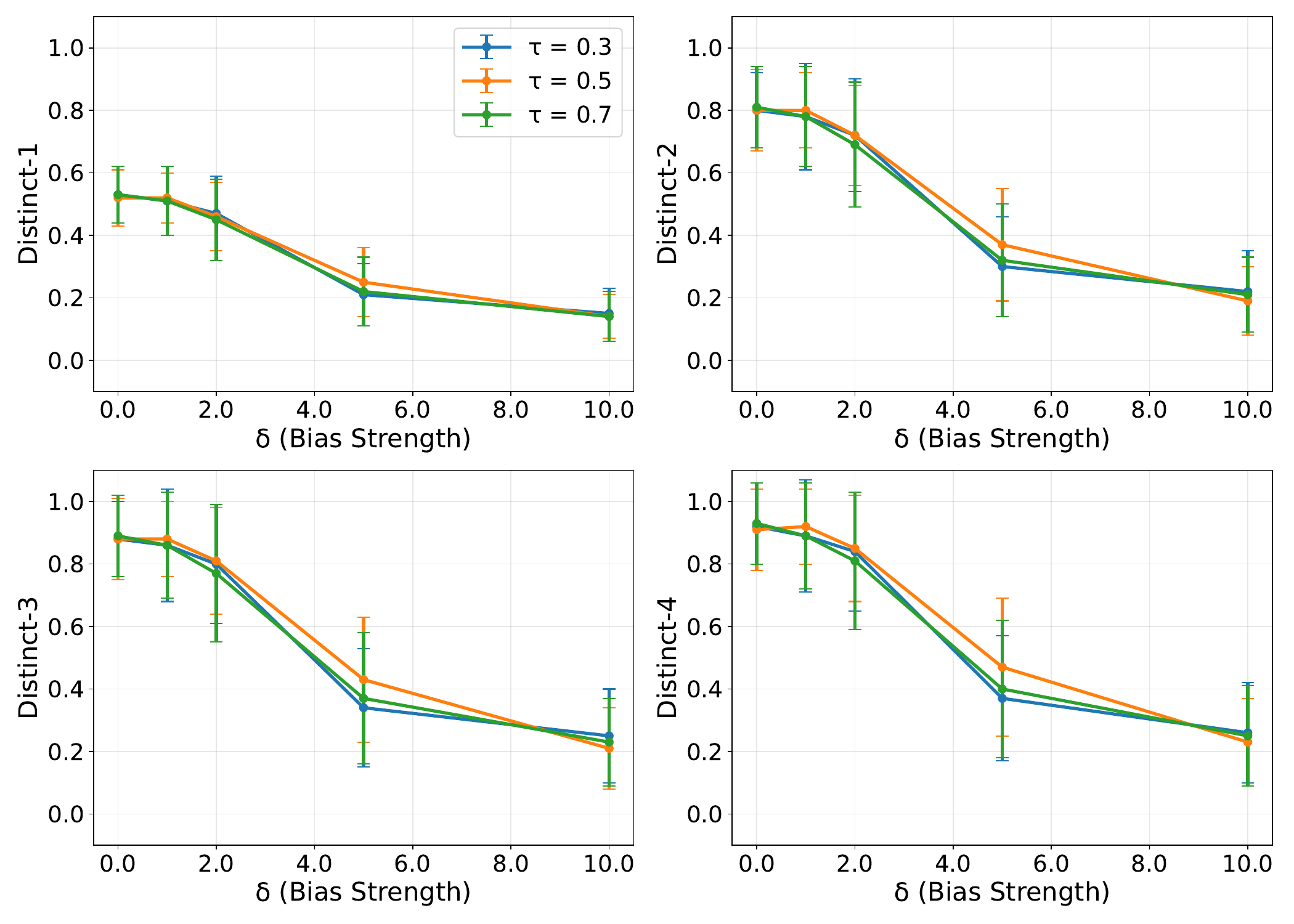}}
\caption{Lexical diversity (Distinct-N) vs. bias strength $\delta$. Moderate $\delta$ values maintain diversity, while very high strengths lead to increased repetition.}
\label{distinctVdelta}
\end{minipage}
\end{center}
\end{figure}

\textbf{Detection-Quality Trade-off.}
Figures~\ref{detectionVdelta} and~\ref{distinctVdelta} illustrate the core trade-off in watermark strength. As $\delta$ increases, detection scores rise monotonically across all $\tau$ values, reaching $z$-scores above 24 at $\delta = 10.0$ (Figure~\ref{detectionVdelta}). However, higher strengths gradually reduce lexical diversity (Figure~\ref{distinctVdelta}), with the most notable changes occurring between $\delta = 2.0$ and $\delta = 5.0$. This pattern underscores the importance of moderate watermark strengths for balancing detectability and output quality. Notably, $\tau$ has minimal influence on either detection performance or diversity, indicating that semantic alignment can be optimized independently of other quality metrics. Additional visualizations covering the full parameter space are provided in~\ref{parameter_space}.

\textbf{Practical Recommendations.}
For most deployment scenarios, we recommend $\delta = 2.0$-$3.0$, which sits just before the diversity inflection point in Figure~\ref{distinctVdelta} and well past the detection saturation zone in Figure~\ref{detectionVdelta}. In this range, TBW consistently achieves strong detection performance ($z$-score $> 10$) while preserving high lexical variety (Distinct-N $> 0.7$). A similarity threshold of $\tau = 0.7$ provides optimal semantic coherence without sacrificing detection capability.

\textbf{Overall Comparison.}
Across all metrics, TBW strikes the most favorable balance relative to prior watermarking schemes. In terms of accuracy, TBW's maximum $z$-score detection consistently achieves near-perfect ROC-AUC ($>0.99$), remaining competitive with more expensive multi-pass schemes. On robustness, TBW substantially outperforms lightweight methods such as SynthID and KGW under both lexical perturbation and full-text paraphrasing, closing much of the gap to heavyweight approaches (e.g., EXP, ITS-Edit) without their quality degradation. Finally, in efficiency, TBW introduces negligible generation overhead, matching production-ready systems while delivering robustness gains that prior efficient schemes cannot achieve. Taken together, these results underscore TBW's practicality for real-world deployment.

\begin{figure*}[t]
    \centering
    \includegraphics[width=2.0\columnwidth]{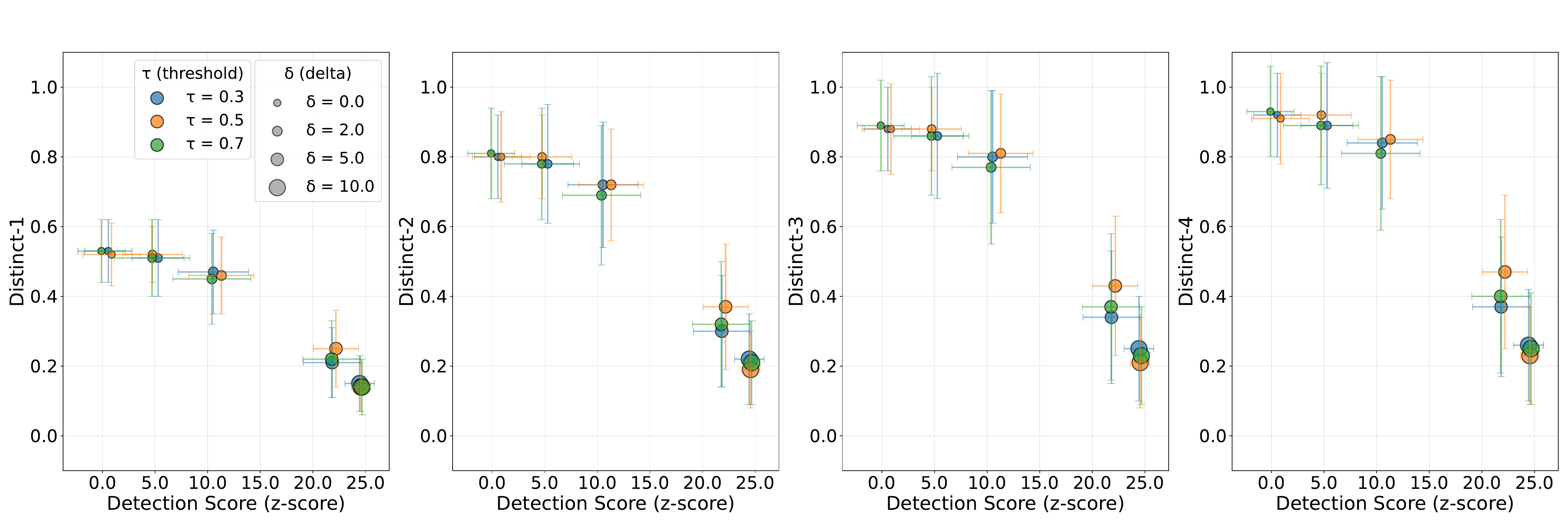}
    \caption{Distinct-N scores vs. detection scores across parameter combinations. The relationship demonstrates consistent quality degradation as watermark strength increases, with minimal threshold influence.}
    \label{scatter_plot}
\end{figure*}

\begin{figure*}[t]
    \centering
    \includegraphics[width=2.0\columnwidth]{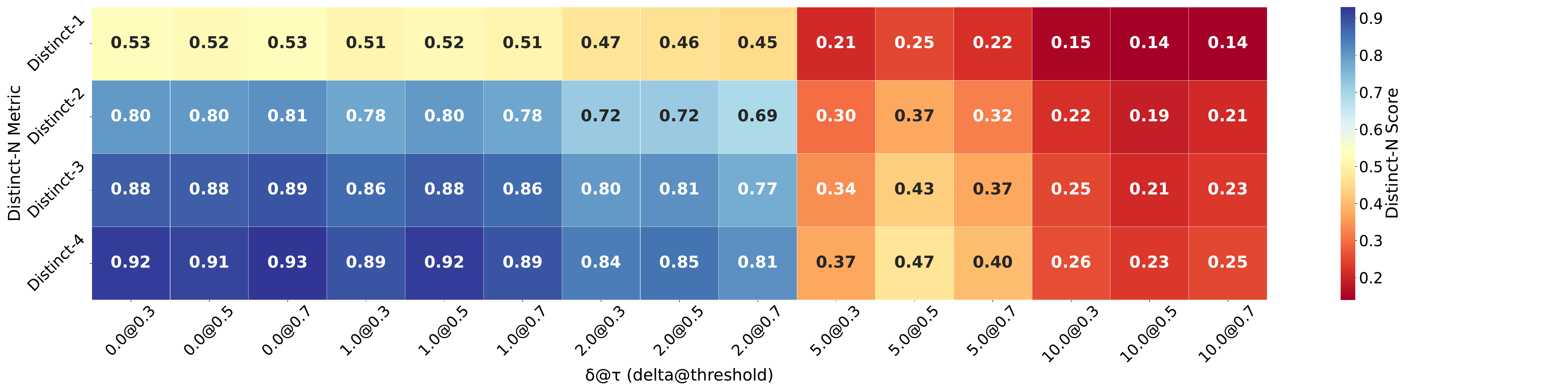}
    \caption{Comprehensive heat map of Distinct-N metrics across bias strength $\delta$ and similarity threshold $\tau$ combinations.}
    \label{heatmap_plot}
\end{figure*}

\subsection{Parameter Space Analysis}\label{parameter_space}
Figures~\ref{scatter_plot} and~\ref{heatmap_plot} provide complementary views of the complete parameter space. The scatter plots in Figure~\ref{scatter_plot} reveal that as detection scores increase with higher $\delta$ values, lexical diversity consistently decreases across all Distinct-N metrics, with threshold values showing negligible variation. This relationship is further confirmed by the heat map in Figure~\ref{heatmap_plot}, which illustrates that lower bias strengths ($\delta \leq 2.0$) maintain high Distinct-N scores ($>0.7$) across all thresholds, while aggressive watermarking ($\delta \geq 5.0$) results in substantial quality degradation (Distinct-N $< 0.5$). The vertical consistency in the heat map reinforces that similarity threshold $\tau$ can be selected based on semantic considerations without significantly affecting lexical diversity.

\subsection{Task-Specific Evaluation on Summarization}\label{task_specific_ablation}
To evaluate TBW's performance on downstream tasks beyond general text generation, we conduct a targeted experiment on scientific document summarization. This evaluation addresses a critical question for practical deployment on whether topic-based watermarking maintains effectiveness when applied to domain-specific, constrained generation tasks that require both semantic coherence and factual accuracy.

\subsubsection{Experimental Setup} We use the TLDR dataset~\cite{cachola-etal-2020-tldr}, which consists of arXiv abstracts paired with reference one-sentence summaries, providing a summarization task within well-defined scientific domains. For this evaluation, we define five domain-specific topics aligned with the dataset's content: \{\texttt{physics}, \texttt{chemistry}, \texttt{mathematics}, \texttt{biology}, and \texttt{computer science}\}.

Using \textsc{OPT-6.7B}, we condition the model on scientific abstracts and prompt it to generate single-sentence summaries of $50 \pm 5$ tokens. This constrained generation setting tests whether TBW can maintain watermark detectability while preserving task-specific quality in a domain where precision and semantic accuracy are paramount. We evaluate summarization quality using ROUGE-1, ROUGE-2, and ROUGE-L metrics, which measure n-gram overlap and longest common subsequence matching between generated and reference summaries, providing standard benchmarks for summarization performance. We employ our standard similarity threshold $\tau = 0.7$ aligned with our main evaluations.

\subsubsection{Quality and Robustness Results} Table~\ref{tab:summarization_rouge} presents the summarization quality comparison between watermarked and non-watermarked outputs. TBW summaries demonstrate superior performance across all ROUGE metrics compared to non-watermarked baselines.
\begin{table}[h]
\small
\centering
\begin{tabular}{lcc}
\toprule
\textbf{Metric} & \textbf{TBW} & \textbf{Non-Watermarked} \\
\midrule
ROUGE-1 & $\mathbf{0.190 \pm 0.111}$ & 0.158 ± 0.098 \\
ROUGE-2 & $\mathbf{0.044 \pm 0.073}$ & 0.018 ± 0.035 \\
ROUGE-L & $\mathbf{0.131 \pm 0.075}$ & 0.112 ± 0.070 \\
\bottomrule
\end{tabular}
\caption{Summarization quality comparison on TLDR dataset using \textsc{OPT-6.7B}. Bold indicates better performance.}
\label{tab:summarization_rouge}
\end{table}

The consistent quality improvements across all ROUGE metrics suggest that TBW's topic-aware token selection provides benefits for domain-specific tasks. We attribute these improvements to TBW's semantic biasing mechanism, which guides token selection toward domain-appropriate vocabularies during generation. By favoring tokens semantically aligned with scientific topics, the watermarking process promotes technical terminology and conceptual relationships that better match reference summary patterns.

\begin{table}[!htbp]
\centering
\begin{tabular}{lc}
\toprule
\textbf{Detection Metric} & \textbf{Value} \\
\midrule
AUC-ROC & 0.946 \\
TPR@1\%FPR & 0.860 \\
TPR@10\%FPR & 0.880 \\
\bottomrule
\end{tabular}
\caption{TBW detection performance on scientific summarization task using \textsc{OPT-6.7B}.}
\label{tab:summarization_detection}
\end{table}

Table~\ref{tab:summarization_detection} demonstrates that these quality improvements do not compromise watermark detectability. TBW maintains robust detection performance with AUC-ROC of 0.946 and high TPR values at stringent false positive thresholds, confirming reliable detection capabilities suitable for production deployment on specialized tasks.

\subsection{Scaling the Number of Topics}\label{ablation:num_topics}
This section provides full experimental details and additional visualizations for the topic scalability analysis presented in \S\ref{topic_scalability}.

\subsubsection{Experimental Setup.}
Unless noted here, we follow the protocol used in the main experiments. We use \textsc{Gemma-7B} because its larger vocabulary is more representative of deployed LLMs than \textsc{OPT-6.7B}, and because scaling $K$ increases partition granularity. We sample 100 prompts from C4, taking the first 100 tokens as the input prompt and generating $200\pm5$ tokens. For watermarking we apply a logit bias $\delta=2.0$ and a semantic threshold $\tau=0.5$ when constructing topic token lists. These settings are relaxed relative to the main experiments to compensate for smaller per-topic coverage at larger $K$ (i.e., admitting more tokens per topic while avoiding over-biasing). Detection uses the maximum-$z$ scheme from the main paper. For text quality we report BERTScore F1~\cite{zhang2020bertscoreevaluatingtextgeneration}, following prior work~\cite{zhang2024remarkllmrobustefficientwatermarking}.

We adopt generic, broad-coverage topic inventories designed to be distinct rather than near-synonymous, starting from the four topics used in the main experiments:
\begin{itemize}
\item $K=4$: \{\texttt{animals}, \texttt{technology}, \texttt{sports}, \texttt{medicine}\}
\item $K=8$: $K{=}4$ $+$ \{\texttt{politics}, \texttt{entertainment}, \texttt{education}, \texttt{finance}\}
\item $K=16$: $K{=}8$ $+$ \{\texttt{science}, \texttt{law}, \texttt{food}, \texttt{travel}, \texttt{environment}, \texttt{religion}, \texttt{fashion}, \texttt{history}\}
\item $K=32$: $K{=}16$ $+$ \{\texttt{art}, \texttt{military}, \texttt{gaming}, \texttt{literature}, \texttt{parenting}, \texttt{space}, \texttt{transportation}, \texttt{psychology}, \texttt{agriculture}, \texttt{housing}, \texttt{cryptocurrency}, \texttt{architecture}, \texttt{economics}, \texttt{fitness}, \texttt{relationships}, \texttt{mythology}\}
\end{itemize}
As in the main method, tokens are assigned to a topic if their embedding similarity exceeds $\tau$; unassigned tokens are round-robined to balance coverage.

\subsubsection{Results}

\begin{figure}[!htbp]
\begin{center}
\centerline{\includegraphics[width=\columnwidth]{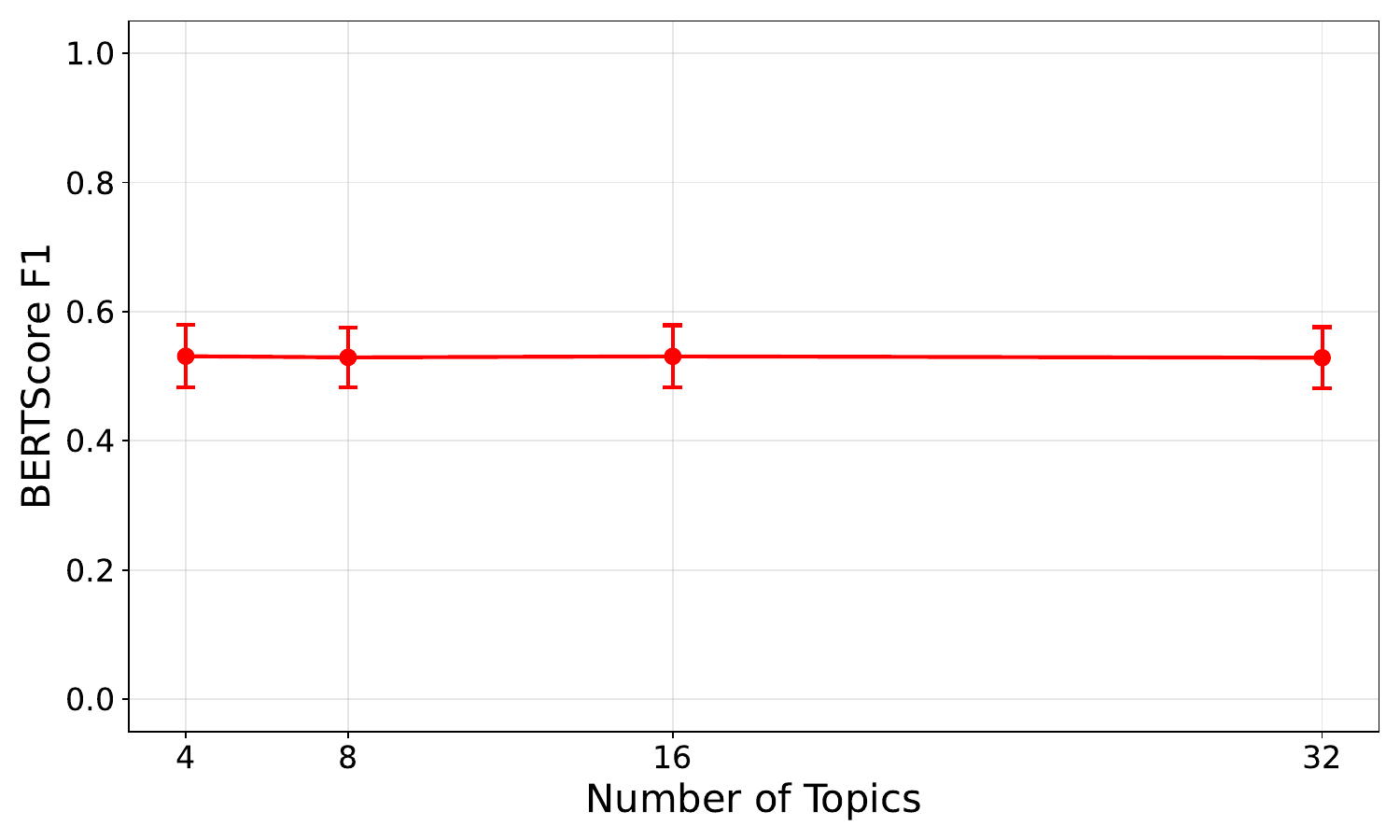}}
\caption{Text quality vs. number of topics. BERTScore F1 remains flat as $K$ increases, indicating no quality degradation. Error bars denote $\pm$s.d.}
\label{bertscore}
\end{center}
\end{figure}

\textbf{Text quality vs. $K$.}
Using the same setup, we assess BERTScore F1 as a text-quality metric~\cite{zhang2024remarkllmrobustefficientwatermarking}. Figure~\ref{bertscore} shows no degradation as $K$ increases with scores remaining flat with low variability. This indicates that relaxing $\tau$ to $0.5$, in order to maintain reasonable per-topic coverage, together with a moderate $\delta{=}2.0$, preserves fluency and semantics.

\begin{figure}[!htbp]
\begin{center}
\centerline{\includegraphics[width=\columnwidth]{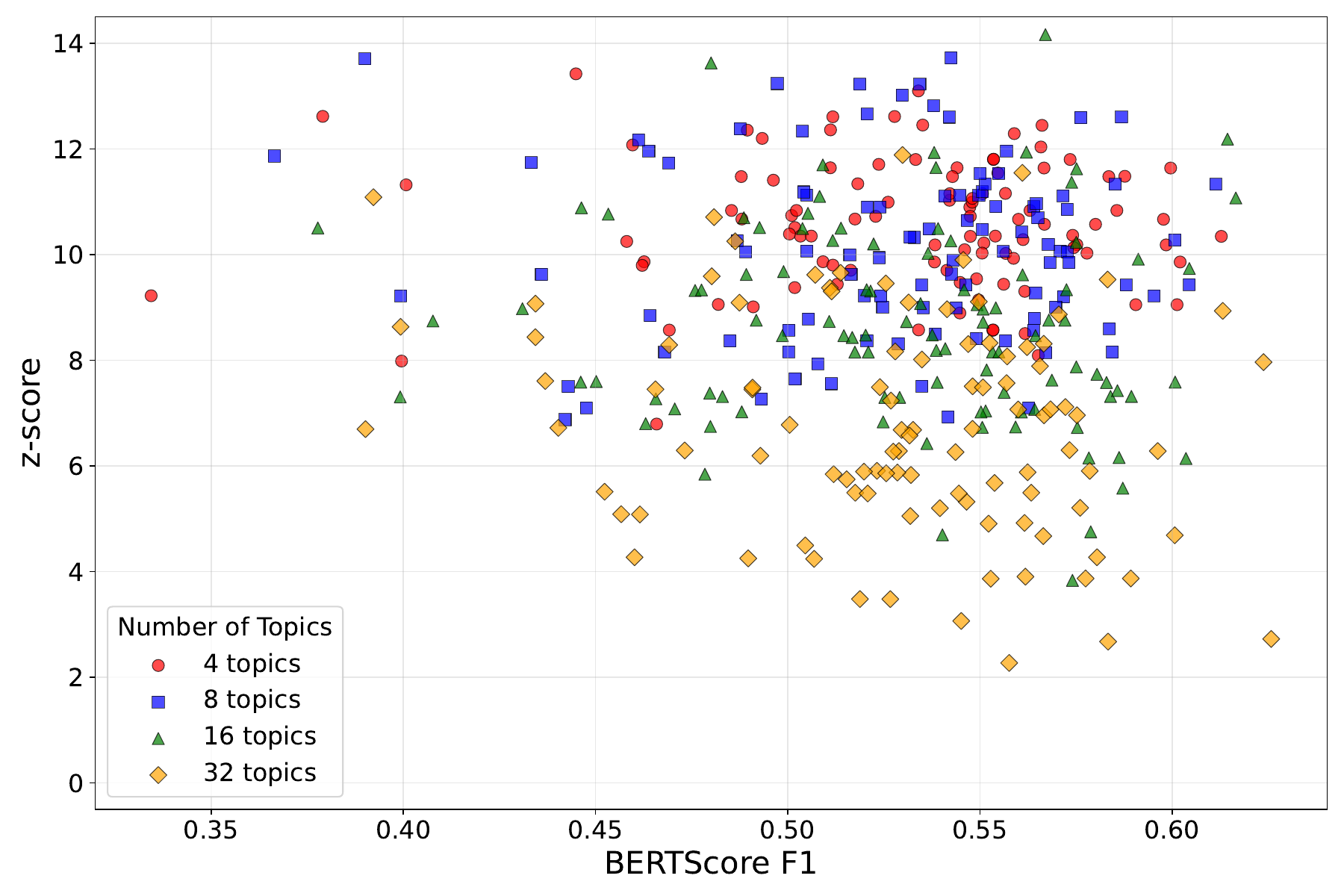}}
\caption{Detection vs. quality trade-off. Per-sample $z$-score (max-$z$) vs. BERTScore F1 across all $K$. Points cluster in $z\in[8,12]$, F1$\in[0.50,0.60]$; $K=32$ shifts modestly lower in $z$ without affecting F1.}
\label{multi_scatters}
\end{center}
\end{figure}

\textbf{Joint view quality vs. strength.} 
To visualize the trade-off directly, Figure~\ref{multi_scatters} scatters per-sample $z$-score against BERTScore F1 across all $K$. Most points cluster in $z\in[8,12]$ and F1$\in[0.50,0.60]$, with the $K=32$ points shifted modestly lower in $z$ (consistent with Figure~\ref{zcore_lists}) but without a quality penalty. Larger $K$ thus modestly weakens the detectable signature while keeping quality stable.

\begin{table*}[t]
\centering
\scriptsize
\begin{tabular}{|p{0.4cm}|p{1.5cm}|p{3cm}|p{3cm}|p{3cm}|p{3cm}|}
\hline
\textbf{idx} & \textbf{Model} & \textbf{Prompt} & \textbf{Real Completion} & \textbf{No Watermark} & \textbf{Watermarked} \\
\hline
27 & \textsc{OPT-6.7B} & [...] is for those updated engines to be able to comply with pollution standards.

The Environment and Public Works panel advanced the measure to the floor of the Senate by voice vote. Carper, the panel's ranking Democrat, 
& had initially authored the bill more than a decade ago with former Ohio Sen. George Voinovich (R).

"With today's vote, we're one step closer to making sure this bipartisan tradition, imagined and inc [...truncated] 
& said the bill would provide an incentive for more than \$1 billion in investments in diesel engine retrofits and would help reduce harmful emissions and improve air quality.
The bill would expand the DERA program to include [...truncated]
& described DERA as a "crucial issue that has failed to get sufficient attention from our colleagues in recent years despite significant environmental and public health concerns."
He explained that about 1 billion tons of [...truncated] \\
\hline
177 & \textsc{OPT-6.7B} & [...] start building a diesel version of the Grand Cherokee next year, its first diesel passenger vehicle in several years.

That confirms a statement by Chrysler CEO Sergio Marchionne from last April that the popular Jeep
& crossover would get a diesel option, which he reiterated in October.
The new clean-diesel sport-utility vehicle was buried in an announcement that Chrysler would add a third production shift at its Jeep[...truncated] 
& 
Grand Cherokee would get a diesel version.

In April, he told reporters "I'm very confident that we're going to build a diesel Grand Cherokee."
Diesels have been slow to take off in the U.S. But the need for [...truncated] 
& Grand Cherokee, which accounts for nearly 30 percent of the U.S. market, will have diesel capability when they launch, and that they already have plans to add diesel four-door Jeeps next year, with more diesel [...truncated] \\
\hline
456 & \textsc{GEMMA-7B} & [...] of minimalism.

For prints isn't something expected from this storied house of minimalism. Monochrome is more its style. Those prints were subtle yet eye-opening on many levels. They showed Mr. Costa was again
& pushing the boundaries of minimalism, this time by proposing a riveting, inspired kind of glam minimalism.
& thinking outside the box, and not just in terms of the printed pieces but in the silhouettes and the new mixing of materials.

The designer created a collection that reflected a New York City woman's [...truncated] 
& taking risks at what could become his signature look (I say signature as the man always seems to hit just the right button). The bold color against these lush materials like herm wool or cashmere looked luxurious but at [...truncated] \\
\hline
233 & \textsc{GEMMA-7B} & [...], it was never feasible.
In brief, the report cited by Axios suggested that the only way to truly secure the next generation of wireless networks, on which critical infrastructure like self-driving cars will rely, against snooping by China
& and others, would be for the government to build that network itself.
There are several things wrong with this idea. You probably thought of a couple before you even got to this sentence.
That would[...truncated] 
& is for the government to build its own network. The idea is that the government would operate the network, rather than private companies, and place it under strict security protocols.
The idea was put forth, according to the report[...truncated] 
& or other hostile powers is if the federal government operates the service itself (there were some suggestions as well to use military bases as anchors for these connections.) As you can imagine, given how many people[...truncated] \\
\hline
\end{tabular}
\caption{Text generation examples comparing non-watermarked and TBW watermarked outputs across \textsc{OPT-6.7B} and \textsc{GEMMA-7B} models with $\delta = 3.0$, $\tau = 0.7$.}
\label{tab:watermark_examples}
\end{table*}

\section{Attack Configuration Details}\label{attack_config}

\subsection{Paraphrasing Attacks}\label{details_on_paraphrasing_attacks}

\textbf{PEGASUS.} We employ PEGASUS~\cite{pegasus}, a sequence-to-sequence model pre-trained on unlabeled text for abstractive summarization. For paraphrasing evaluation, we configure the model with \texttt{num\_sequences=1} and \texttt{num\_beams=3} to generate single paraphrased outputs with controlled beam search diversity.

\noindent\textbf{DIPPER.} We utilize DIPPER~\cite{dipper}, a paraphrase generation model built by fine-tuning T5 for controllable text rewriting. Following the model's control code specification~\cite{dipperCode}, we set lexical diversity to \texttt{lexical=20} and syntactic order diversity to \texttt{order=40}, corresponding to high lexical and syntactic variation (L80-O60 in the original paper's notation).

\subsection{Lexical Perturbation Attacks}\label{details_on_perturbation_attacks}

We implement combination attacks that uniformly distribute perturbations across three edit types: insertion, deletion, and substitution. For each perturbation percentage $p$, we calculate the total number of edits as $\lfloor p \times \frac{n}{100} \rfloor$ where $n$ is the text length, then allocate edits equally across operations (insertion: $\lfloor \text{total}/3 \rfloor$, deletion: $\lfloor \text{total}/3 \rfloor$ , substitution: remainder).

\textbf{Random Perturbations.} Words are selected uniformly at random from the text. Insertions use high-frequency terms from the Reuters corpus~\cite{lewis2004rcv1}, deletions remove arbitrary tokens, and substitutions replace words with random high-frequency alternatives.

\textbf{Targeted Perturbations.} Following the intuition that watermarks may concentrate in semantically important tokens, we target words with Part of Speech (POS) tags indicating nouns, verbs, adjectives, and adverbs (tags beginning with 'N', 'V', 'J', 'R'). Substitutions utilize WordNet~\cite{fellbaum1998wordnet} synonyms when available, while insertions and deletions preferentially operate on this linguistically important subset.

\section{Watermarked Text Examples}\label{examples}
To provide insight into the practical impact of TBW on text generation quality, we present representative examples comparing watermarked and non-watermarked outputs across both evaluated models in Table~\ref{tab:watermark_examples}. These samples demonstrate that TBW maintains semantic coherence and natural language flow while embedding detectable watermark signals. Examples are selected from the C4 evaluation dataset using our standard experimental configuration ($\delta=3.0$, $\tau=0.7$).

\end{document}